\documentclass{article}
\usepackage{tikz}

\usepackage{amsmath, amssymb, amsthm}
\usepackage{algorithm}
\usepackage{algpseudocode}
\usepackage{float, graphicx, multirow,subcaption, setspace}
\usepackage{comment}
\usepackage{url}
\usepackage{enumitem}
\usepackage{tikz}
\usepackage{bm, bbm}
\usetikzlibrary{shapes.geometric, positioning}
\usetikzlibrary{quotes, angles}
\usepackage{rotating}
\usepackage{hyperref}
\usepackage{natbib}
\usepackage[normalem]{ulem}
\usepackage{cleveref}

\definecolor{SkyBlue}{RGB}{14, 118, 188}
\definecolor{BrightRed}{RGB}{223, 82, 78}
\definecolor{Green638}{RGB}{165,255,118} 
\definecolor{FoamGreen}{RGB}{25, 200, 150}

\definecolor{RevisedRed}{RGB}{238, 51, 119}

\newlist{assumenum}{enumerate}{1}
\setlist[assumenum]{label=\textbf{A\theassumenumi}, ref=A\theassumenumi}
\newcounter{assumption}
\crefname{assumenumi}{Assumption}{Assumptions}
\Crefname{assumenumi}{Assumption}{Assumptions}

\newlist{assumenumx}{enumerate}{1}
\setlist[assumenumx]{label=\textbf{Ax\theassumenumxi}, ref=Ax\theassumenumxi}

\newcommand\ci{\perp\!\!\!\perp} 

\newcommand\numberthis{\addtocounter{equation}{1}\tag{\theequation}} 

\newcommand{\E}{\mathbb{E}} 
\def\P{\mathbb{P}} 

\newcommand{\calC}{\mathcal{C}}

\newcommand{\calE}{\mathcal{E}}
\newcommand{\calB}{\mathcal{B}}

\newcommand{\bcalE}{\boldsymbol{\calE}}

\newcommand{\cutset}{\calC}

\newcommand{\ind}[1]{\mathbbm{1}\left( #1 \right)} 

\newcommand{\late}{\textrm{LATE}}
\newcommand{\clate}{\textrm{CLATE}}
\newcommand{\itt}{\textrm{ITT}}
\newcommand{\yobs}{Y^{\textrm{obs}}}
\newcommand{\robs}{R^{\textrm{obs}}}
\newcommand{\normaldist}[2]{\mathcal{N}\left(#1,#2\right)} 
\newcommand{\truncnormaldist}[3]{\mathcal{N}_{#1}\left(#2, #3\right)} 

\newcommand{\by}{\bm{y}}
\newcommand{\bx}{\bm{x}}

\newcommand{\bY}{\bm{Y}}
\newcommand{\bX}{\bm{X}}
\newcommand{\bC}{\bm{C}}

\newcommand{\btheta}{\boldsymbol{\theta}}
\newcommand{\bTheta}{\boldsymbol{\Theta}}

\newcommand{\bxi}{\boldsymbol{\xi}}

\theoremstyle{plain}

\theoremstyle{definition}

\newtheorem{ex}{Example}

\theoremstyle{plain}

\usepackage{fullpage, parskip}
\usepackage{array}
\newcolumntype{L}[1]{>{\raggedright\let\newline\\\arraybackslash\hspace{0pt}}m{#1}}
\newcolumntype{C}[1]{>{\centering\let\newline\\\arraybackslash\hspace{0pt}}m{#1}}
\newcolumntype{R}[1]{>{\raggedleft\let\newline\\\arraybackslash\hspace{0pt}}m{#1}}

\hypersetup{pdfborder = {0 0 0.5 [3 3]}, colorlinks = true, linkcolor = BrightRed, citecolor = SkyBlue}

\newcommand{\includesupp}{1} 
\newcommand{\switchref}[2]{%
  \if\includesupp1%
    #1%
  \else%
    #2%
  \fi%
}
\newcommand{\revise}[1]{\textcolor{black}{#1}}
\newcommand{\reviseTwo}[1]{\textcolor{black}{#1}}

\title{Heterogeneous Treatment Effect Estimation under Noncompliance in the Illinois Workplace Wellness Study with Bayesian Tree Ensembles}
\author{Jared D. Fisher\thanks{Dept. of Statistics, Brigham Young University, \texttt{fisher@stat.byu.edu}}, David W. Puelz\thanks{University of Austin, \texttt{dpuelz@uaustin.org}}, Sameer K. Deshpande\thanks{Dept. of Statistics, University of Wisconsin--Madison, \texttt{sameer.deshpande@wisc.edu}}}
\date{\today}

\begin{document}

\maketitle
\begin{abstract}
Estimating varying treatment effects in randomized trials with noncompliance is inherently challenging since variation comes from two separate sources: variation in the impact itself and variation in the compliance rate. In this setting, existing flexible machine learning methods are sensitive to the weak instruments problem and can yield unstable estimates of heterogeneity when run repeatedly with different initialization or random seeds. Our main methodological contribution is to present a Bayesian Causal Forest model for binary response variables in scenarios with noncompliance.  By repeatedly imputing individuals' compliance types, we can flexibly estimate heterogeneous treatment effects among compliers.  Simulation studies demonstrate the usefulness of our approach when compliance and treatment effects are heterogeneous.  We use this method to detect and analyze heterogeneity in the treatment effects in the Illinois Workplace Wellness Study, which not only features heterogeneous and one-sided compliance but also several binary outcomes of interest.  We focus on three outcomes one year after intervention. We confirm a null effect on the presence of a chronic condition, discover meaningful heterogeneity in the impact of the intervention on metabolic parameters though the average effect is null in classical partial effect estimates, and find heterogeneity in the intervention's effect on individuals' perception of management prioritization of health and safety.
\end{abstract}

\newpage
\section{Introduction}
\label{sec:intro}
\subsection{The effectiveness of workplace wellness programs}
Many employers pay their employees to adopt and maintain healthy habits like regular exercise, dieting, or other wellness practices. 
This pay often comes in the form of rebates of healthcare premiums or compensation in addition to salary.
To employers, healthier employees may be more productive, be happier in their jobs, and incur fewer healthcare costs.
On this view, workplace wellness programs offer potentially high return on investment, benefiting both employer and employees.

To examine the practical effects of workplace wellness programs, a randomized controlled trial was conducted at the University of Illinois at Urbana-Champaign \citep{qje,jama}. 
A workplace wellness program, called iThrive, was offered to the treatment group while a control group was also monitored. 
12,459 benefits-eligible employees were invited to be part of the study, with 4,834 joining the study. 
3,300 of these were randomly assigned to the treatment group and invited to participate in iThrive. 
Financial awards were given for participating in the program, though not all invitees actually participated. 
The other 1,534 subjects were not invited to the program and thus comprise the control group. 
\cite{qje} found that iThrive had no significant effect on 40 out of 42 outcomes related to healthcare spending, job productivity, and health behaviors measured 12 and 24 months after the start of iThrive.
Participation in iThrive did, however, significantly increase the likelihood that participants received a health screening and believed that their managers prioritized employee health and safety.
\cite{jama} extended \cite{qje}'s analysis, finding that iThrive had an insignificant effect on several biometric outcomes, diabetes, hypertension, and hospital visits with two exceptions: participants were more likely to have a primary care physician and to have positive beliefs about their own health.
In this paper, we examine the extent to which the apparent null effects are a product of offsetting positive and negative effects within subgroups of the study population.
That is, we examine the potential heterogeneity of iThrive's \revise{effects}.

\subsection{Noncompliance and treatment effect heterogeneity}
Randomized controlled trials with noncompliance exist somewhere between observational studies and randomized experiments.
The intervention is indeed randomly assigned but is imperfectly taken by the subjects. 
Two common approaches to deal with noncompliance are instrumental variable techniques and principal stratification. 
For trials with noncompliance, the random assignment acts as the instrument if it is only correlated with outcomes through the actual receipt of treatment. 
Principal stratification refers to identifying different subgroups of interest based on post-randomization variables, such as the actual receipt of treatment.
Formalized by \citet{frangakis2002principal}, these subgroups, called principal strata, can have different treatment effects; 
see \citet{page2015principal} for an overview of methods for estimating these subgroup effects.

Treatment effect heterogeneity can also complicate the detection of population treatment effects. 
It could be the case that some subjects have a positive treatment effect, while others have an opposite, negative treatment effect, resulting in a null population average effect.
These two halves of the population likely differ in important ways, and the variables that describe differences in effects are called moderators. 
When the moderators are not known \textit{a priori}, they must be discovered while estimating the treatment effect heterogeneity.
In recent years, tree-based machine learning methods have emerged as popular and effective approaches for this task. 
Three, in particular, stand out. 
\citet{hill2011bayesian} proposed a two-step process: first one uses \revise{Bayesian Additive Regression Trees} \citep[BART;][]{bart} to estimate the conditional expectation of the outcome given treatment and covariates; and second, one estimates causal effects with differences in evaluations of the estimated conditional expectation function.
\citet{bcf} argued that their Bayesian causal forests (BCF) model, which jointly estimates the prognostic function and treatment effect function with separate BART ensembles, provides more transparent, targeted, and effective regularization of the treatment effects.
Finally, \citet{wagerathey2018_jasa} \revise{introduced a random forests-based estimator termed ``causal forests'' or ``generalized random forests'' (GRF) \citep[see also][]{athey2019_obs}, which displays favorable theoretical properties under certain strong \citep[arguably unrealistic; see ][]{cattaneo2025honest} assumptions and has been quite popular in the empirical heterogeneous treatment effect literature.}

\subsection{Our contributions}

We propose a BCF-based approach for estimating heterogeneous effects of a binary treatment on binary outcomes from randomized trials with one-sided noncompliance.
Our work adds to a growing literature on heterogeneous effect estimation with instruments, principal stratification, and/or noncompliance.
This literature includes \citet{grf}, which presents a random forests-based IV estimator in \S7.
\cite{johnson2022detecting} first identifies potential effect modifiers with interpretable machine learning methods and then tests for modifier's effect on the outcome with matching. 
\citet{mcculloch2021causal} and \citet{bcfiv} both present BART-based extensions of the popular \reviseTwo{two-stage} least squares approach for IV regression.
Before proceeding, we pause to highlight three recent works that also jointly estimate latent principal strata and treatment effects with BART.
\citet{chen2024bayesian} jointly model latent survival class with nested probit BART models and a continuous outcome using BART. 
 \citet{kimZigler2024} jointly model a continuous intermediate variable and a continuous outcome variable with BCF-like models.  
\cite{garraza2024combining} jointly model compliance strata with nested probit BART models and a binary outcome also using probit BART models. 
Whereas this last work separately estimates effects within each principal strata separately, we ``borrow strength'' across the strata. 

In this paper, we jointly model the binary outcome and compliance status of all subjects in the iThrive study.
\revise{In simulation studies, our joint modeling approach produced better point estimates and uncertainty intervals for the} conditional local average treatment effects than other leading methods, which are based on the ratio of the conditional intention-to-treat effect on the outcome to the that of treatment receipt \citep[i.e., a localized Wald estimator;][]{wald1940fitting}.

We found that subjects in the Illinois Workplace Wellness Study who (i) complied with the invitation to participate in the workplace wellness program and (ii) did not self-report high blood pressure, cholesterol, or blood glucose in 2016 were much more likely to self-report high values of these measures in 2017 than those who did not self-report high values in 2016.
This suggests that participation in the programs made compliant subjects more aware of their own health conditions.
We also found heterogeneity in the effect of the program on employees' view about management's health and safety priorities.

This paper proceeds as follows.
In \Cref{sec:conditional_identification}, we define the causal estimand of interest and state the necessary assumptions needed to identify it from the observed data.
We present our estimation strategy in \Cref{sec:estimation} and report the results of several simulation studies comparing our approach to existing methods in \Cref{sec:simulations}.
In \Cref{sec:real_world}, we re-analyze the data from the Illinois Workplace Wellness Study.
We conclude in \Cref{sec:discussion} with a discussion of the strengths and limitations of our re-analysis and outline several methodological extensions.

\section{The Illinois Workplace Wellness Study}
\label{sec:application}
\revise{Workplace wellness programs involve employers incentivizing employees to participate in behaviors that support health and wellness. These can include smoking cessation programs, lifestyle and behavioral coaching, and weight-loss programs \citep{kff2024ehbs}. 
The employer's goal is typically to lower medical costs while increasing both employees' productivity and well-being. }

\revise{The Affordable Care Act \citep{ACA2010} is often thought of as the catalyst for the increased prevalence of workplace wellness programs as it raised the maximum limit of financial incentives employees can offer \citep{jama}.  
In 2024, a large survey by \cite{kff2024ehbs} found that 54\% of small firms and 79\% of large firms offering health benefits have some kind of wellness program, where large firms have 200 or more workers. 
Others have found this number closer to 90\% \citep{Wieczner2013WSJ}.}

\revise{
Despite the widespread prevalence of these programs, empirical evidence of their efficacy is mixed. 
In a meta-analysis, \cite{baicker2010workplace} found cost savings in both medical costs and absenteeism. 
\cite{gowrisankaran2013hospital} analyzed the impact of a large employer's new wellness program, which saw a decrease in hospitalizations but no decrease in costs within two years. 
In another meta-analysis, \cite{baxter2014relationship} found that while there appears to be a positive \reviseTwo{return on investment}, higher quality studies report smaller financial returns. 
\cite{gubler2018doing} studied a small employer's wellness program and found that workers that improved their health also improved their productivity. 
\cite{levy2019workplace} found that subjects' health outcomes improved during the 10-week program but no significant difference in healthcare expenses one year later. 
\cite{song2021health} analyzed a three-year program, finding that while self-reported health behaviors improved, there was minimal change in health or economic outcomes. 
}

\revise{We focus on the Illinois Workplace Wellness Study, a large randomized controlled trial conducted at the University of Illinois at Urbana-Champaign \citep{qje, jama}. 
The study offers two major opportunities for analysis. 
First, its design and scale make it highly relevant for evaluating workplace wellness programs. 
Second, it contains a rich set of covariates, outcomes, and follow-up survey measures.}

\revise{The details of the experiment are as follows. 
A workplace wellness program, called iThrive, was offered to the treatment group while a control group was also monitored. 
12,459 benefits-eligible employees were invited to be part of the study, with 4,834 joining the study. 
3,300 of these were randomly assigned to the treatment group and invited to participate in iThrive,  though not all invitees actually participated. 
Organizers offered financial awards to the treated subjects for completing the iThrive program and conducted multiple annual survey waves. 
The other 1,534 subjects were not invited to the program and thus comprise the control group. 
Subjects had equal probability of being assigned to the treatment group: $3300/4834 \approx 0.683$. 
Subjects were chosen within strata that were defined by five variables of interest \citep[Appendix D.1.2]{qje}.} 

\revise{\cite{qje} analyzed healthcare spending, job productivity, and health behaviors for 12 and 24 months after the start of the program. 
They found that iThrive had no significant effect on 40 out of 42 outcomes related to healthcare spending, job productivity, and health behaviors measured 12 and 24 months after the start of iThrive.
The two significant effects of participation in iThrive increases in the chances that participants received a health screening and increases in the probability that subjects believe that their managers prioritized employee health and safety.
\cite{jama} extended \cite{qje}'s analysis, finding that iThrive had an insignificant effect on several biometric outcomes, diabetes, hypertension, and hospital visits with two exceptions: participants were more likely to have a primary care physician and to have positive beliefs about their own health.}

\revise{Given the large investment of time and money that both employers and employees are putting into workplace wellness programs, there should be a way that best benefits peoples health and wellness, which would naturally benefit employers by lowering healthcare costs and increasing employee productivity. 
The aforementioned  studies prompt important questions.
It makes sense that improving wellness should decrease healthcare costs, but typically studies look at overall effects within a 1-3 year window and often see only small impacts. 
How much do these effects \reviseTwo{vary}  across type of employer and type of wellness program? How heterogeneous is the impact of wellness programs across different people?  If these effects are indeed heterogeneous, what implications are there for developing different types of programs to help different of people?}

\revise{In this paper, we examine the extent to which the apparent null effects reported in the Illinois Workplace Wellness studies are a product of offsetting positive and negative effects within subgroups of the study population.
That is, we examine the potential heterogeneity of iThrive's effect on health outcomes. }

\revise{We study the anonymized public-use data from the Illinois Workplace Wellness Study, which are available online.\footnote{Data and codebooks from iThrive studies are available at \href{https://github.com/reifjulian/illinois-wellness-data}{https://github.com/reifjulian/illinois-wellness-data}.}   
\switchref{\Cref{tab:covariate_tab}}{Table E1 in the Appendix} gives summary statistics of the covariates, which were measured in the baseline survey administered in 2016. 
These include demographics (e.g., age, race, sex); health behavior and health history (e.g., smoking status and chronic conditions); and measures of job satisfaction.
All are binary except for age, which was discretized for privacy (less than 37, 37--49, greater than 49 years old).}

\revise{We focused on estimating the effect of participation in iThrive on three binary indicator outcomes that were measured in the 2017 follow-up survey: (i) whether subjects reported at least one chronic condition; (ii) whether subjects reported having high blood pressure, cholesterol, or blood glucose levels; and (iii) whether subjects believed that their health and safety was prioritized by their managers. 
See \switchref{\Cref{tab:outcome_tab}}{Table E2 in the Appendix} for summary statistics about these outcomes. 
Note that only 3,567 of the 4834 subjects participating in the study completed the 2017 follow-up survey.}
 
\revise{The public-use dataset also recorded whether subjects were invited to iThrive and whether they completed an online health risk assessment in 2016, which was required before they could participate in iThrive.
Following \citet{qje}, we take completion of this assessment as a proxy for participation in the iThrive.
Of the 3,330 subjects invited to iThrive, 1,848 completed the assessment, such that only around half of the subjects assigned to treatment were considered as having received treatment. 
Hence, the methodology used for analyzing this study builds on the literature around noncompliance.}

\section{Identification, heterogeneity, and noncompliance}
\label{sec:conditional_identification}
For each subject $i = 1, \ldots, n,$ we observe (i) a $p$-dimensional covariate vector $\bx_{i};$ (ii) a binary indicator $A_{i}$ recording whether they were invited to participate in the workplace wellness program ($A_{i} = 1$) or not ($A_i=0$); (iii) a binary indicator $R^{\text{obs}}_{i}\in \{0,1\}$ recording whether they actually participated in the program or not; and (iv) a binary outcome $Y^{\text{obs}}_{i} \in \{0,1\}.$
We wish to estimate how the effects of participating in the workplace wellness program on several binary health outcomes vary with respect to $\bx.$ 
In doing so, we hope to identify who benefits the most from workplace wellness programs and by how much. 

For each subject $i$, let $R_{i}(1)$ and $R_{i}(0)$ denote two potential binary ``received treatments'' under each treatment assignment.
Using the terminology from \citet{angrist1996identification}, we can stratify our study population into four groups based on the value of $\{R_{i}(0), R_{i}(1)\}$: (i) never-takers $\{0,0\}$; (ii) compliers $\{0,1\}$; (iii) defiers $\{1,0\}$; and (iv) always-takers $\{1,1\}$.
In the workplace wellness study, subjects were not able to participate in iThrive without an invitation.
Thus, for every subject $i,$ we have $R_{i}(0) = 0,$ which implies that there are only never-takers and compliers in our study.
\revise{Like \citet{chib2008analysis}, we} introduce the binary indicator $C_{i},$ which encodes whether subject $i$ is a complier ($C_{i} = 1$) or a never-taker ($C_{i} = 0$). 
This means that $C_i = R_i(1)$ and that $R_i(0)$ provides no information about subject $i$'s compliance status. 
Finally, for each pair $(A_i, R_i(A_i)),$ we can define a potential outcome $Y_i(A_i, R_i(A_i))$ to be the outcome observed if a subject was assigned treatment $A_i$ and received treatment $R_i(A_i).$
Each subject in our study has three potential outcomes: (i) the outcome that would be observed if they were invited to participate and did participate in the workplace wellness program $Y_i(1,1)$; (ii) the outcome if they were invited to participate but ultimately did not participate $Y_i(1,0)$; and (iii) the outcome if they were not invited to participate $Y_i(0,0).$
Implicit in our notation is the standard ``stable unit treatment value assumption'' (SUTVA; hereafter \Cref{assum:sutva}).
\Cref{assum:onesided} encodes our one-sided noncompliance assumption.

\begin{assumenum}
\item{\textit{(SUTVA)}: \revise{Following \citet[\S 23.3]{imbens2015causal}, there are no alternate forms of the treatment received and there is no interference. So, each subject's potential outcomes do not vary with other subjects' treatment assignments.
\label{assum:sutva}}}
\item{\textit{(One-sided Noncompliance)}: for all subjects $i$, $R_{i}(0) = 0$. \label{assum:onesided}}
\end{assumenum}
\Cref{assum:onesided} is stronger than the conventional monotonicity assumption from the principal stratification literature \citep[\S 24.5.2]{imbens2015causal} as it also precludes always-takers. 

Momentarily leaving aside issues of heterogeneity, to determine whether the workplace wellness program has an effect on health outcomes, we could 
na\"{i}vely compare the average responses amongst the treated and control subjects:
\begin{equation}
\label{eq:naive1}
\frac{\sum_{i}{\yobs_{i}\times \ind{A_{i} = 1}}}{\sum_{i}{\ind{A_{i} = 1}}} - \frac{\sum_{i}{\yobs_{i} \times \ind{A_{i} = 0}}}{\sum_{i}{\ind{A_{i} = 0}}}.
\end{equation}
Although the comparison is unconfounded due to the random treatment assignment, it can be misleading because not every subject assigned to treatment actually receives it.
If the workplace wellness program truly improved health outcomes, then the presence of non-compliers in the treatment arm may deflate the effect estimated by \Cref{eq:naive1}.
Put differently, the comparison suggested by \Cref{eq:naive1} will produce an unbiased estimate of the \emph{intention-to-treat} effect $\itt_{Y}$ (i.e., the effect of randomization to treatment) but will generally yield a biased estimate of the actual effect of treatment received.

To overcome this bias, it is tempting to consider the average response amongst treated subjects who actually received treatment to the average response amongst the control group:
 \begin{equation}
\label{eq:naive2}
\frac{\sum_{i}{\yobs_{i} \times \ind{\text{$A_{i} = 1$ \& $\robs_{i} = 1$}}}}{\sum_{i}{\ind{ \text{$A_{i} = 1$ \& $R_{i} = 1$}}}} - \frac{\sum_{i}{\yobs_{i} \times \ind{A_{i} = 0}}}{\sum_{i}{\ind{A_{i} = 0}}}.
\end{equation}
Though intuitive, the estimator in \Cref{eq:naive2} remains unsatisfying in face of noncompliance, as it implicitly assumes that all control subjects would have participated in the wellness program had they been invited.
Instead, the ideal comparison is between (i) treated subjects who received treatment and (ii) control subjects who would have received treatment had they been assigned treatment.
Unfortunately, we cannot directly carry out the ideal comparison, as it involves reasoning about counterfactual outcomes that would have been observed under a potentially different received treatment. There are, however, methods to estimate this average treatment effect for just the compliers, termed the local average treatment effect (\late) or the complier average causal effect. 

We aim to estimate the \emph{conditional} local average treatment effect ($\clate$) as a function of the covariates: $\clate(\bx) = \E[Y_{i}(1,1) - Y_{i}(0,0) \vert C_{i} = 1, \bX = \bx].$

To identify $\clate(\bx),$ we make the following, relatively standard, assumptions that apply for all subjects $i$ and covariate vectors $\bx.$
\begin{assumenum}[start=3]
\item{\textit{(Exclusion Restriction)}:  Assignment to treatment has no effect on the outcome except through the receipt of treatment: $Y_i(0,0) = Y_i(1,0)$. 
\label{assum:exclusion}\setcounter{assumption}{\value{assumenumi}}
}

\item{\textit{(Unconfoundedness)}: Given $\bx$, potential outcomes and received treatments are independent of assigned treatment: $A_i \ci [R_i(0), R_i(1), Y_i(0,0),  Y_i(1,0),  Y_i(1,1) ] \vert \bX_{i} = \bx.$\label{assum:unconfounded_cond}\setcounter{assumption}{\value{assumenumi}}
}

\item{\textit{(Existence of Compliers)}:    $ 0 < \P(C_i = 1|\bX_i = \bx)$.
\label{assum:compliers_cond}\setcounter{assumption}{\value{assumenumi}}
}

\item{\textit{(Overlap)}:  $0 < \P(A_i = 1|\bX_i = \bx) < 1$.
\label{assum:overlap_cond}\setcounter{assumption}{\value{assumenumi}}
}
\end{assumenum}

\Cref{assum:exclusion} encodes the belief that assignment to treatment affects the outcome only through the actual receipt of treatment. 
\Cref{assum:compliers_cond,assum:overlap_cond} allow estimation to be meaningful, as there are at least some compliers and some units assigned to each treatment arm for all covariate vectors $\bx.$

\revise{\Cref{assum:onesided,assum:unconfounded_cond,assum:overlap_cond} hold trivially for our application. 
Specifically, \Cref{assum:onesided} holds as the treatment could not be accessed by subjects in the control group. 
\Cref{assum:unconfounded_cond,assum:overlap_cond} hold as the treatment was randomly assigned to subjects with equal probability. 
However, we include these assumptions here as they are needed for the formal derivation of our analytical approach.
To the extent that our analysis strategy could be applied in other settings with one-sided noncompliance, it is important to evaluate the appropriateness of these assumptions.
This is especially true in observational studies, where the assumption of unconfoundedness requires careful justification.
We note further that although one-sided noncompliance is uncommon outside of randomized trials, there are examples (see, e.g., \cite{frolich2013identification} and \cite{kennedy2020efficient}).
}

\revise{If we knew the compliance status of each subject, then under \Cref{assum:sutva,assum:onesided,assum:exclusion,assum:unconfounded_cond,assum:compliers_cond,assum:overlap_cond} (as shown in \switchref{\Cref{app:proof_clate}}{Appendix A}), we can identify the $\clate$ as
\begin{equation}
\label{eq:conditional_late_id}
\clate(\bx) = \E[\yobs_{i} \vert C_{i} = 1, A_{i} = 1, \bX_{i} = \bx] - \E[\yobs_{i} \vert C_{i} = 1, A_{i} = 0, \bX_{i} = \bx].
\end{equation}
}
\Cref{eq:conditional_late_id} suggests a natural estimation strategy: we can separately regress the observed outcome onto the covariates amongst the compliers in the treatment arm and amongst those subjects in the control arm who would have complied with treatment had they been assigned to treatment.
Unfortunately, we only observe $\bC^{(1)} = \{C_{i}: A_{i} = 1\}$, the compliance statuses in the treatment arm, but not $\bC^{(0)} = \{C_{i}: A_{i} = 0\},$ the counterfactual compliance statues amongst controls. 
So, we cannot operationalize \Cref{eq:conditional_late_id}, by, for instance, regressing $\yobs_i$ onto $A_i$ and $\bX_i$ using data from the observed and would-be compliers. 

There are two main ways to overcome this challenge.  
A very common approach is based on Wald estimators, which are ratios of different intention-to-treat estimators (\Cref{sec:wald}).
The other approach, which we pursue and detail in \Cref{sec:estimation}, imputes the missing $\bC^{(0)}.$ 

\subsection{The Wald estimator}
\label{sec:wald}
 $\clate(\bx)$ can be estimated with an extension of a Wald estimator \citep{wald1940fitting}, defined as the ratio of two intention-to-treat (\itt) estimates. 
We sketch the derivation below and defer full details to \switchref{\Cref{sec:app_wald}}{Appendix A.3}. 

Under \Cref{assum:sutva} and \Cref{assum:unconfounded_cond}, we can identify $\itt_R$ and $\itt_{Y}$:
\begin{align}
\begin{split}
\label{eq:constant_itt_id}
\itt_{R}(\bx) &:=\E[R_{i}(1)  - R_{i}(0) \vert \bX_{i} = \bx] \\ 
&= \E[\robs_{i} \vert A_{i} = 1, \bX_{i} = \bx] - \E[\robs_{i} \vert A_{i} = 0, \bX_{i} = \bx] \\
\itt_{Y}(\bx) &:= \E[Y_{i}(1, R_{i}(1)) - Y_i(0,R_i(0)) \vert \bX_{i} = \bx] \\
&= \E[\yobs_{i} \vert A_{i} = 1,\bX_{i} = \bx] - \E[\yobs_{i} \vert A_{i} = 0, \bX_{i} = \bx].
\end{split}
\end{align}
Our assumptions allow us to write $\clate(\bx) = \itt_{Y}(\bx)/\itt_{R}(\bx).$ 
If we did not condition on $\bx$, and thereby estimate the local average treatment effect instead of $\clate$, this ratio is identical to the estimator from a two-stage least squares instrumental variables regression with no covariates \citep[\S 4.1.2]{angrist2009mostly}.
Thus, by \Cref{assum:onesided} and the fact that $\yobs_{i}, \robs_{i} \in \{0,1\}$ (as shown in \switchref{\Cref{sec:app_wald}}{Appendix A.3}),
\begin{equation}
\label{eq:conditional_late_iv_estimate}
    \clate(\bx) =  
    \frac{\P(\yobs_{i} = 1 \vert A_{i} = 1, \bX_{i} = \bx) - \P(\yobs_{i} = 1 \vert A_{i} = 0, \bX_{i} = \bx)}{\P(\robs_{i} = 1  \vert A_{i} = 1,  \bX_{i} = \bx)}.
\end{equation}

There are many ways to implement the estimator in \Cref{eq:conditional_late_iv_estimate} by separately modeling the numerator and denominator.
To estimate the numerator, we could use any number of causal machine learning approaches, e.g., \citet{hill2011bayesian}, \citet{kunzel2019metalearners}, \citet{grf}, \citet{bcf}. 
Regardless of how we estimate the numerator, however, the estimator in \Cref{eq:conditional_late_iv_estimate} becomes  \revise{very sensitive to small variations in the denominator whenever} $\P(\robs_{i} = 1 \vert A_{i} = 1, \bX_{i} = \bx)$ is near zero.
So, estimating $\clate(\bx)$ for individuals who are unlikely to be compliers can be fraught. 
This situation is termed the ``weak instrument problem'' \revise{and it is reminiscent of the challenges faced by IPW approaches when there is limited overlap (viz., dividing by very small probabilities.).}

\subsection{The weak instrument problem and the local compliance rate}
\label{sec:weak_instrument}
In instrumental variables analyses, an important consideration is the relevance of the instrument, i.e., the strength of its relationship with the treatment. 
For us, the instrument is the assignment to treatment $A_i$ and the treatment is $R_i(A_i)$. 
If $A_i$ has no causal effect on $R_i(A_i)$, then clearly $\itt_R(\bx) = \E[R_i(1) - R_i(0)\vert \bX_{i} = \bx] = 0$  and the Wald estimator is undefined. 
Under the aforementioned assumptions, as $R_i(1) = C_i$ and $R_i(0)=0$, we see that this denominator is equivalent to the compliance rate $\P(C_i=1\vert \bX_{i} = \bx)$. Thus, a low compliance rate is synonymous with a weak instrument. 

The weak instrument problem is well known \citep{andrews2019weak}, and a compliance rate near zero would be an obvious indicator of a challenged experiment. 
However, we are not only interested in the experiment's compliance rate as heterogeneity in compliance rate may mean that, while the global compliance rate is reasonable, the local compliance rate near certain values of $\bX$ may present a locally weak instrument. 
In Section~\ref{sec:estimation}, we propose an alternative that does not use the standard Wald estimator but instead models compliance like \cite{ratkovic2014strengthening} and \cite{imbens2015causal}. 
\revise{Then, in \Cref{sec:simulation_weak},} we examine, via simulation, what happens when heterogeneous compliance rates yield locally weak instruments.

\section{Estimating the $\clate$ with BART}
\label{sec:estimation}
\revise{To estimate $\clate(\bx)$ in Equation \ref{eq:conditional_late_id}, we essentially must estimate the response surface $\E[Y^{\text{obs}}_i \vert C_i = 1, A_i = a, \bX_i = \bx] = \P(\yobs_{i} = 1 \vert C_{i} = c, A_{i} = a, \bX_{i} = \bx)$} \emph{without observing $\bm{C},$ the full set of compliance statuses.}
\revise{To this end, we specify a \emph{joint} model for $(\yobs, \bC)$ that will allow us to (i) impute the missing compliance statuses and then (ii) conditionally estimated the response surface function $\E[\yobs_{i} \vert C_{i} = c, A_{i} = a, \bX_{i} = \bx].$
Formally, we decompose $\bC = \bC^{(0)} \cup \bC^{(1)}$ into two parts, $\bC^{(0)},$ which is the \emph{observed} compliance statuses for those assigned to the treated group, and $\bC^{(1)},$ which is the \emph{unobserved} compliance statuses for those assigned to control.}
At a high level, 
\revise{letting $\Phi$ denote the standard normal cumulative distribution function,} we model
\begin{align}
\begin{split}
\label{eq:general_model}
\P(Y_{i} = 1 \vert C_{i} = c_{i}, A_{i} = a_{i}, \bX_{i} = \bx_{i}) &= \Phi\left[f(\bx_{i}, c_{i}, a_{i})\right] \\
\P(C_{i} = c_{i} \vert \bX_{i} = \bx_{i}) &= \Phi\left[\eta(\bx_{i})\right],
\end{split}
\end{align}
where $f$ is a to-be-estimated function of covariates, compliance status, and treatment assignment and $\eta$ is a to-be-estimated function of covariates.
To fit this model, we could specify priors for the unknown functions $f$ and $\eta$ and compute posterior summaries using a high-level Gibbs sampler that iterates between (i) imputing the unobserved compliance statuses for those subjects in the control arm conditionally given the observed outcomes, $f$ and $\eta$; and (ii) sampling from the posteriors of $f$ and $\eta$ conditionally fixing the observed and imputed compliance statues and the data. 
\revise{This is similar to the approach of \cite{chib2008analysis}.}

If we, for instance, expressed both $f$ and $\eta$ with regression tree ensembles, the model would be, in some sense, a natural extensions of \citet{hill2011bayesian}'s BART-based approach for estimating heterogeneous causal effects to the one-sided noncompliance setting.
Although the resulting model is quite general, it does not directly parametrize the estimand of interest, $\clate(\bx).$
Indeed, under~\eqref{eq:general_model}, we compute $\clate(\bx) = \Phi[f(\bx, 1, 1)] - \Phi[f(\bx, 1,0)].$
By placing BART priors on $f,$ we explicitly shrink every evaluation $f(\bx, c,a)$ towards zero.
The implied regularization of $\clate(\bx)$ is much less clear in this model.

So, motivated by the arguments of \citet{bcf}, instead of working with the very general formulation in Equation~\eqref{eq:general_model}, we decompose $f$ into three components
$$
f(\bx, c, a) = \mu(\bx) + c \times \mu_{c}(\bx) + c \times a \times \tau(\bx).
$$
\revise{
The function $\mu(\bx)$ captures variability in outcomes of the never-takers:  $ \P(Y = 1 \vert C = 0, A = a, \bX = \bx) =   \Phi\left[\mu(\bx)\right].$
Likewise, the probability that $Y = 1$ among untreated compliers is $\Phi\left[\mu(\bx) + \mu_{c}(\bx)\right].$
So, $\mu_{c}(\bx)$ captures differences between compliers and never-takers in the control arm. 
Similarly, $\tau(\bx)$ captures differences in outcome probabilities between treated and untreated compliers.}

This additive decomposition allows for different levels of regularization for different groups comparisons. 
Regularization on $\mu_c$ encourages the model to model outcomes of untreated compliers similar to never-takers' outcomes. 
Similarly, regularization on $\tau$ encourages treated compliers' outcome probabilities ($\Phi[\mu(\bx) + \mu_{c}(\bx)+ \tau(\bx)]$)  to be similar to those of untreated compliers ($\Phi[\mu(\bx) + \mu_{c}(\bx)]$), which essentially allows us to regularize $\clate(\bx)$. 
Moreover, using a first-order approximation, we can bound $\clate(\bx) = \Phi[\mu_{0}(\bx) + \mu_{c}(\bx) + \tau(\bx)] - \Phi[\mu_{0}(\bx) + \mu_{c}(\bx)]$ by
\begin{equation}
\label{eq:clate_bound}
\clate(\bx) \approx \tau(\bx) \times \phi\left(\mu_{0}(\bx) + c \times \mu_{c}(\bx) \right)\leq (2\pi)^{-1/2}\lvert \tau(\bx)\rvert.
\end{equation}

In summary, we model
\begin{align}
\label{eq:main_model}
\begin{split}
\P(Y_{i} = 1 \vert C_{i} = c_{i}, A_{i} = a_{i}, \bX_{i} = \bx_{i}) &= \Phi\left[\mu(\bx_{i}) + c_{i} \times \mu_{c}(\bx_{i}) + c_{i} \times a_{i} \times \tau(\bx_{i})\right] \\
\P(C_{i} = 1 \vert \bX_{i} = \bx_{i}) &= \Phi\left[\eta(\bx_{i})\right].
\end{split}
\end{align}
We further express each of $\mu,$ $\mu_{c},$ $\tau,$ and $\eta$ with a regression tree ensemble and compute a joint posterior distribution over the missing compliance statuses and the tree ensembles.  
\revise{Since we do not expect a constant compliance rate, we model the compliance rate as  $ \Phi \left[\eta(\bx)\right]$ and allow the regression tree ensemble to detect variability in the compliance rate; such an ensemble would naturally regularize toward the overall compliance rate. }

\revise{We note that different covariates $\bx$ may be used in each of the four BART functions of interest. However, in our application, we wish to explore possible sources of heterogeneity in the CLATE, and thus we include all variables in each function. Should certain variables not be useful for certain functions, the priors we choose encourage selection so that irrelevant covariates are not used the corresponding tree ensemble.}

\subsection{Prior regularization}
\label{sec:prior_reg}
Before specifying the prior and describing our posterior sampling strategy, we assume without loss of generality that all continuous covariates are re-scaled to the interval [0,1] and introduce some additional notation.  
A regression tree $(T, \calB)$ is comprised of a binary decision tree $T$, which consists of collection of internal nodes and a collection of terminal or \textit{leaf nodes}, and a collection $\calB$ of scalar \textit{jumps}, one for each leaf.
Each internal node of $T$ is associated with a decision rule $\{X_{j} \in \calC\}.$ 
When $X_{r}$ is continuous variable, $\calC$ is a half-open interval of the form $[0, c)$ and when $X_{j}$ is categorical, $\calC$ is a subset of the possible levels of that variable.
Given $T,$ we can associate each $\bx$ with a single leaf by tracing a path down the tree by following the decision rules.
Specifically, whenever the path encounters the rule $\{X_{j} \in \calC\},$ it proceeds to the left (resp.\ right) if $x_{j} \in \calC$ (resp.\ $x_{j} \not\in \calC).$
This way, the decision tree $T$ partitions the covariate space into disjoint regions, one for each leaf.
Denoting the jump associated with leaf $\ell$ as $\beta_{\ell},$ the pair $(T, \calB),$ represents a piece-wise constant function that returns $\beta_{\ell}$ for all $\bx$ in the region associated with leaf $\ell.$

For compactness, we use $\calE^{(\mu)} = \{(T_{m}^{\mu}, \calB_{m}^{(\mu)}): m = 1, \ldots, M^{(\mu)}\}$ to denote the ensemble of $M^{(\mu)}$ regression trees used to approximate $\mu.$
We similarly define $\calE^{(\mu_{c})},$ $\calE^{(\tau)},$ and $\calE^{(\eta)}$ for the remaining ensembles.
Like \citet{bcf} and \citet{vcbart}, we specify independent priors for each ensemble in $\bcalE$ and place independent and identical priors on the regression trees within each ensemble.
For ease of exposition, we only describe the regression tree prior for $\calE^{(\mu)}$ as the priors for the remaining ensembles are analogous. 
The regression tree prior consists of three parts: (i) a marginal prior over the structure of the decision tree; (ii) a conditional prior over the decision rules given the tree structure; and (iii) a conditional prior for the jumps $\calB$ given the decision tree.

We can describe the tree structure prior with a branching process in which nodes are added sequentially so that whenever a new node is added at depth $d,$ two children at depth $d+1$ are attached to it with probability $0.95(1 + d)^{-2}.$ 
This prior places overwhelming prior probability on trees of depth five or less. 

Conditional on the tree structure, we sequentially draw decision rules at each non-terminal node in two steps.
 First, we draw the splitting variable index $j \sim \text{Multinomial}(\theta^{(\mu)}_{1}, \ldots, \theta^{(\mu)}_{p}).$
Then, conditional on $j,$ we set $\cutset$ to be a random subset of $\mathcal{A}(j),$ the set of available $X_{j}$ values at the current non-terminal node. 
If $X_{j}$ is continuous, we set $\cutset = [0,c)$ where $c$ is drawn uniformly from the interval $\mathcal{A}(j).$
Otherwise, if $X_{j}$ is categorical, we set $\cutset$ to be a random non-trivial subset of the discrete set $\mathcal{A}(j).$
Following \citet{linero2018dart}, we model $(\theta_{1}^{(\mu)}, \ldots, \theta_{p}^{(\mu)})\sim \text{Dirichlet}(\xi^{(\mu)}/p, \ldots, \xi^{(\mu)}/p)$ and $\xi^{(\mu)}/(\xi^{(\mu)} + p) \sim \text{Beta}(0.5, 1).$ 
\revise{This prior encourages sparsity in which covariates are used. 
However, \cite{linero2018dart} and \cite{vcbart} find that this sparsity-inducing prior does not negatively affect model fit in ``dense'' settings with many relevant covariates.}

Finally, conditional on the decision tree $T$, we respectively place independent normal priors with mean $\beta_{0}^{(\mu)}/M^{(\mu)}$ and variance $\sigma^{2 (\mu)}/M^{(\mu)}$ on the jumps in $\calB,$ where $\beta_{0}^{(\mu)}$ and $\sigma^{(\mu)}$ are fixed constants.
We specify similar priors for the other ensembles $\calE^{(\mu_{c})},$ $\calE^{(\tau)},$ and $\calE^{(\eta)}.$
We recommend using $50$ trees in each ensemble (i.e., \revise{$M^{(\mu)} = M^{(\mu_{c})} = M^{(\tau)} = M^{(\eta)} = 50$}) and the following hyper-parameter settings: $(\beta^{(\mu)}_{0}, \sigma^{(\mu)}) = (\Phi^{-1}(\overline{y}), 1.5)$; $(\beta^{(\eta)}_{0}, \sigma^{(\eta)}) = (\Phi^{-1}(\overline{c}), 1.5)$; $(\beta^{(\mu_{c})}_{0}, \sigma^{(\mu_{c})}) = (0, 0.5);$ and $(\beta^{(\tau)}_{0}, \sigma^{(\tau)}) = (0, 0.5)$ where $\overline{y}$ is the average response in the whole dataset and $\overline{c}$ is the observed compliance rate among treated subjects. 
These choices imply the following marginal priors: $\mu(\bx) \sim \normaldist{\Phi^{-1}(\overline{y})}{1.5^{2}}$; $\eta(\bx) \sim \normaldist{\Phi^{-1}(\overline{c})}{1.5^{2}}$; $\mu_{c}(\bx) \sim \normaldist{0}{0.5^{2}};$ and $\tau(\bx) \sim \normaldist{0}{0.5^{2}}$.

In other words, we shrink each subjects' compliance and outcome probabilities towards the overall averages observed in the dataset, \revise{through $\eta$ and $\mu$ respectively. 
The smaller prior variances for $\mu_{c}$ and $\tau$ reflect the belief that untreated compliers are similar to never-takers and treated compliers. 
\revise{The prior on $\tau$} directly affects our estimation of $\clate(\bx)$.}
Based on \Cref{eq:clate_bound}, our prior places approximate 68\% probability on the event that $\lvert \clate(\bx)\rvert < 0.2.$
In the context of the workplace wellness studies, a 20-percentage point effect is quite large, making the implied prior on $\clate(\bx)$ fairly weakly informative.

\revise{
While there are other justifiable hyperparameter choices,  we find that several other specifications do not improve model performance.
For instance, increasing $M^{(\eta)}$ to allow for a more complex model of compliance,  decreasing $M^{(\tau)}$ for a simpler $\clate(\bx),$ and varying the prior jump variances worsened the predictive fit.
See \switchref{\Cref{sec:simulation_trees,sec:simulation_variances}}{Appendices C.2.1 and C.2.2}).
}

\subsection{Posterior computation}
\label{sec:posterior_computation}
We use a Gibbs sampler to simulate draws from the joint posterior distribution $p(\bm{C}^{(0)}, \bcalE, \bTheta, \bxi, \vert \bY, \bm{C}^{(1)}),$ where $\bTheta = \{\btheta^{(\mu)}, \btheta^{(\mu_{c})}, \btheta^{(\tau)}, \btheta^{(\eta)}\}$ is the collection of prior splitting probabilities for each ensemble and $\bxi = \{\xi^{(\mu)}, \xi^{(\mu_{c})}, \xi^{(\tau)}, \xi^{(\eta)}\}$ are the corresponding prior hyperparameters. 
We provide a high-level overview of the sampler here and present a detailed derivation in \switchref{\Cref{app:gibbs_sampler}}{Appendix B}.
Each iteration of our sampler involves four steps.
In the first step, we draw the missing compliance statuses $\bm{C}^{(0)}$ conditionally given the ensembles $\bcalE$ and $\bY.$
The missing $C_{i}$'s are conditionally independent of one another given $\bcalE$ and $\bY$ (see \switchref{\Cref{eq:compliance_post_density,eq:compliance_conditional}}{Equations (B1) and (B2) in Appendix B}). 

After drawing the missing $\bm{C}^{(0)}$, we sequentially update each regression tree in $\bcalE$ in the second step of our sampler.
To facilitate these updates, we introduce latent utilities $\tilde{Y}_{i}$ and $\tilde{C}_{i}$ such that $Y_{i} = \ind{\tilde{Y}_{i} \geq 0}$ and $C_{i} = \ind{\tilde{C}_{i}  \geq 0}.$
Conditional on these latent utilities, we sweep over the trees in $\bcalE,$ updating them one at a time while holding the others fixed.
Each individual tree update proceeds in two steps.
First, we draw a new tree structure from its \emph{marginal} distribution using a Metropolis-Hastings step in which proposals are drawn by randomly growing or pruning the current tree structure.
Then, we draw a new collection of jumps $\calB$ from its conditional distribution given the new tree structure and all other regression trees. 
Essentially, these tree updates are a weighted version of \citet{bart}'s original Bayesian backfitting strategy that was also used in \cite{bcf} and \citet{vcbart}. 
After updating each regression tree in $\bcalE,$ we sample new values of each vector in $\bTheta$ conditionally on $\boldsymbol{\eta}$ using a conjugate Dirichlet-Multinomial update before drawing new values of $\boldsymbol{\xi}$ using an independence Metropolis step.

\reviseTwo{
In the applied analysis, we ran 10 Markov chains for $n_{\textrm{iter}}$ total iterations, discarding the first $n_{\textrm{burn}}=1000$ iterations of each chain as burn-in and retaining every $n_{\textrm{thin}}$th post-burn-in draw. 
We iteratively increased $n_{\textrm{iter}}$ and adjusted $n_{\textrm{thin}}$ so that (i) we obtained 10,000 retained posterior draws across all chains and (ii) the maximum $\widehat R$ across all subjects' $\clate(\bx_i)$ evaluations was less than 1.05. We chose to monitor the pointwise $\clate(\bx_i)$ values because they are the building blocks for the overall and subgroup treatment-effect summaries. We use the $\widehat R<1.05$ criterion as a practical screening rule for these pointwise diagnostics, rather than as the final diagnostic standard for reported estimands. 
For the final scalar estimands, namely the overall LATEs, subgroup LATEs, and subgroup contrasts, we assessed convergence using the stricter recommendations of \citet{vehtari2021rank}: $\widehat R<1.01$, together with bulk and tail effective sample sizes greater than 400.  
In the applied analysis, 250,000 post-burn-in iterations per chain yielded diagnostics for which most reported estimands had $\widehat R<1.01$, all had $\widehat R<1.02$, and all had bulk and tail effective sample sizes greater than 400.
For the simulation studies, we used four chains and retained 1000 post-burn-in draws per chain. We chose run lengths so that the vast majority of pointwise $\clate(\bx_i)$ diagnostics satisfied $\widehat R<1.05$. This led to 5000 post-burn-in iterations per chain for the simple univariate simulations and 50,000 post-burn-in iterations per chain for the higher-dimensional simulation studies.
}

\section{Simulations}
\label{sec:simulations}
We investigated the operating characteristics of our proposed approach (hereafter \texttt{BCF-LATE}) using three synthetic data simulation studies.
We compared \texttt{BCF-LATE} with two \revise{conditional versions of the Wald estimator (\Cref{eq:conditional_late_iv_estimate}): 
\texttt{GRF} and  \texttt{Wald-BART}.  
\texttt{GRF} is the generalized random forest of \cite{grf}, which is implemented with \texttt{instrumental\_forest} from the \textsf{R} package \textbf{grf} with \texttt{tune.parameters} option set to \texttt{``all"}.  
\texttt{Wald-BART} uses two separate probit BART models: one for the Wald estimator's numerator $\itt_{Y}(\bx)$ and the other for the denominator $\itt_{R}(\bx)$. 
The posterior samples of the Wald estimator are obtained by dividing each $\itt_{Y}(\bx)$ sample by a $\itt_{R}(\bx)$ sample.
}
We specifically assessed how well each method could (i) predict $\clate(\bx_i)$ for each simulated subject and (ii) quantify its uncertainty about those predictions.
For each study, we generated $n$ observations $(\bx_{1}, A_{1}, \robs_{1}, \yobs_{1}), \ldots,$ $(\bx_{n}, A_{n}, \robs_{n}, \yobs_{n})$ as follows.
First,  covariates $\bx_{i} \in [-1,1]^{p}$ are uniformly drawn.
Then random treatment assignment $A_{i},$ treatment receipt $\robs_{i},$ and outcome $\yobs_{i}$ are drawn as 
\begin{align*}
    A_i &\sim \textrm{Bernoulli}(0.5) 
    & 
    C_i &\sim \textrm{Bernoulli}\left(p^c_{i}\right) 
    & 
    p^c_{i} &= \Phi\left[\eta(\bx_{i})\right] 
    \\
    \robs_{i} &= A_{i}C_{i} 
    & 
    \yobs_{i} &\sim \textrm{Bernoulli}\left(p^y_{i}\right)
    & 
    p_{i}^y &= \Phi[\mu(\bx_{i}) + C_ i\mu_c(\bx_i) + A_iC_i\tau(\bx_{i})].
\end{align*}
We used different functions $\mu$, $\mu_c$, $\tau$, and $\eta$ in each simulation study. 
When visualizing the results in \Cref{fig:simple,fig:varycompliance,fig:WeakInstrument}, variability/uncertainty is shown by shading 0.65, 1.3, and 2 standard errors around the estimates from frequentist methods and, for Bayesian methods, shading the 50\%, 80\%, and 95\% point-wise posterior credible intervals.  
Likewise, in \Cref{fig:hard,fig:sim-binary}, to show variability across simulations, we shade the 50\%, 80\%, and 95\% intervals across simulation replications. 

\revise{As we know the true $\clate(\bx_i)$ values from the data-generating processes, we used multiple metrics to compare model performance. For point accuracy, our primary metric is root mean squared error (RMSE). For the true values $t_i = \clate(\bx_i)$, and the predicted values $\hat{t}_i = \widehat{\clate}(\bx_i)$, 
\begin{equation}
\mathrm{RMSE} = \sqrt{\frac{1}{n} \sum_{i=1}^{n} (t_i - \hat{t}_i)^2}.
\label{eq:rmse}
\end{equation}
For some simulations we also look at bias through the integrated absolute bias (IAB) with the trapezoidal-rule approximation for $i=1,...,n$ ordering by increasing $x_i$: 
\begin{equation}
\mathrm{IAB} =  \sum_{i=1}^{n-1} 
\frac{ \left| t_i - \hat{t}_i \right| 
     + \left| t_{i+1} - \hat{t}_{i+1} \right| }{2}
     \bigl( x_{i+1} - x_{i} \bigr).\label{eq:iab}
\end{equation}
}

\revise{For uncertainty quantification/interval accuracy, we look at interval width and coverage of \revise{point-wise} intervals. 
For a $100(1-\alpha)\%$ interval for $\clate(\bx_i)$ with lower bound $l_i$ and upper bound $u_i$, we define the average interval width and coverage as
\begin{align}
\mathrm{AIW} &= \frac{1}{n} \sum_{i=1}^{n} (u_i - l_i) & \mathrm{Coverage} = \frac{1}{n} \sum_{i=1}^{n} \ind{l_i \le t_i \le u_i}.
\end{align}
We also compute the interval score \citep{gneiting2007score}, which is a proper scoring rule for assessing uncertainty intervals that combines the ideas of width and coverage, 
\begin{equation}
\mathrm{IS}_{\alpha} = \frac{1}{n} \sum_{i=1}^{n} (u_i - l_i) 
+ \frac{2}{\alpha}(l_i - t_i)\,\ind{t_i < l_i} 
+ \frac{2}{\alpha}(t_i - u_i)\,\ind{t_i > u_i}.
\label{eq:interval_score}
\end{equation}
We computed interval scores using the \texttt{wis()} function in the \textsf{R} package \textbf{scoringutils} and averaged the interval scores across all simulated observations. 
}

\subsection{Visualizing the weak instrument problem}
\label{sec:simulation_weak}

\begin{figure}
    \centering
    \includegraphics[width=5.3in]{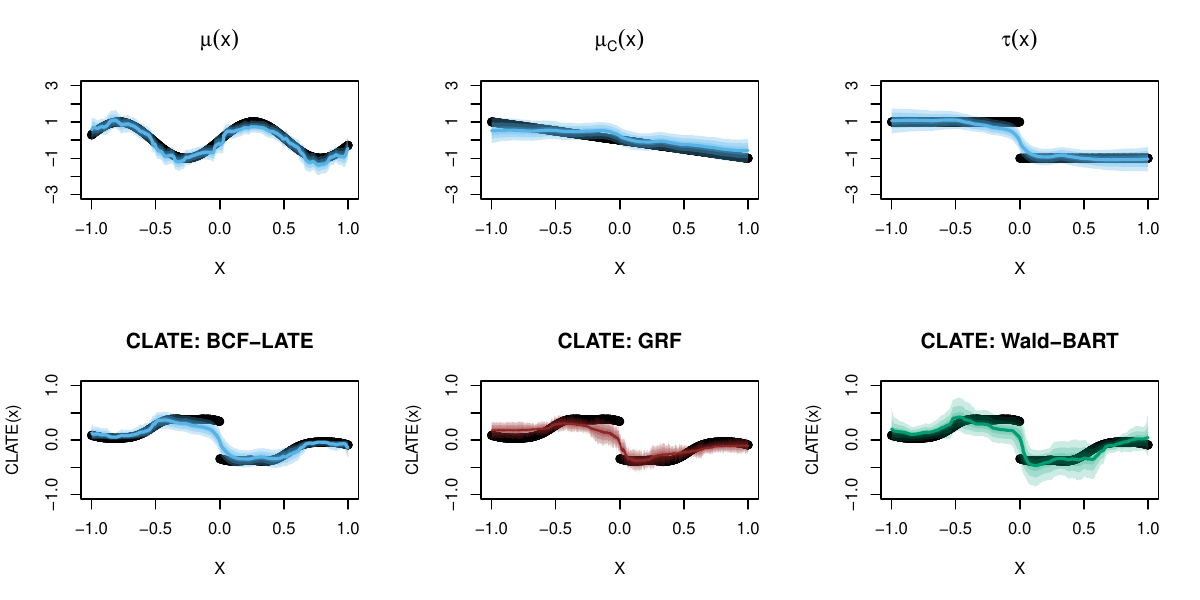}
    \caption{Comparison of different methods' estimation of heterogeneous $\clate(x)$ when a univariate data-generating process has a constant, large compliance rate.}
    \label{fig:simple}
\end{figure}

To visualize the differences between each method, we set $p = 1$ for our first simulation study.
For this study, we drew each $\bx_{i}$ uniformly from $[-1,1]$ and generated the data using
\begin{align*}
    \eta(x) &= 0 &
    \mu(x) &= \sin(6x) &
    \mu_c(x) &= -x &
    \tau(x) &= 2 \times \ind{x \le 0} - 1.  
\end{align*}
The top row of \Cref{fig:simple} shows\revise{, for a single simulation replicate,} each of $\mu(x), \mu_{c}(x)$ and $\tau(x)$ with \texttt{BCF-LATE}'s posterior mean (solid line) and shaded regions for the pointwise posterior credible intervals. 
\texttt{BCF-LATE} recovers each of $\mu(x), \mu_{c}(x),$ and $\tau(x)$ reasonably well;  across 100 replications, the 95\% pointwise credible intervals for $\mu(x)$ and $\mu_{c}(x)$ had more than the nominal 95\% frequentist coverage and the intervals for $\tau(x)$ covered about \reviseTwo{87.6\%} of evaluations on average.
Although $\tau(x)$ takes only two values, the function of interest, $\clate(x)$ is more complicated (bottom row of Figure~\ref{fig:simple}).
In this setting with a moderate, homogeneous compliance rate, all three methods performed similarly, with \texttt{BCF-LATE} enjoying a slight advantage.
\revise{Averaging over }100 simulation replications, the average root mean square errors (RMSE) for evaluating $\clate(x_i)$ were \reviseTwo{0.105} (\texttt{BCF-LATE}), \reviseTwo{0.108} (\texttt{Wald-BART}), and \reviseTwo{0.118} (\texttt{GRF}). 
The average IABs were \reviseTwo{0.144} (\texttt{BCF-LATE}), \reviseTwo{0.162} (\texttt{Wald-BART}), and  \reviseTwo{0.185} (\texttt{GRF}),  
and the average coverage of the 95\% uncertainty intervals were \reviseTwo{86.9}\% (\texttt{BCF-LATE}), \reviseTwo{98.0}\% (\texttt{Wald-BART}), and \reviseTwo{84.7}\% (\texttt{GRF}). 
While \texttt{Wald-BART}'s coverage is much closer to the target 95\%, we note that its intervals are much wider than the other methods' intervals, as seen in \Cref{fig:simple}. 

\revise{All methods struggled with estimating $\clate(x)$ around the discontinuity at $x=0$. 
Although \texttt{Wald-BART}'s larger intervals seem appropriate at this point, we note that \texttt{Wald-BART} returns very wide intervals over the range of $x,$ reflecting the central difficulty in constructing valid, nonparametric intervals for $\clate(\bx)$ \citep{athey2016recursive,smucler2025note}.}
 
While it is reassuring that \texttt{BCF-LATE} performs well when all subjects have the same, moderately large compliance probability, the heterogeneous compliance setting is arguably more relevant.
Recall from \Cref{sec:conditional_identification}, that we observe three types of subjects, each with their own outcome model:
    \begin{align}
        \P(Y_i=1|C_i, A_i, \bX_i = \bx) = 
        \begin{cases}
        \Phi[\mu(\bx)] & \text{if $C_i = 0$} \\
        \Phi[\mu(\bx) + \mu_{c}(\bx)] & \text{if $C_i = 1$ and $A_i = 0$}\\
        \Phi[\mu(\bx) + \mu_{c}(\bx) + \tau(\bx)] & \text{if $C_i = 1$ and $A_i = 1$}.
        \end{cases}
        \label{eq:observed_prognostics}
    \end{align}
Intuitively, we might expect to estimate $\mu(x)$ well when there is a large number of never-takers with $x_{i} \approx x.$
Similarly, because there is no uncertainty about compliance status in the treatment arm, we would expect to estimate $\mu(x) + \mu_{c}(x) + \tau(x)$ well whenever there is a large number of invited compliers with $x_{i} \approx x.$
So, if there are reasonably large sets of never-takers and invited compliers with $x_{i} \approx x$ we would expect \texttt{BCF-LATE} to also estimate their difference ($\mu_{c}(x) + \tau(x)$) well.
To estimate $\clate(x)$ though, we must disentangle $\mu_{c}(x)$ from $\tau(x).$
This is best done when the compliance rate when $\P(C_{i} = 1 \vert x_{i} = x)$ is not too small and there are large numbers of compliers with $x_{i} \approx x.$

\begin{figure}
    \centering
        \includegraphics[width=5.3in]{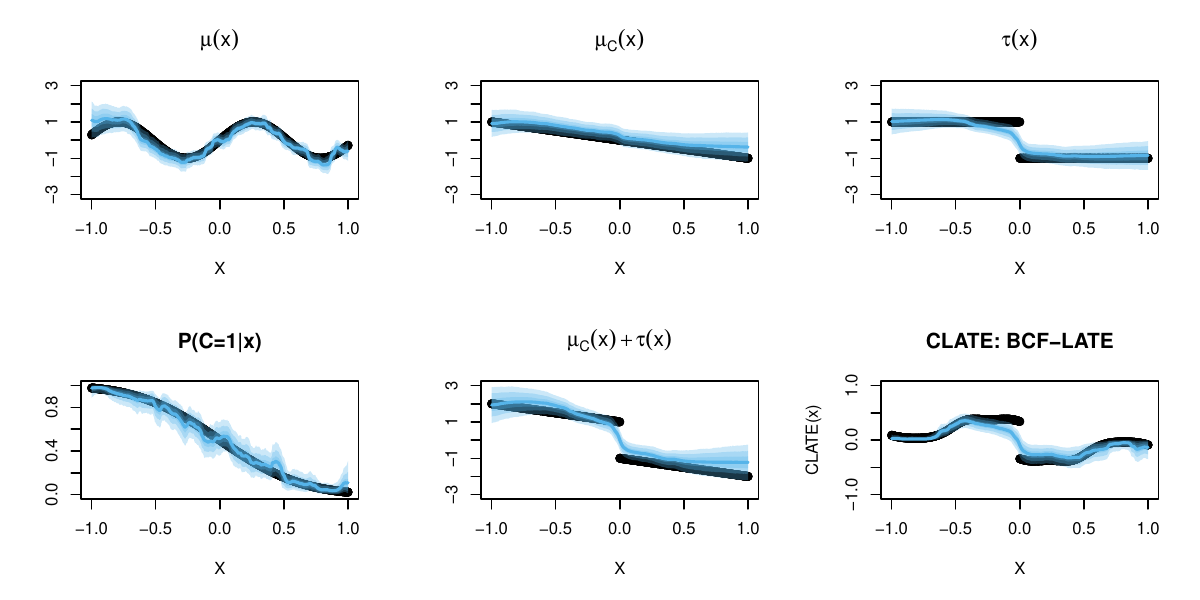}
    \caption{BCF-LATE fits for a univariate data-generating process with heterogeneous outcome and heterogeneous compliance. 
    }
    \label{fig:varycompliance}
\end{figure}

To check this intuition, we re-ran the simulation study using the same data generating functions as before but this time with $\eta(x) = -2x.$
\Cref{fig:varycompliance} shows how \texttt{BCF-LATE} recovers each of $\mu(x)$, $\mu_{c}(x),$  $\tau(x),$ $\P(C = 1 \vert x),$ $\mu_{c}(x) + \tau(x),$ and $\clate(x).$
As anticipated, when $x$ is small and the compliance rate is close to one (yielding few never-takers), the uncertainty bands of $\mu(x)$ widen in the top left pane. 
However, when there is a reasonable local sample size of never-takers (i.e., when the compliance rate is not close to 1), the function is well-estimated \revise{by the posterior mean} and the uncertainty intervals are tighter.
Similarly, for more central values of $x$ yielding non-extreme compliance probabilities, \texttt{BCF-LATE} produced shorter uncertainty intervals for $\mu_{c}(x) + \tau(x).$
For $x$ close to $\pm1,$ which produce extreme compliance probabilities, \texttt{BCF-LATE} is much more uncertain about this function.

\begin{figure}
    \centering
\includegraphics[width = 1.35in]{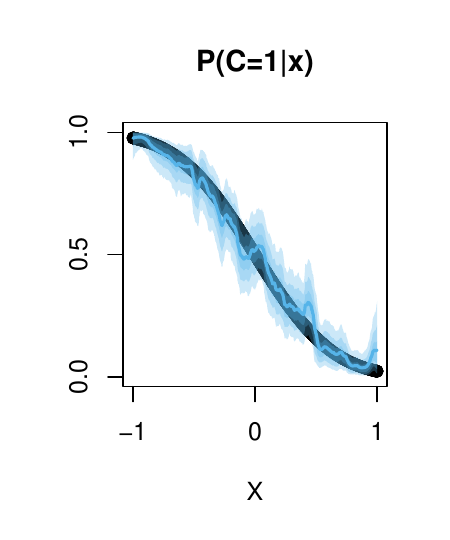}
\includegraphics[width = 1.35in]{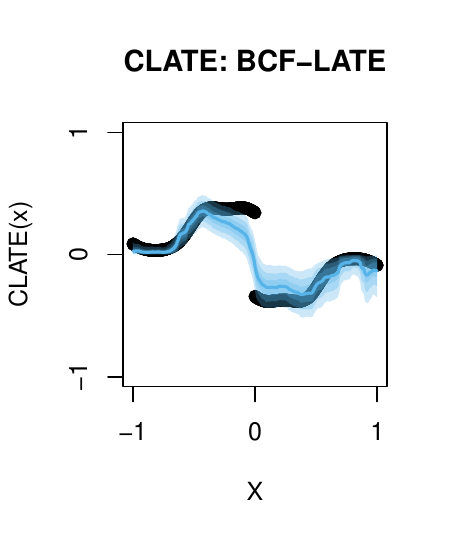}
\includegraphics[width = 1.35in]{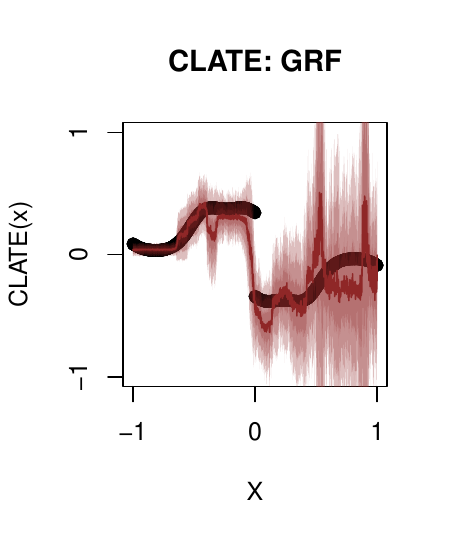}
\includegraphics[width = 1.35in]{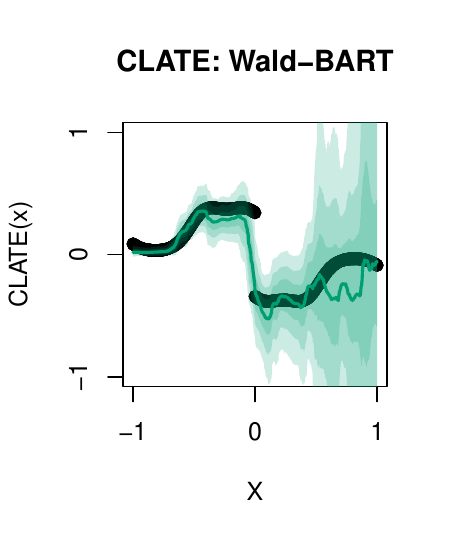}
    \caption{
    Comparison of different methods' estimation of heterogeneous $\clate(x)$ from a univariate data-generating process. The compliance rate varies such that the instrument is weak when $x$ is large.}
    \label{fig:WeakInstrument}
\end{figure}

\Cref{fig:WeakInstrument} compares \texttt{BCF-LATE}'s ability to estimate $\clate(x)$ in the presence of a weakening instrument with those of \texttt{GRF} and \texttt{Wald-BART}.
As expected, all three methods do well when the compliance rate is large.
However, as $x$ increases and the compliance probability drops, we see that the uncertainty intervals for \texttt{GRF} and \texttt{Wald-BART} become more erratic and inflated, \revise{such as the spike in uncertainty near $x=0.5$. 
Visually it further appears that \texttt{BCF-LATE} produced more accurate estimates of
$\clate(x)$ when the compliance probability was low, though like the homogeneous compliance example in \Cref{fig:simple}, \texttt{BCF-LATE} does not look as accurate around $x=0$ as the other two methods. 
However, across 100 simulation replications, 
\texttt{BCF-LATE}'s average RMSE was \reviseTwo{0.114} and average IAB was \reviseTwo{0.157}, which were smaller than \texttt{GRF}'s \reviseTwo{(0.354, 0.388)} and \texttt{Wald-BART}'s \reviseTwo{(0.883, 0.714)}.  
Coverage of 95\% intervals for subjects' $\clate(x_i)$ is still challenging,  with \texttt{BCF-LATE}  at \reviseTwo{85.8}\% and \texttt{GRF} at  \reviseTwo{82.5}\%, though the wide intervals of \texttt{WaldBART} achieved \reviseTwo{97.8}\% coverage.}

\subsection{Data generating process with continuous covariates}
\label{sec:sim_complicated}

\begin{figure}[ht!]
\centering
\begin{subfigure}{2.7in}
\centering
\includegraphics[width = 2.7in]{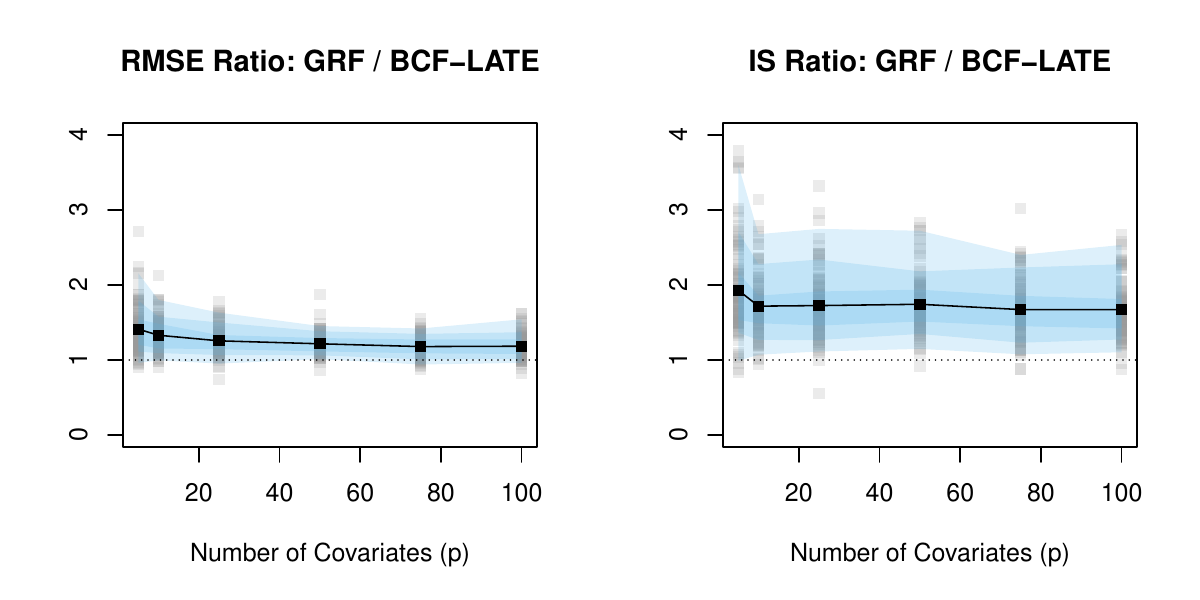}
\caption{}
\label{fig:sim-hardDGP-increasingP}
\end{subfigure}
\begin{subfigure}{2.7in}
\centering
\includegraphics[width=2.7in]{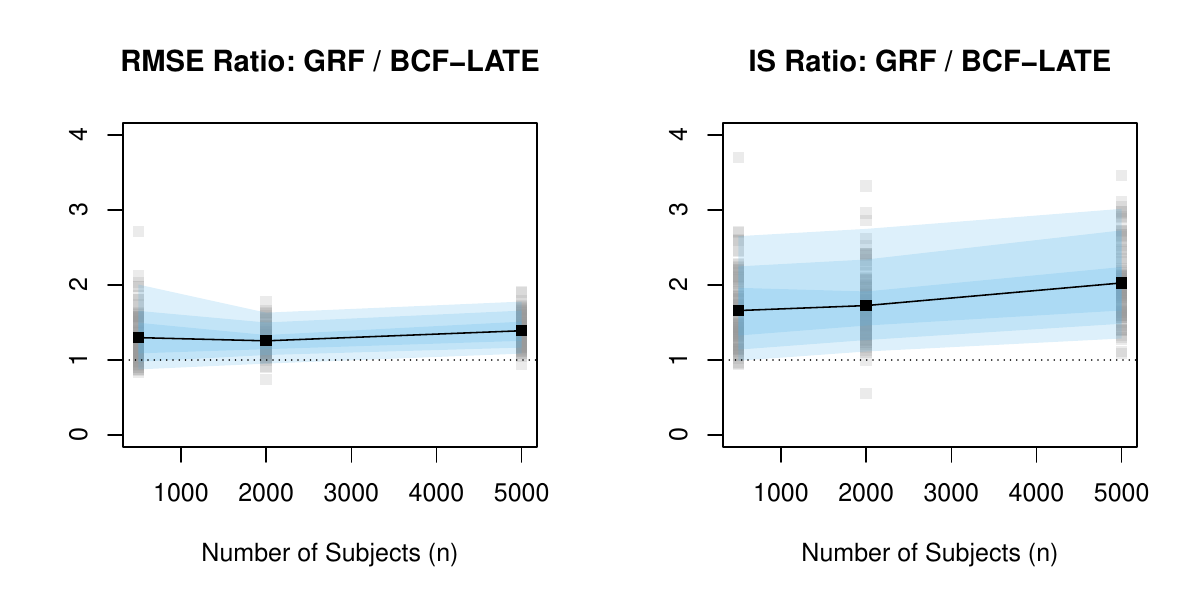}
\caption{}
\label{fig:sim-hardDGP-increasingN}
\end{subfigure}
\caption{Comparison of \texttt{BCF-LATE} and \texttt{GRF} across 100 replications of the complex DGP when $n=2000$ is fixed and $p$ increases (a) and when $p=25$ is fixed and $n$ increases (b). Each simulation is represented with a transparent gray square. 
}
\label{fig:hard}
\end{figure}

\revise{For} a more detailed study, we generated data with more complicated functions:
\begin{align*}
    \eta(\bx) &= \exp(x_3)-x_1-x_2-x_4-x_5  \quad 
    &
    \mu(\bx) &= 2 \times \ind{x_2+x_5 >1} -x_3
    \\
    \mu_C(\bx) &= x_1x_3 - x_4 
    &  
    \tau(\bx) &= \sin(\pi x_1 x_2)  -  (x_3 - 0.5)^2 + .1x_4 - .2x_5.
\end{align*}
For each combination of $n \in \{500,2000,5000\}$ and $p \in \{5,10,25,50,75,100\}$, we generated 100 synthetic datasets of size $n$. 
Figure~\ref{fig:hard} compares the performance of \texttt{GRF} to \texttt{BCF-LATE} as $p$ and $n$ increase.
In the figures, values greater (resp. less) than one indicate that \texttt{BCF-LATE} performs better (resp. worse) than \texttt{GRF}. 
The gap between \texttt{GRF} and \texttt{BCF-LATE} diminishes as $p$ increases while keeping $n$ fixed and the gap increases  as $n$ increases while keeping $p$ fixed; see \switchref{\Cref{tab:sim-hardDGP-increasingP,tab:sim-hardDGP-increasingN}}{Tables C3 and C4 in Appendix C}. 
\revise{For every combination of $n$ and $p,$ at least \reviseTwo{87}\% of simulations saw \texttt{BCF-LATE} have} smaller RMSE and interval scores than \texttt{GRF}.

\subsection{Data generating process with binary covariates}
\label{sec:simulation_binary}

As the wellness program we are studying largely consist of binary covariates, we also present such a simulation. 
We generate independent Bernoulli covariates $\mathbf{x}_i \in \{0,1\}^p$  for $i = 1,\ldots,n$ and $j = 1,\ldots,p \ge 3$, where $Pr(x_{ij} = 1) = \frac{j}{p+1}$  so that the covariates have different levels of balance. 
This results in $2^p$ different possible combinations of the p binary covariates; let k be an index of the $1,...,2^p$ different possible combinations.  
Compliance status for each combination is a random Bernoulli draw with probability $\Phi(\eta_k)$, where $\eta_k \sim N(0,1)$.  
Potential outcomes are likewise are random Bernoulli draws where the probability of $Y_i = 1$ is $\Phi(\mu_k), \, \Phi(\mu_k + \mu_{ck}),  \text{ or } \Phi(\mu_k + \mu_{ck} + \tau_k)$ for never-takers, untreated compliers, and treated compliers respectively. 
The components are randomly drawn for each $k$ as follows: $\mu_k \sim N(-.5, .3^2)$, $\mu_{ck} \sim N(0,.2^2)$, $\tau_k \sim N(0,.15^2)$. 
This yields expected potential outcomes for each group $k$ that range from a bottom decile of 0.15 to a top decile of 0.5, with an average of 0.31. 
$\mu_{ck}$ builds in the  expectation that compliers and never-takers are different even without the treatment, as $SD(\mu_{ck}) = 0.2$. 
$\tau_k$ builds in the expectation that there is no average treatment effect as it is mean zero, but the individual $\clate(\bx_i)$'s are nonzero, with standard deviation 0.05.  
Treatment assignment is randomized as $A_i \sim \mathrm{Bernoulli}(0.5)$, and the received treatment equals $R_i = C_i A_i$. 

\begin{figure}[th!]
\centering
\begin{subfigure}{2.7in}
\centering
\includegraphics[width = 2.7in]{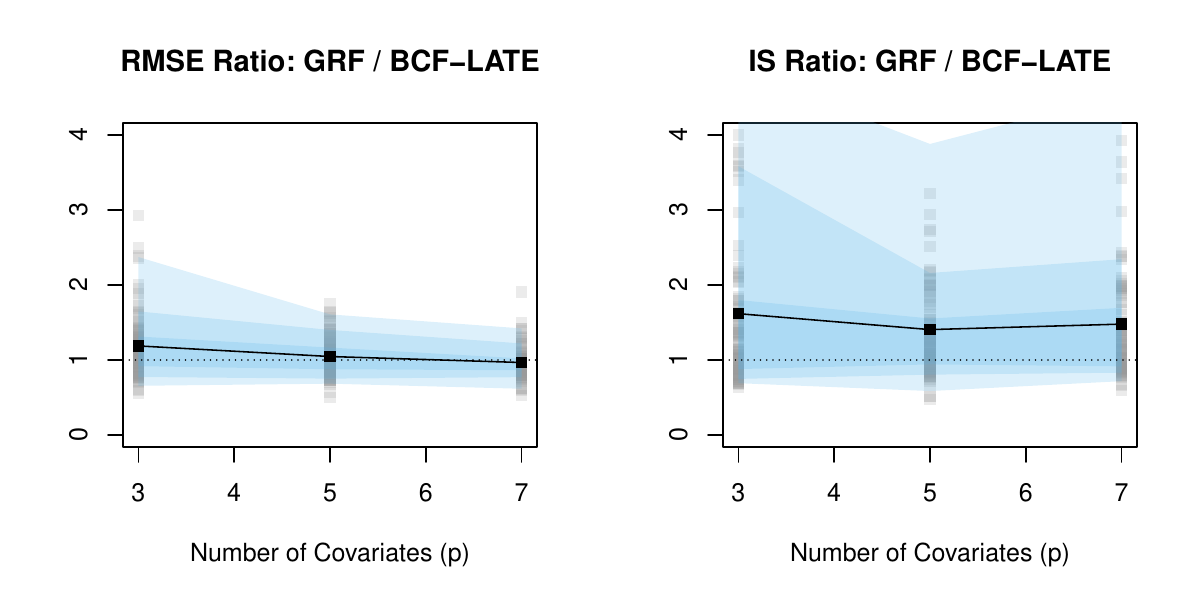}
\caption{}
\label{fig:sim-binary-increasingP}
\end{subfigure}
\begin{subfigure}{2.7in}
\centering
\includegraphics[width=2.7in]{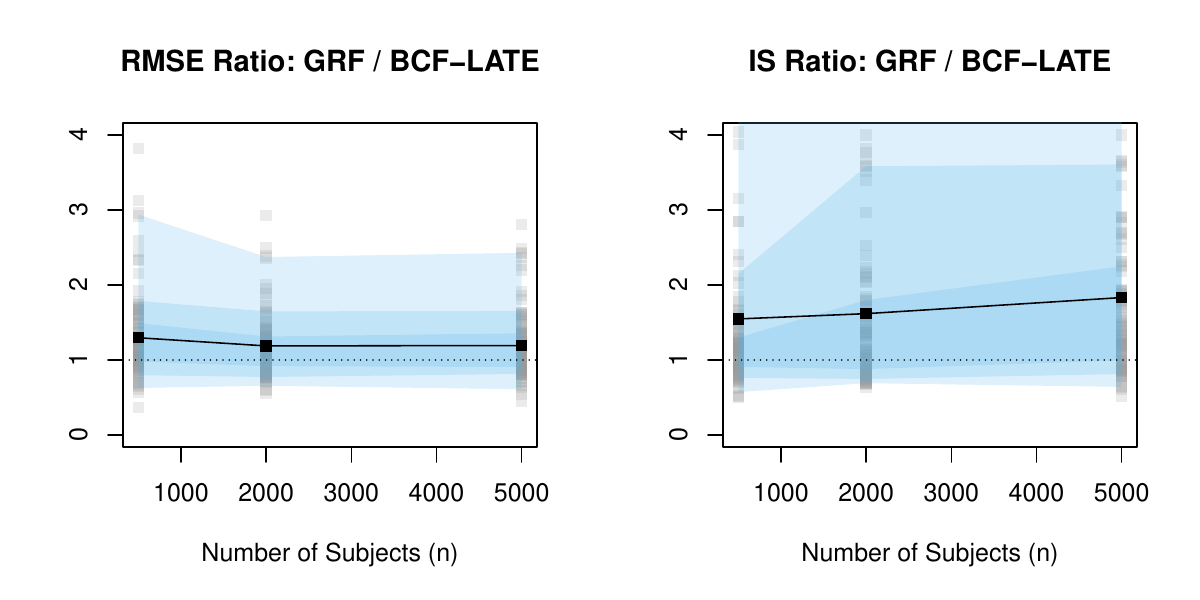}
\caption{}
\label{fig:sim-binary-increasingN}
\end{subfigure}
\caption{ \revise{Comparison of \texttt{BCF-LATE} and \texttt{GRF} across 100 replications of the potential-outcome-based DGP when $n=2000$ is fixed and $p$ increases (a) and when $p=3$ is fixed and $n$ increases (b). Each simulation is represented with a transparent gray square.} 
}
\label{fig:sim-binary}
\end{figure}

100 simulations are conducted for combinations of $p \in \{3,5,7\}$ and $n \in \{500, 2000, 5000\}$. 
Results are shown in \Cref{fig:sim-binary} and in \switchref{\Cref{sec:simulation_binary_appendix}}{Appendix C.3}.  
We see that both RMSE and interval scores are similar for \texttt{GRF} and \texttt{BCF-LATE}, where \texttt{BCF-LATE}'s metrics typically are better (lower) on average than \texttt{GRF}'s, as seen when the ratios are larger than 1. 
\texttt{BCF-LATE} had better interval scores in \reviseTwo{about 60}\% of simulations, \reviseTwo{ranging from 59\% to 74\%, which appears to stay fairly consistent across changes in $p$ but increases with $n$}. 
\reviseTwo{For $p=3$, \texttt{BCF-LATE}'s edge in RMSE reduces as $n$ increases, with better RMSEs in 72\% of simulations with n=500 dropping to 59\% when $n=5000$. For $n=2000$,  \texttt{GRF}'s RMSEs stay fairly consistent as $p$ increases (roughly 0.065 on average across all simulation settings), whereas \texttt{BCF-LATE} starts with an advantage with an average RMSE of 0.056 when $p=3$ that increases to 0.064 and 0.067 as $p$ increases to 5 and 7 respectively, yielding a slight edge to \texttt{GRF} at $p=7$ where \texttt{GRF}'s average RMSE is smaller in 63\% of simulation replicates (detailed in Table C6).}

We conclude from our simulation studies that \texttt{BCF-LATE} is at least as capable as --- and often better than --- \texttt{GRF} in estimating the conditional local average treatment effects $\clate(\bx)$ in terms of point and interval estimation.

\section{Results for Wellness Study}
\label{sec:real_world}

In this section, we apply \texttt{BCF-LATE} to public use data from the Illinois Workplace Wellness Study. 
For each outcome, we ran 10 chains of \texttt{BCF-LATE} for \reviseTwo{251,000} total iterations each.
After discarding the first \reviseTwo{1,000 iterations} of each chain as ``burn-in,'' we \reviseTwo{every 250th draw from each chain which yielded} 10,000 MCMC draws of $\clate(\bx_{i})$ for each subject $i.$ 
In each outcome's model, the lowest estimated compliance rate across all subject is \reviseTwo{0.486,  while most subjects' estimates are between 0.6 and 0.7.} 
Hence, there are not severely weak instruments/compliance rates in this study.

\begin{figure}
\centering
\begin{subfigure}{1.8in}
\centering
\includegraphics[width = 1.8in]{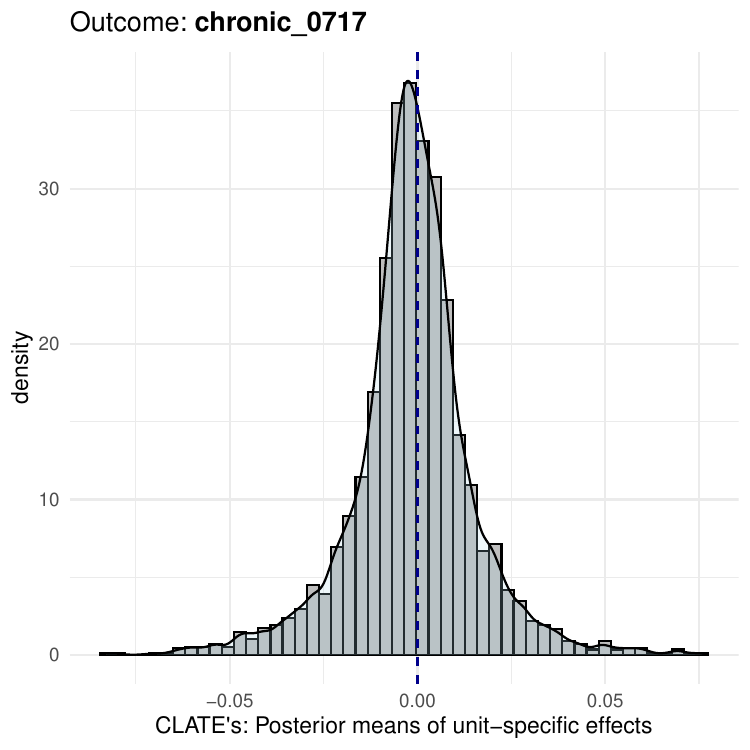}
\caption{}
\label{fig:chronic_means}
\end{subfigure}
\begin{subfigure}{1.8in}
\centering
\includegraphics[width = 1.8in]{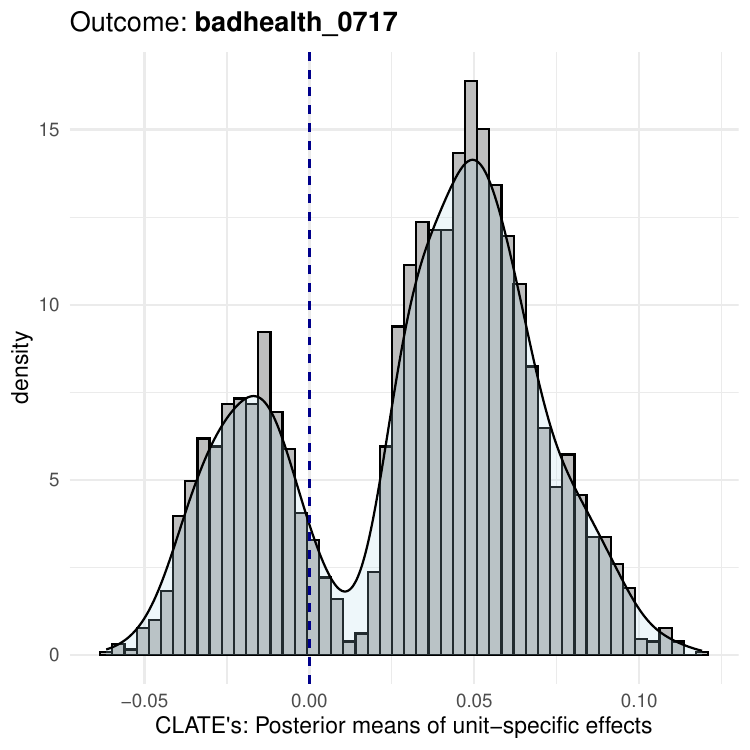}
\caption{}
\label{fig:badhealth_means}
\end{subfigure}
\begin{subfigure}{1.8in}
\centering
\includegraphics[width = 1.8in]{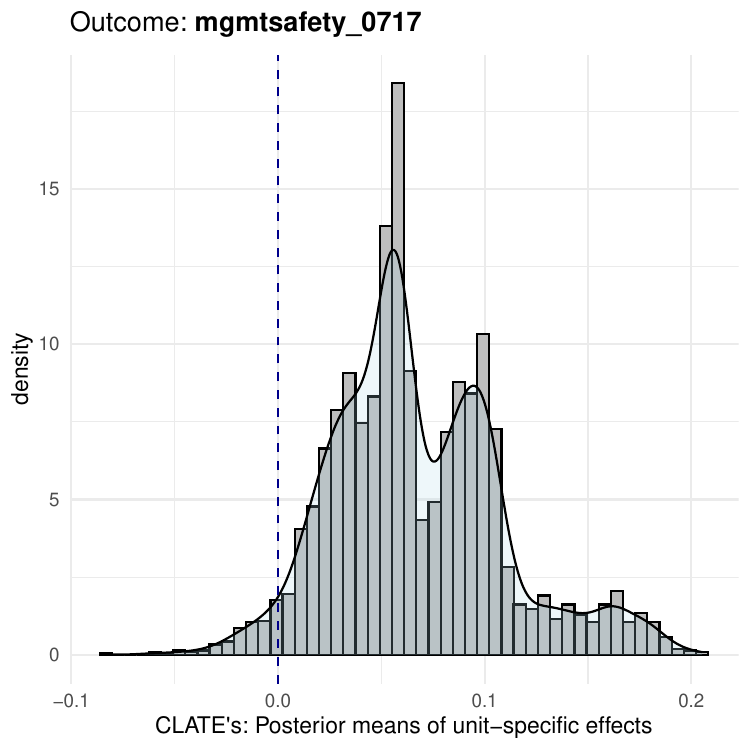}
\caption{}
\label{fig:mgmtsafety_means}
\end{subfigure}
\caption{Posterior mean estimates of $\clate(\bx_{i}$) across our sample.}
\label{fig:posterior_means}
\end{figure}

\Cref{fig:posterior_means}  shows the posterior means of $\clate(\bx_{i})$ for each outcome. 
To better summarize the heterogeneity visible in these estimates, we fit a classification tree to the posterior mean estimates $\E[\clate(\bx_{i}) \vert \by]$ using the covariates $\bx_{i}.$ 
Often called ``fitting the fit,'' this increasingly popular summarization step produces a parsimonious description of the $\clate(\bx_i)$ by splitting on covariates describing subgroups with the most meaningful heterogeneity \citep[see, e.g.,][]{hahn2015decoupling, surpuelz, fishmonotonic, bolfarine2024decoupling}.  
In other words, by fitting the fit, we can automatically identify interesting subgroups which may have significant positive or negative effects. \revise{We can then analyze the posterior distribution of the average of the $\clate(\bx_{i})$'s within each subgroup, hereafter called the subgroup CLATE.}
 
\begin{figure}
\centering
\begin{subfigure}{1.8in}
\centering
\includegraphics[width = 1.8in]{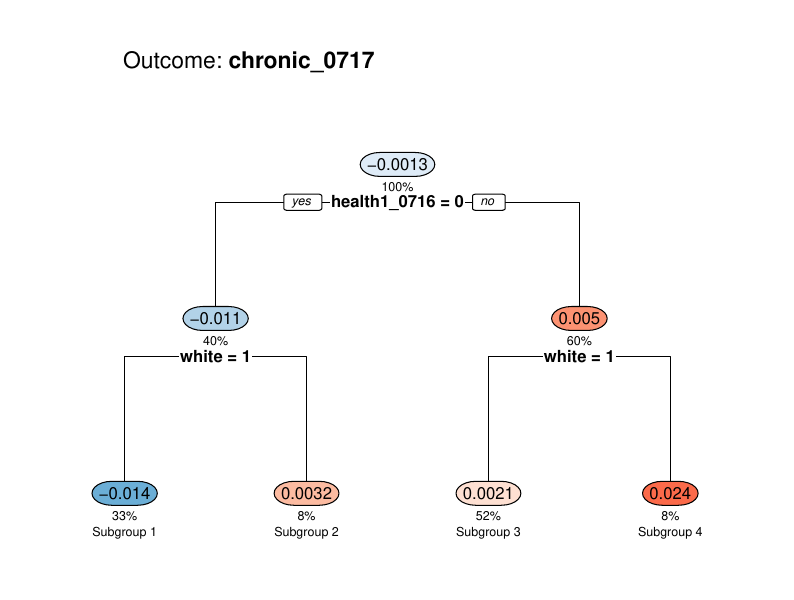}
\caption{}
\label{fig:chronic_tree}
\end{subfigure}
\begin{subfigure}{1.8in}
\centering
\includegraphics[width = 1.8in]{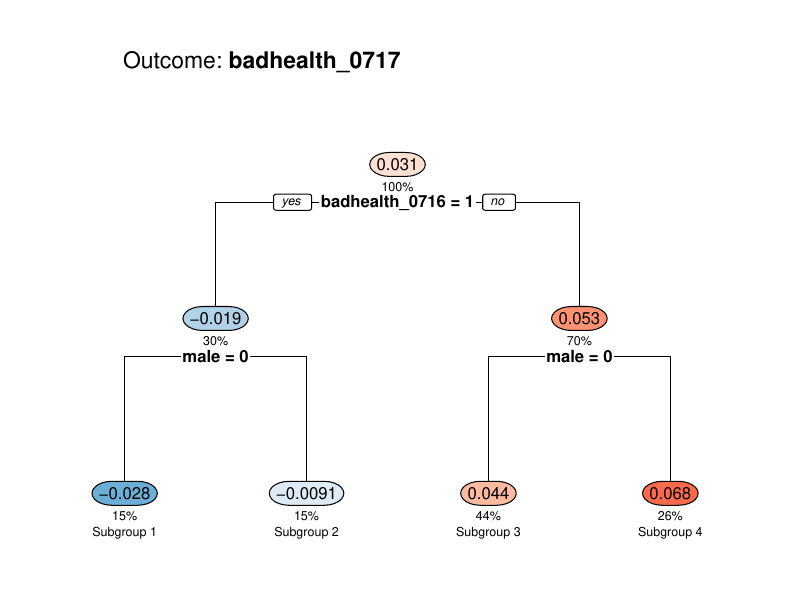}
\caption{}
\label{fig:badhealth_tree}
\end{subfigure}
\begin{subfigure}{1.8in}
\centering
\includegraphics[width = 1.8in]{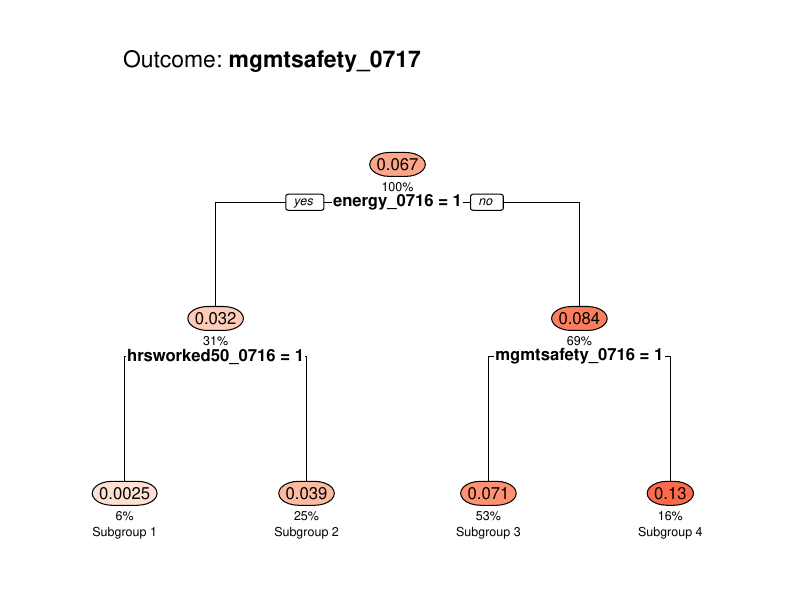}
\caption{}
\label{fig:mgmtsafety_tree}
\end{subfigure}
\caption{Trees summarizing the heterogeneity in fitted CLATE estimates. Numbers within nodes are the subgroup average CLATE estimates. 
Numbers below the nodes report the proportion of the sample in each subgroup.}
\label{fig:posterior_trees}
\end{figure}

\begin{figure}
\centering
\begin{subfigure}{1.8in}
\centering
\includegraphics[width = 1.8in]{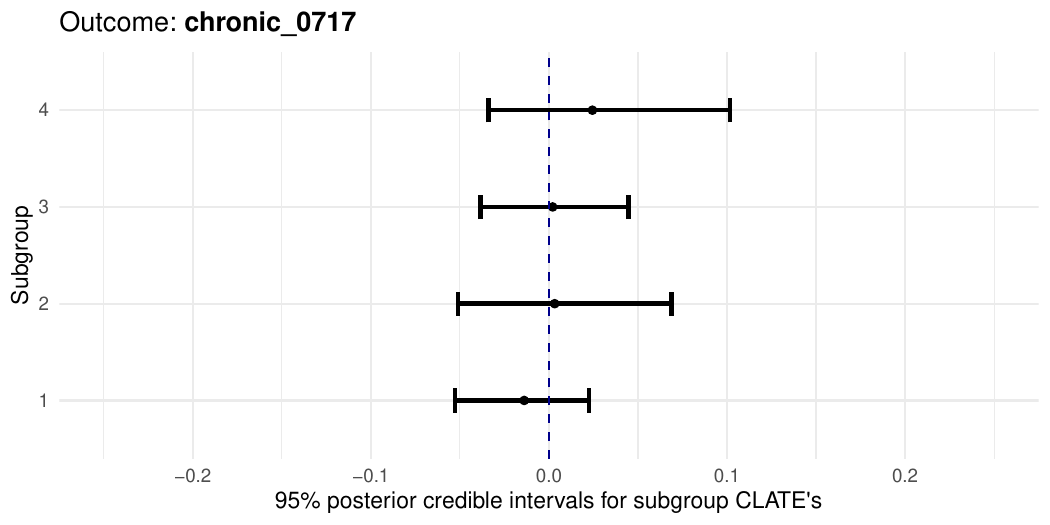}
\caption{}
\label{fig:chronic_sg}
\end{subfigure}
\begin{subfigure}{1.8in}
\centering
\includegraphics[width = 1.8in]{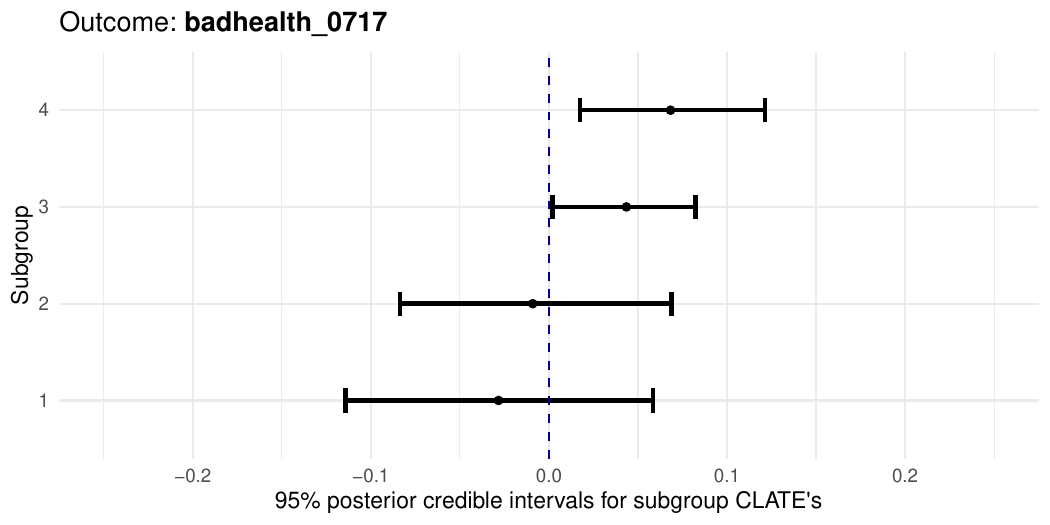}
\caption{}
\label{fig:badhealth_sg}
\end{subfigure}
\begin{subfigure}{1.8in}
\centering
\includegraphics[width = 1.8in]{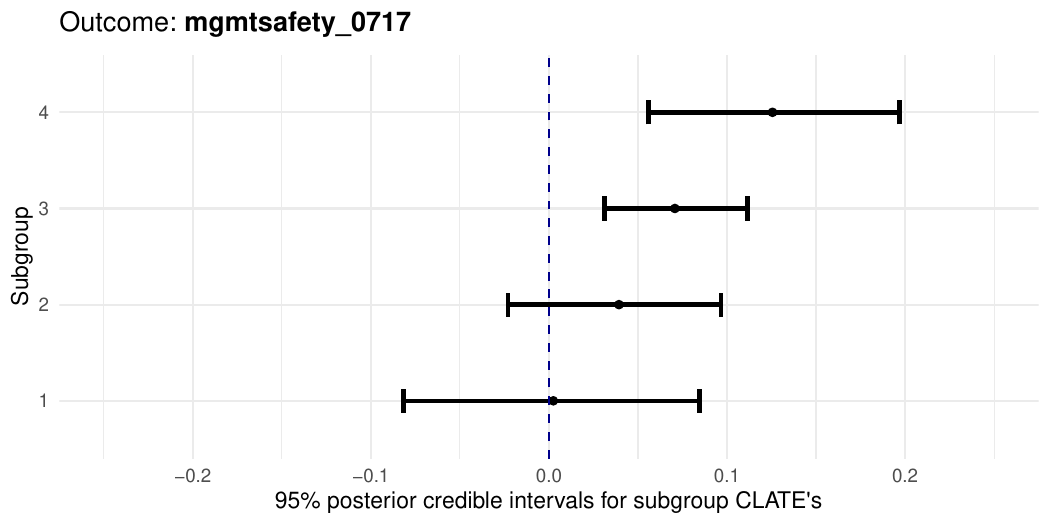}
\caption{}
\label{fig:mgmtsafety_sg}
\end{subfigure}
\caption{Credible intervals for the subgroup CLATEs corresponding to leaf nodes in the summary trees of Figure~\ref{fig:posterior_trees}.}
\label{fig:subgroup_intervals}
\end{figure}

\textbf{Self-report of chronic conditions}.  \revise{We first look at the outcome of self-reporting at least one chronic condition in 2017 (\texttt{chronic\_0717}). 
In \Cref{fig:chronic_means}, we observe a single prominent mode around a negligible treatment effect: the overall sample LATE has a posterior mean of \reviseTwo{-0.13} percentage points. 
The summary tree in \Cref{fig:chronic_tree} splits on covariates in symmetric positive and negative directions.  
It first splits on whether subjects  
\reviseTwo{ 
rated their health as excellent or very good at baseline, and those who did not are assigned to the left branch while those who did are assigned to the right branch.
Among those who did not report excellent or very good baseline health, iThrive participation decreased their chances of self-reporting a chronic condition in 2017 by -1.07 percentage points, on average.
} 
The far-left leaf node, labeled Subgroup 1, corresponds to a subgroup of white individuals who did not rate their health as excellent or very good at baseline. 
Within this subgroup, iThrive participation led to a \reviseTwo{1.4} percentage point average decrease in the changes of reporting at least one chronic health condition in 2017.
In contrast, among non-white subjects who rated their baseline health as excellent or very good (i.e., the far-right leaf labeled Subgroup 4 in \Cref{fig:chronic_tree}), iThrive participation led to a \reviseTwo{2.44} percentage point \textit{increase} in reporting at least one chronic health condition in 2017. However, we find these effects to not be significantly different from 0. 
\Cref{fig:chronic_sg} shows the 95\% posterior credible intervals for the subgroup CLATEs corresponding to the bottom nodes of \Cref{fig:chronic_tree}.
}
There is considerable posterior probability on both positive and negative effects within each subgroup and there is substantial overlap in the intervals across the subgroups.
Consistent with \citet{qje}'s findings, our results suggests that iThrive's impact on chronic conditions is essentially null and that there is not much heterogeneity in its effect.

\textbf{Metabolic parameters.}
\Cref{fig:badhealth_means} shows the subgroup CLATE estimates of  the effect of iThrive participation on the chances that participants reported high cholesterol, blood pressure, or blood glucose in 2017 (\texttt{badhealth\_0717} in the dataset, hereafter called ``metabolic parameters''). These tend to be positive, and the overall sample LATE is positive (\reviseTwo{3.1} percentage points) and borderline-\reviseTwo{significant} (posterior probability of being positive is \reviseTwo{0.9537}).  
\reviseTwo{Furthermore, \Cref{fig:badhealth_means} shows some heterogeneity in the posterior means: there are separate modes on both sides of zero.}
In fact, these two modes correspond to the first split in \Cref{fig:badhealth_tree}, which was on the variable \texttt{badhealth\_0716}.
This variable is equal to 1 if subjects self-reported high metabolic parameters at baseline in 2016 and is equal to 0 otherwise. 
The two modes in \Cref{fig:badhealth_means}  correspond to the subgroups formed by this first split: the positive mode aligns with those who said their cholesterol, blood pressure, and blood glucose levels were ``low'' or ``normal'' \citep{qje}. 
These subjects were, in some sense, optimistic about their health status at baseline.

\revise{The second level of splits in this tree both split on gender, yielding four subgroups, with males having hiring effect estimates. \Cref{fig:badhealth_sg} shows the posterior distributions for these four subgroups. 
Standing out is Subgroup 4: compliers who were previously optimistic at baseline (first split) and male (second split) had a \reviseTwo{6.8} percentage point increase in reporting high metabolic parameters in 2017. 
This effect is significant with \reviseTwo{0.9959} posterior probability of being positive and visualized by Subgroup 4's interval in Figure~\ref{fig:badhealth_sg}.} 

\revise{We can also inquire of our posterior distribution whether the first level split is significant (i.e. split only on \texttt{badhealth\_0716} and not on \texttt{male}). 
Among these subjects that did not report high metabolic parameter in 2016, the estimated subgroup CLATE was estimated as \reviseTwo{5.3} percentage points (as shown in \Cref{fig:badhealth_tree}'s middle row, right node), with \reviseTwo{0.9941} posterior probability  of being positive, versus those that did report such in 2016 whose subgroup CLATE (estimated to be \reviseTwo{-1.9} percentage points) has only a  \reviseTwo{0.3098} posterior probability of being positive \reviseTwo{and thus not significantly different than zero}.
Furthermore, this heterogeneous effect is evidenced through the posterior distribution of the difference in \reviseTwo{these two subgroups'} CLATEs, which \reviseTwo{has 0.9259} probability on the positive axis.} 
These results suggests that (i) the two modes visible in \Cref{fig:badhealth_means} \reviseTwo{reasonably} represent subgroups experiencing distinctly different effects of iThrive participation; and (ii) the dominant positive mode represents a real, non-zero treatment effect. 

\revise{At face value, the increase in self-reported high blood pressure and related metabolic conditions among wellness-program participants appears counterintuitive. 
Yet, the subgroup driving this result consists of individuals who had not reported such conditions in the preceding year. 
A more credible interpretation is that participation in iThrive increased health awareness and thus detection of pre-existing conditions, rather than worsened metabolic health. 
This interpretation is consistent with theoretical \citep{rosenstock1988social} and empirical \citep{dalbo2017lack} evidence of divergences between perceived and actual health status.} 
 
\textbf{Management's prioritization of health \& safety.} 
\Cref{fig:mgmtsafety_means} also displays \revise{heterogeneity, as it looks like a mixture distribution with perhaps at least four subgroups.} 
iThrive \revise{participation's effect on \texttt{mgmtsafety\_0717} has a positive estimates in each of the subgroups identified by the summary tree in \Cref{fig:mgmtsafety_tree}, and in \Cref{fig:mgmtsafety_sg} the 95\% credible intervals for the average effect in two out of the four subgroups  are entirely to the right of zero.}
This suggests that participating in iThrive increased the chances that subjects in these subgroups believed management prioritized their health and safety.

The summary tree splits on energy level, \revise{working more than 50 hours}, and baseline belief that management prioritized employee well-being.   
Among individuals who (i) did not self-report lots of energy at baseline; (ii) \reviseTwo{did not} believe their managers prioritized their wellbeing at baseline (i.e., \revise{Subgroup 4 in} \Cref{fig:mgmtsafety_tree}),  iThrive participation increased the chances that they believed management prioritized their well-being by 13 percentage points. 
 \revise{
 Like with the metabolic parameters outcome, if we instead look at only the first split in the tree (i.e. combine Subgroup 1 with 2, and Subgroup 3 with 4), the group with lots of energy in 2016 has an estimated subgroup CLATE of \reviseTwo{3.2} percentage points, with \reviseTwo{0.8595} posterior probability of being positive. The group that reported not having lots of energy has estimated effect size of \reviseTwo{8.4}  percentage points, with $\approx 100$\% posterior probability of being positive.  The difference between these two subgroup CLATEs has a \reviseTwo{0.9665}  posterior probability of being greater than zero, indicating significant heterogeneity in their treatment effects.}
This result is in line with the analysis in \cite{qje}, where they find a significant first-year treatment effect on the management prioritization outcome.  
\texttt{BCF-LATE} bolsters this finding by uncovering heterogeneity among the subjects' effect sizes and showing which individuals within the sample drive this positive effect. 

\textbf{Comparison to \texttt{GRF}}.  We conclude by comparing our results to those obtained using \texttt{GRF}.
Broadly speaking, our results were qualitatively similar: there is little evidence suggesting participation in iThrive had a markedly positive or negative effect, on the \revise{health} outcomes considered, \emph{on average}. 
\texttt{GRF} and \texttt{BCF-LATE} did, however, uncover somewhat different estimates of the heterogeneity.  
\switchref{\Cref{fig:sens_mgmtsafety}}{Figure D1 in the Supplementary Materials} demonstrates a prime example. 
For the perception of management's prioritization of employee well-being outcome, while \texttt{BCF-LATE} suggests that at least some different subgroups of importance, if not many, \texttt{GRF} suggests effectively no heterogeneity.

We found moreover that, compared to \texttt{BCF-LATE}, the results obtained using \texttt{GRF} were highly sensitive to the hyperparameter tuning, initialization, and randomization seed.
\switchref{\Cref{fig:sens_mgmtsafety,fig:sens_badhealth} in \Cref{app:grf_comp}}{Figures D1 and D2 in the Supplementary Materials} show the histograms of $\clate(\bx_{i})$ estimates for the metabolic parameter and perception of management's prioritization of employee well-being outcomes obtained from ten different runs of both \texttt{GRF} and \texttt{BCF-LATE}, i.e. each only differed by the random seed used for computation, and we see much more variability in estimates from  \texttt{GRF}. 
Moreover, in other analyses, we have observed large variation in \texttt{GRF}'s estimates of the  $\clate(\bx_{i})$'s  as we varied the number and range of tunable hyperparameter values.
In some runs, we obtained three modes, suggestive of important heterogeneity, while in others we obtained only one mode, suggesting little meaningful heterogeneity.
The output of \texttt{BCF-LATE} was much more stable across repeated runs and appears less sensitive to operational ``researcher degrees of freedom.''

\section{Discussion}
\label{sec:discussion}
In this paper we have proposed an extension of the Bayesian Causal Forest model (BCF) to account for one-sided noncompliance when outcomes of interest are binary. 
Our approach includes modeling compliance jointly with the outcome using nonlinear functions from Bayesian Additive Regression Trees \citep[BART;][]{bart}. 
We have shown that by modeling compliance, our method performs better than traditional methods when compliance rates are low/the instrument is weak. 

Applying our method to the Illinois Workplace Wellness Study \citep{qje,jama}, we find that the conditional local average treatment effects vary for certain binary outcomes. Specifically, we find heterogeneity in the impact of the workplace wellness intervention on compliers' rates of self-reported health issues. 
Compliers who did not report high blood pressure, cholesterol, or glucose levels at baseline were more likely to report high metabolic parameter levels the following year if they were randomized to the wellness plan. 
Similarly, we find that while  employees' views about management's priority on health and safety have a significant local average treatment effect like in \cite{qje}, we also see that there are interesting subgroups with varying magnitudes of effects. 

\revise{It would be interesting to extend our method to settings when the model assumptions do not hold. First,} \Cref{assum:sutva} implies that there is no interference across units.
There may, however, be peer effects in which employees encourage each other to participate in iThrive.
Estimating such peer effects is the subject of on-going work by the authors of \citet{qje}.
It would be interesting to extend that work to explore treatment effect heterogeneity in the presence of interference. 

\revise{Second, our work here has been for one-sided noncompliance when the outcome is binary (\Cref{assum:onesided}). 
If there are no defiers as per the standard monotonicity assumption, two-sided noncompliance introduces three latent strata, necessitating a multinomial model.
\citet{chen2024bayesian} and \citet{garraza2024combining} both used a series of nested probit models to impute the latent strata.
Rather than fitting multiple models, it would be interesting to use \citet{linero2024generalized}'s generalized BART to model strata membership.}

\revise{Third, while it is common to assume that an instrument only affects the outcome through the treatment, the exclusion restriction in \Cref{assum:exclusion} could be violated for some outcomes.  
In the Illinois Workplace Wellness Study, while it is reasonable to assume that the invitation to the wellness program alone, without program participation, could not affect my awareness of metabolic parameters, this could be the case for an individual's beliefs about their management's prioritization of their health and safety. 
Recent work around LATE without the exclusion restriction has been done by bounding the LATE \citep{spieker2022bounding} or estimating it conservatively \citep{markovich2025estimating}. 
Future work could include adapting these methods to BART-based approaches when the exclusion restriction is in question. 
}

\revise{Lastly, \Cref{assum:unconfounded_cond} holds trivially in this study as the treatment was randomly assigned and propensity scores of the form $Pr(A_i = 1|\bX_i = \bx)$ are the same for all subjects. 
However, should conditional unconfoundedness hold in an observational study that does have variation in propensity scores, then following  \cite{bcf}, it may make sense to include propensity scores in $\mu(\bx)$ and $\mu_c(\bx)$ but not in $\tau(\bx)$, but these choices would need further investigation. 
}  

\section{Significance Statement}
\label{sec:significance}
\revise{
Workplace wellness programs are common but their true effectiveness remains unclear, partly because many employees offered such programs do not participate. This paper introduces a Bayesian machine learning approach that estimates how treatment effects vary across individuals when participation is incomplete by jointly modeling compliance and outcomes. 
In the Illinois Workplace Wellness Study, a large randomized wellness trial, the method reveals that the program increased health awareness in certain subgroups, despite minimal average effects. 
These results thus demonstrate that while wellness programs may not improve baseline health measures within a handful of years (the usual study length), participants may become more aware of their own health status, allowing for improved long-term habits and decisions. 
}

\section*{Acknowledgements}
We would like to thank Julian Reif, David Molitor, and Damon Jones for their advice with the data from their works \citep{qje,jama}.  We would also like to thank Jared Murray and Avi Feller for their ideas and support on the early variants of this project and Hyunseung Kang for several helpful discussions.  

The third author was supported in part by the University of Wisconsin–Madison, Office of the Vice Chancellor for Research and Graduate Education with funding from the Wisconsin Alumni Research Foundation.

{
\small
\bibliographystyle{apalike}
\bibliography{bcflate_refs}
}
\newpage
\appendix
\renewcommand{\thefigure}{\thesection\arabic{figure}}
\renewcommand{\thetable}{\thesection\arabic{table}}
\renewcommand{\theequation}{\thesection\arabic{equation}}

\setcounter{figure}{0}
\setcounter{equation}{0}
\setcounter{table}{0}
\section{CLATE Derivations}
\label{app:proof_clate}
Recall that the conditional local average treatment effect ($\clate$) is
$$
\clate(\bx) = \E[Y_{i}(1,1) - Y_{i}(0,0) \vert C_{i} = 1, \bX = \bx].
$$
This notation is made possible by \switchref{\Cref{assum:sutva,assum:onesided}}{Assumptions A1 and A2}.  

\subsection{Observables}
\label{sec:app_observables}
To show the identification of the estimators in the paper, we begin by formalizing the observed values. Potential outcome notation is made possible by \switchref{\Cref{assum:sutva}}{Assumption A1}. 
\begin{align}
    \robs_i &= A_i R_i(1) + (1-A_i)R_i(0) \label{eq:robs_raw}
    \\
    \yobs_i &= A_iY_i(1,R_i(1)) + (1-A_i)Y_i(0,R_i(0)) \label{eq:yobs_raw}
\end{align}
With the latter assumptions, these can be simplified to other useful quantities.

First, note that the definition of complier via \switchref{\Cref{assum:onesided}}{Assumption A2} means $C_i = R_i(1)$, which confirms that complier-status is latent when $A_i = 0$.    
Combined with \switchref{\Cref{assum:onesided}}{Assumption A2}, where $R_i(0) = 0$, we see that
\begin{equation}\robs_i = A_i C_i + (1-A_i)(0) = A_iC_i. \label{eq:robs}\end{equation}

Second, we note that $Y_i(1,R_i(1)) = C_iY_i(1,1) + (1-C_i)Y_i(1,0)$, such that \switchref{\Cref{assum:sutva,assum:onesided}}{Assumptions A1 and A2} also lead to 
        \begin{align*} \yobs_i 
        &= A_iY_i(1,R_i(1)) + (1-A_i)Y_i(0,R_i(0))
        \\&= A_i[ C_iY_i(1,1) + (1-C_i)Y_i(1,0)] + (1-A_i)Y_i(0,R_i(0))
        \\&= A_i C_i Y_i(1,1) + A_i(1-C_i) Y_i(1,0) + (1-A_i)Y_i(0,0) 
        \numberthis \label{eq:yobs_pre}
        \end{align*}
and, as such, it is simple to show that the three summands above point to the three quantities we can most easily estimate with observables: 
        \begin{align}
            \E[\yobs_i \vert A_i = 1, C_i = 1, \bX = \bx] &= \E[Y_i(1,1) \vert A_i = 1, C_i = 1, \bX = \bx] \label{eq:Eyobs11} \\
           \E[\yobs_i \vert A_i = 1, C_i = 0, \bX = \bx] &=  \E[Y_i(1,0) \vert A_i = 1, C_i = 0, \bX = \bx] \label{eq:Eyobs10} \\
            \E[\yobs_i \vert A_i = 0, \bX = \bx] &= \E[Y_i(0,0) \vert A_i = 0, \bX = \bx]. \label{eq:Eyobs0}
        \end{align}

With these preliminaries, we expand these statements with the next two assumptions.         
\switchref{\Cref{assum:exclusion}}{Assumption A3} yields $Y_i(0,0) = Y_i(1,0)$, which changes \Cref{eq:yobs_pre}  to
\begin{align*}
        \yobs_i  &= A_i C_i Y_i(1,1) + A_i(1-C_i) Y_i(0,0) + (1-A_i)Y_i(0,0)  
        \\&= A_i C_i Y_i(1,1) + (1-A_iC_i) Y_i(0,0). \numberthis \label{eq:yobs}
\end{align*}
\switchref{\Cref{assum:exclusion}}{Assumption A3} also augments \Cref{eq:Eyobs10} into:
        \begin{align*}
           &\E[\yobs_i \vert A_i = 1, C_i = 0, \bX = \bx] = \E[Y_i(1,0) \vert A_i = 1, C_i = 0, \bX = \bx] 
           \\&= \E[Y_i(0,0) \vert A_i = 1, C_i = 0, \bX = \bx].
        \end{align*}
\switchref{\Cref{assum:unconfounded_cond}}{Assumption A4} means many of the above equations' expectations of potential outcomes hold without being conditioned on assignment $A_i$: 
        \begin{align}
            \E[\yobs_i \vert A_i = 1, C_i = 1, \bX = \bx] &= \E[Y_i(1,1) \vert   C_i = 1, \bX = \bx] \label{eq:y11} \\
           \E[\yobs_i \vert A_i = 1, C_i = 0, \bX = \bx] &=  \E[Y_i(0,0) \vert   C_i = 0, \bX = \bx] \label{eq:y10} \\
            \E[\yobs_i \vert A_i = 0, \bX = \bx] &= \E[Y_i(0,0) \vert   \bX = \bx]. \label{eq:y00}
        \end{align}

\subsection{Identification with imputed compliance types}
\label{sec:app_clate}
Recall \switchref{\Cref{eq:conditional_late_id}}{Equation (3)}, which we reproduce here: 
$$\clate(\bx) = \E[\yobs_{i} \vert C_{i} = 1, A_{i} = 1, \bX_{i} = \bx] - \E[\yobs_{i} \vert C_{i} = 1, A_{i} = 0, \bX_{i} = \bx].$$
In this section we will derive this statement using the properties derived in \Cref{sec:app_observables}. 
We first recall the definition of CLATE, which can be split into two expectations
\begin{align*}
\clate(\bx) &= \E[Y_{i}(1,1) - Y_{i}(0,0) \vert C_{i} = 1, \bX = \bx] 
\\&= \E[Y_{i}(1,1)  \vert C_{i} = 1, \bX = \bx]  - \E[ Y_{i}(0,0) \vert C_{i} = 1, \bX = \bx]
\end{align*}
and we note that the first expectation can be identified with \Cref{eq:y11}.

The second expectation can be found by employing \Cref{eq:yobs}
\begin{align*}
&\E[\yobs_{i} \vert C_{i} = 1, A_{i} = 0, \bX_{i} = \bx] 
\\&=\E[A_i C_i Y_i(1,1) + (1-A_iC_i) Y_i(0,0) \vert C_{i} = 1, A_{i} = 0, \bX_{i} = \bx] 
\\&=\E[ Y_i(0,0) \vert C_{i} = 1, A_{i} = 0, \bX_{i} = \bx] 
\end{align*}
and by \switchref{\Cref{assum:unconfounded_cond}}{Assumption A4}, we can remove $A_i=0$ as a condition such that
\begin{equation}
\E[\yobs_{i} \vert C_{i} = 1, A_{i} = 0, \bX_{i} = \bx] = \E[ Y_i(0,0) \vert C_{i} = 1, \bX_{i} = \bx]. 
\end{equation}
This quantity has not been previously discussed as it is not identified from observables. 
However, \texttt{BCF-LATE} imputes compliance types for control subjects ($A_i = 0$) within each MCMC iteration, and thus $\E[\yobs_{i} \vert C_{i} = 1, A_{i} = 0, \bX_{i} = \bx]$ can be reasonably modeled.


\subsection{Identification of the conditional Wald estimator}
\label{sec:app_wald}
The standard estimator for the local average treatment effect for a binary instrument $A_i$ is the Wald estimator \citep{wald1940fitting}, which does not require assuming or imputing compliance types for the untreated observations. 
The traditional Wald estimator is a ratio of the $\itt_Y$ to $\itt_R$ \citep{imbens2015causal}

\begin{equation}
    \frac{\itt_Y}{\itt_R} := \frac{\E[Y_{i}(1, R_{i}(1)) - Y_i(0,R_i(0))  ] }{\E[R_{i}(1)  - R_{i}(0) ] }, 
    \label{eq:wal_definition_marginal}
\end{equation}
which similarly has a version conditional on covariates $\bm{x}$
\begin{equation}
    \frac{\itt_Y(\bx)}{\itt_R(\bx)} := \frac{\E[Y_{i}(1, R_{i}(1)) - Y_i(0,R_i(0)) \vert \bX_{i} = \bx] }{\E[R_{i}(1)  - R_{i}(0) \vert \bX_{i} = \bx] }. 
    \label{eq:wald_definition_conditional}
\end{equation}

We first show that $\itt_R(\bx)$ can be identified with $ \E[\robs_i | A_i=1,\bX_{i} = \bx] - \E[\robs_i|A_i=0,\bX_{i} = \bx]$ using \Cref{eq:robs_raw} and \switchref{\Cref{assum:unconfounded_cond}}{Assumption A4}.

\begin{align*}
    &\E[\robs_i | A_i=1,\bX_{i} = \bx] - \E[\robs_i|A_i=0,\bX_{i} = \bx]
     \\&= \E[A_iR_i(1) + (1-A_i)R_i(0) | A_i=1,\bX_{i} = \bx]
    \\& \quad 
    - \E[A_iR_i(1) + (1-A_i)R_i(0)|A_i=0,\bX_{i} = \bx]
    \\& = \E[ R_i(1) | A_i=1,\bX_{i} = \bx] - \E[ R_i(0)|A_i=0,\bX_{i} = \bx]
    \\& =\E[  R_i(1) \vert \bX_{i} = \bx]  - \E[  R_i(0) \vert \bX_{i} = \bx]
    \\& =\itt_R
\end{align*}

$\itt_Y(\bx)$ can be identified with $ \E[\yobs_i \vert A_i=1,\bX_{i} = \bx] - \E[\yobs_i \vert A_i=0,\bX_{i} = \bx]$ using \Cref{eq:yobs_raw} and \switchref{\Cref{assum:unconfounded_cond}}{Assumption A4}.

\begin{align*}
    &\E[\yobs_i | A_i=1,\bX_{i} = \bx] - \E[\yobs_i|A_i=0,\bX_{i} = \bx]
    \\& =\E[  A_iY_i(1,R_i(1)) + (1-A_i)Y_i(0,R_i(0)) | A_i=1,\bX_{i} = \bx] 
    \\& \quad 
    - \E[  A_iY_i(1,R_i(1)) + (1-A_i)Y_i(0,R_i(0))|A_i=0,\bX_{i} = \bx]
    \\& =\E[  Y_i(1,R_i(1)) | A_i=1,\bX_{i} = \bx] - \E[   Y_i(0,R_i(0))|A_i=0,\bX_{i} = \bx]
    \\& =\E[ Y_i(1,R_i(1)) \vert \bX_{i} = \bx ]  - \E[ Y_i(0,R_i(0)) \vert \bX_{i} = \bx]
    \\& =\itt_Y.
\end{align*}

Using \switchref{\Cref{assum:sutva,assum:onesided,assum:exclusion,assum:unconfounded_cond,assum:compliers_cond,assum:overlap_cond}}{Assumptions A1-A6}, we can now show  that $\clate(\bx)$ is identified by the conditional Wald estimator. 
Specifically,
$$
\clate(\bx) = \E[Y_{i}(1,1) - Y_{i}(0,0) \vert C_{i} = 1, \bX = \bx].
$$
is identified by a ratio of the above expectations on observables: 
\begin{equation}
 \frac{\itt_Y(\bx)}{\itt_R(\bx)}  = \frac{\E[\yobs_i \vert A_i=1 ,\bX_{i} = \bx] - \E[\yobs_i \vert A_i=0,\bX_{i} = \bx]}{\E[\robs_i \vert A_i=1,\bX_{i} = \bx] - \E[\robs_i \vert A_i=0,\bX_{i} = \bx]}. \label{eq:wald_ident_conditional}
 \end{equation}

First note the following simplifications of the expectations in the denominator which use \Cref{eq:robs}
\begin{align*}
    &\E[\robs_i \vert A_i = 0,\bX_{i} = \bx] = \E[A_i C_i \vert A_i = 0,\bX_{i} = \bx] =   0 
    \\&\E[\robs_i \vert A_i = 1,\bX_{i} = \bx] 
    = \E[A_i C_i \vert A_i = 1,\bX_{i} = \bx] = \E[C_i \vert A_i = 1,\bX_{i} = \bx] 
    \\&= \E[C_i\vert\bX_{i} = \bx] = \P(C_i = 1\vert\bX_{i} = \bx) = \pi_C (\bx) 
\end{align*}
where the second line uses the fact that $C_i = R_i(1)$ and \switchref{\Cref{assum:unconfounded_cond}}{Assumption A4}, i.e. $A_i \ci R_i(1)\vert\bX_{i} = \bx$, which together yield $A_i \ci C_i\vert\bX_{i} = \bx$. 
Thus, the conditional compliance rate $\pi_C(\bx) = \P(C_i = 1\vert\bX_{i} = \bx)$ is something we can identify.

Now for the first expectation in the numerator of \Cref{eq:wald_ident_conditional}, 
\begin{align*}
    &\E[\yobs_i \vert A_i = 1 ,\bX_{i} = \bx] 
    \\&= \E\left[A_i C_i Y_i(1,1) + (1-A_iC_i) Y_i(0,0) \vert A_i = 1 ,\bX_{i} = \bx\right]
    \\ &= \E\left[ C_i Y_i(1,1) + (1-C_i) Y_i(0,0) \vert A_i = 1,\bX_{i} = \bx \right]
    \\ &= \E\left[ C_i Y_i(1,1) +  (1-C_i) Y_i(0,0)\vert  \bX_{i} = \bx  \right] 
    \\ &= \P(C_i = 0\vert  \bX_{i} = \bx)\E\left[ C_i Y_i(1,1) +  (1-C_i) Y_i(0,0) \vert C_i = 0 ,  \bX_{i} = \bx \right] 
    \\& \quad + \P(C_i = 1\vert  \bX_{i} = \bx)\E\left[ C_i Y_i(1,1) +  (1-C_i) Y_i(0,0) \vert C_i = 1 ,  \bX_{i} = \bx \right]
    \\ &= \P(C_i = 0 \vert  \bX_{i} = \bx )\E\left[ Y_i(0,0) \vert C_i = 0 ,  \bX_{i} = \bx \right] 
    \\& \quad + \P(C_i = 1\vert   \bX_{i} = \bx )\E\left[ Y_i(1,1)  \vert C_i = 1 ,  \bX_{i} = \bx \right]
    \\ &= (1-\pi_C(\bx))\E\left[ Y_i(0,0) \vert C_i = 0  ,  \bX_{i} = \bx \right] + \pi_C(\bx) \E\left[ Y_i(1,1)  \vert C_i = 1  ,  \bX_{i} = \bx \right].
\end{align*}

And for the second expectation in the numerator, we use \Cref{eq:y00}, so that when we put it all together: 
\begin{align*}
    &\frac{\itt_Y(\bx)}{\itt_R(\bx)} \\
    &= \frac{\E[\yobs_i \vert A_i=1,  \bX_{i} = \bx] - \E[\yobs_i \vert A_i=0,  \bX_{i} = \bx]}{\E[\robs_i \vert A_i=1,  \bX_{i} = \bx] - \E[\robs_i \vert A_i=0,  \bX_{i} = \bx]}
    \\&= \frac{1}{\pi_C(\bx) - 0}\left\{    (1-\pi_C(\bx))\E\left[ Y_i(0,0) \vert C_i = 0  ,  \bX_{i} = \bx \right] \right.
    \\& \qquad \left. + \pi_C(\bx) \E\left[ Y_i(1,1)  \vert C_i = 1  ,  \bX_{i} = \bx \right]      - \E[Y_i(0,0)\vert \bX_{i} = \bx ]\right\}
    \\&= \frac{     (1-\pi_C(\bx))\E\left[ Y_i(0,0) \vert C_i = 0  ,  \bX_{i} = \bx \right] }{\pi_C(\bx) } 
    	+ \frac{      \pi_C(\bx) \E\left[ Y_i(1,1)  \vert C_i = 1  ,  \bX_{i} = \bx \right]     }{\pi_C(\bx)  }
    	\\& \qquad - \frac{  \E[Y_i(0,0)\vert \bX_{i} = \bx ]}{\pi_C(\bx)  }
    \\&= \frac{     (1-\pi_C(\bx))\E\left[ Y_i(0,0) \vert C_i = 0  ,  \bX_{i} = \bx \right] }{\pi_C(\bx) }     	+         \E\left[ Y_i(1,1)  \vert C_i = 1  ,  \bX_{i} = \bx \right]     
    	 \\& \qquad - \frac{ \pi_C(\bx) \E[Y_i(0,0)\vert C_i=1, \bX_{i} = \bx ]+(1-\pi_C(\bx))\E[Y_i(0,0)\vert C_i=0,\bX_{i} = \bx ]}{\pi_C(\bx)  }
    \\&=         \E\left[ Y_i(1,1)  \vert C_i = 1  ,  \bX_{i} = \bx \right]      -   \E[Y_i(0,0)\vert C_i=1, \bX_{i} = \bx ]
    \\&=         \E\left[ Y_i(1,1)      -   Y_i(0,0)\vert C_i=1, \bX_{i} = \bx \right]
    \\&=  \clate(\bx).
\end{align*}

Furthermore, as originally presented in \switchref{\Cref{eq:conditional_late_iv_estimate}}{Equation (5)}, for $\yobs_{i}, \robs_{i} \in \{0,1\}$, 
\begin{align*}
    \clate(\bx) 
   & = \frac{\E[\yobs_i \vert A_i=1,  \bX_{i} = \bx] - \E[\yobs_i \vert A_i=0,  \bX_{i} = \bx]}{\E[\robs_i \vert A_i=1,  \bX_{i} = \bx] - \E[\robs_i \vert A_i=0,  \bX_{i} = \bx]}
    \\&=     \frac{\P(\yobs_{i} = 1 \vert A_{i} = 1, \bX_{i} = \bx) - \P(\yobs_{i} = 1 \vert A_{i} = 0, \bX_{i} = \bx)}{\P(\robs_{i} = 1  \vert A_{i} = 1,  \bX_{i} = \bx) - \P(\robs_{i} = 1  \vert A_{i} = 0,  \bX_{i} = \bx)}
    \\&=     \frac{\P(\yobs_{i} = 1 \vert A_{i} = 1, \bX_{i} = \bx) - \P(\yobs_{i} = 1 \vert A_{i} = 0, \bX_{i} = \bx)}{\P(\robs_{i} = 1  \vert A_{i} = 1,  \bX_{i} = \bx)}
\end{align*}
the component probabilities of which be estimated with many different methods.

\setcounter{figure}{0}
\setcounter{equation}{0}
\setcounter{table}{0}
\section{Gibbs sampler derivation}
\label{app:gibbs_sampler}
To derive our Gibbs sampler, we introduce some additional notation.
Given a regression tree $(T, \calB)$ and a point $\bx,$ let $\ell(\bx; T)$ be the leaf of $T$ associated with $\bx.$
Further, let $g(\bx; T, \calB)$ be the evaluation function that returns the jump associated with $\ell(\bx;T);$ that is
$$
g(\bx; T, \calB) = \beta_{\ell(\bx;T)}.
$$
With this notation, we can write
$$
\mu(\bx) = \sum_{m = 1}^{M_{\mu}}{g(\bx; T_{m}^{(\mu)}, \calB_{m}^{(\mu)})}.
$$
We can analogously express each of $\mu_{c}(\bx), \tau(\bx),$ and $\eta(\bx)$ as sums of tree evaluations.
Finally, let $\boldsymbol{\Theta} = \{\btheta^{(\mu)}, \btheta^{\mu_{c}}, \btheta^{(\tau)}, \btheta^{(\eta)}\}$ denote the collection of prior variable splitting probabilities.

\subsection{Sampling latent utilities and missing compliance statuses}
\label{app:gibbs_sampler_latent}
Each iteration of our Gibbs sampler begins by first sampling the missing compliance statuses $\bm{C}^{(0)}$ and then drawing latent utilities $\tilde{Y}_{i}$ and $\tilde{C}_{i}$ for each subject.

The joint conditional density of unobserved compliance statuses for control subjects is given by
\begin{align}
\begin{split}
\label{eq:compliance_post_density}
p(\bm{C}^{(0)} \vert \bY, \bcalE) &\propto \prod_{i: a_{i} = 0}{\left\{ \left[\Phi\left(\mu(\bx_{i}) + \mu_{c}(\bx_{i}) \times c_{i}\right)\right]^{y_{i}}\times \left[1 - \Phi\left(\mu(\bx) + c_{i} \times \mu_{c}(\bx)\right)\right]^{1 - y_{i}} \right\}} \\
~&~\times \prod_{i:a_{i} = 0}{\left\{\Phi(\eta(\bx))^{c_{i}} \times \left[1 - \Phi\left(\eta(\bx_{i})\right)\right]^{1 - c_{i}}\right\}}.
\end{split}
\end{align}
From~\eqref{eq:compliance_post_density} we can conclude that the latent compliance statuses are conditionally independent given $\bcalE$ (which determines the functions $\mu, \mu_{c}, \tau,$ and $\eta$) and the observed outcomes $\bY.$ 
We further compute
\begin{align}
\begin{split}
\label{eq:compliance_conditional}
&\P(C_{i} = 1 \vert Y_{i} = 1, \bcalE) 
\\&\quad = \frac{\Phi\left(\eta(\bx)\right) \times \Phi\left(\mu(\bx_{i}) + \mu_{c}(\bx_{i})\right)}{\Phi\left(\eta(\bx)\right) \times \Phi\left(\mu(\bx_{i}) + \mu_{c}(\bx_{i})\right) + (1 - \Phi\left(\eta(\bx)\right)) \times \Phi\left(\mu(\bx_{i})\right)} 
\\ ~ & ~ 
\\&\P(C_{i} = 1 \vert Y_{i} = 0, \bcalE) 
\\&\quad = \frac{\Phi\left(\eta(\bx)\right) \times \left(1 - \Phi\left(\mu(\bx_{i}) + \mu_{c}(\bx_{i})\right)\right)}{\Phi\left(\eta(\bx)\right) \times \left(1 - \Phi\left(\mu(\bx_{i}) + \mu_{c}(\bx_{i})\right)\right) + (1 - \Phi\left(\eta(\bx)\right)) \times \left(1 - \Phi\left(\mu(\bx_{i})\right) \right)} .
\end{split}
\end{align} 

So, in the first step of each Gibbs sampling iteration, we sample a new value of $\bm{C}^{(0)}$ by independently drawing $C_{i}$'s according to Equation~\eqref{eq:compliance_conditional}.

Having drawn the missing compliance statuses, our data likelihood is now
\begin{align*}
&p(\bY, \bm{C} \vert \bcalE) \propto
\\& \prod_{i = 1}^{n}{\left[\Phi\left(f(\bx_{i}, c_{i}, a_{i}) \right)^{y_{i}} \times \left(1 - \Phi\left(f(\bx_{i}, c_{i}, a_{i})\right)\right)^{1 - y_{i}} \times \Phi\left(\eta(\bx_{i})\right)^{c_{i}} \times \left(1 - \Phi\left(\eta(\bx_{i})\right)\right)^{1 - c_{i}}\right]}. 
\end{align*}
Following \citet{albertchib1993_probit}, we introduce latent utilities $\tilde{Y}_{i}$ and $\tilde{C}_{i}$ such that $Y_{i} = \ind{\tilde{Y}_{i} \geq 0}$ and $C_{i} = \ind{\tilde{C}_{i} \geq 0},$ yielding the augmented likelihood
\begin{align}
\begin{split}
\label{eq:augmented_likelihood}
p(\bY, \bm{C},\bm{\tilde{Y}}, \bm{\tilde{C}} \vert \bcalE) &\propto \exp\left\{-\frac{1}{2}\sum_{i = 1}^{n}{(\tilde{y}_{i} - f(\bx_{i}, c_{i}, a_{i}))^{2}}\right\} \times \prod_{i = 1}^{n}{\ind{Y_{i} = \ind{\tilde{Y}_{i} \geq 0}}} \\
~ &~\times \exp\left\{-\frac{1}{2}\sum_{i = 1}^{n}{(\tilde{c}_{i} - \eta(\bx_{i}))^{2}}\right\} \times \prod_{i = 1}^{n}{\ind{C_{i} = \ind{\tilde{C}_{i} \geq 0}}}.
\end{split}    
\end{align}

From Equation~\eqref{eq:augmented_likelihood}, we conclude that
\begin{align}
\tilde{Y}_{i} &\sim \ind{Y_{i} = 1} \times \truncnormaldist{+}{f(\bx_{i}, c_{i}, a_{i})}{1} + \ind{Y_{i} = 0}\times \truncnormaldist{-}{f(\bx_{i}, c_{i}, a_{i})}{1} \label{eq:tildey} \\
\tilde{C}_{i} &\sim \ind{S_{i} = 1} \times \truncnormaldist{+}{\eta(\bx_{i})}{1} + \ind{C_{i} = 0}\times \truncnormaldist{-}{\eta(\bx_{i})}{1} \label{eq:tildes},
\end{align}
where $\mathcal{N}_{+}$ and $\mathcal{N}_{-}$ respectively denote truncations of a normal distribution to the positive and negative axes. 
The second step of each Gibbs sampling iteration begins by drawing latent utilities $\tilde{Y}_{i}$ and $\tilde{C}_{i}$ for each $i = 1, \ldots, n$ according to Equations~\eqref{eq:tildey} and~\eqref{eq:tildes}.

\subsection{Updating regression trees}
\label{app:gibbs_sampler_trees}

Once we draw the latent utilities $\tilde{\bY}$ and $\tilde{\bC},$ we sweep over the trees in $\calE,$ updating them one at a time conditionally on the fits of all other trees and the latent variables.
Each update involves sampling a regression tree $(T, \calB)$ from a particular conditional distribution whose density is of the form
\begin{equation}
\label{eq:tree_conditional_posterior}
p(T, \calB \vert \tilde{\bY}, \tilde{\bC}, \bcalE^{-}) \propto p(T)p(\calB \vert T)p(\tilde{\bY}, \tilde{\bC} \vert T, \calB, \bcalE^{-}),
\end{equation}
where $\bcalE^{-}$ denotes the collection of all other regression trees. 
We draw trees from such distributions in two steps.
First, we sample the decision tree $T$ from the corresponding marginal distribution.
This is done using a variant of the Metropolis-Hastings approach introduced by \citet{chipman1998} that only implements grow and prune moves.
Then we sample $\calB \vert T$ from its corresponding conditional.

From Equation~\eqref{eq:augmented_likelihood}, that the ``likelihood'' term $p(\tilde{\bY}, \tilde{\bC} \vert T, \calB, \bcalE^{-})$ depends on $\tilde{\bY}$ and $\tilde{\bC}$ through the \emph{full residuals}
\begin{align*}
R_{i}^{(y)} &= \tilde{y}_{i} - f(\bx_{i}, c_{i}, a_{i}) \\ R_{i}^{(c)} &= \tilde{c}_{i} - \eta(\bx_{i}).    
\end{align*}

\textbf{Updating $\calE^{(\mu)}$.} To update the $m$-th tree $(T_{m}^{(\mu)}, \calB_{m}^{(\mu)})$ in $\calE^{(\mu)},$ we set
\begin{equation}
\label{eq:mu_tree_posterior}
p(\tilde{\bY},  \tilde{\bC} \vert T, \calB, \bcalE^{-}) = \exp\left\{-\frac{1}{2}\left[\sum_{i = 1}^{n}{(r_{i}^{(\mu)} - g(\bx_{i}; T, \calB))^{2}}\right]\right\},
\end{equation}
where $r^{(\mu)}_{i}$ is the partial residual
\begin{equation}
\label{eq:partial_residual_mu}
r_{i}^{(\mu)} = R^{(y)}_{i} + g(\bx_{i}; T^{(\mu)}_{m}, \calB_{m}^{(\mu)}).
\end{equation}

\textbf{Updating $\calE^{(\mu_{c})}$.} To update the $m$-th tree $(T_{m}^{(\mu_{c})}, \calB_{m}^{(\mu_{c})})$ in $\calE^{(\mu_{c})},$ we set 
\begin{equation}
\label{eq:muc_tree_posterior}
p(\tilde{\bY}, \tilde{\bC} \vert T, \calB, \bcalE^{-}) = \exp\left\{-\frac{1}{2}\left[\sum_{i = 1}^{n}{(r_{i}^{(\mu_c)} - c_{i} \times g(\bx_{i}; T, \calB))^{2}}\right]\right\}.
\end{equation}
introduce the partial residual
\begin{equation}
\label{eq:partial_residual_mu_c}
r_{i}^{(\mu_{c})} = R^{(y)}_{i} + c_{i} \times g(\bx_{i}; T, \calB).
\end{equation}
Then the conditional posterior density of the $m$-th tree is

\textbf{Updating $\calE^{(\tau)}$.} To update the $m$-th tree $(T_{m}^{(\tau)}, \calB_{m}^{(\tau)})$ in $\calE^{(\tau)},$ we introduce the partial residual
\begin{equation}
\label{eq:partial_residual_tau}
r_{i}^{(\tau)} = R^{(y)}_{i} + c_{i} \times a_{i} \times g(\bx_{i}; T, \calB)
\end{equation}
Then the conditional posterior density of the $m$-th tree is 
\begin{equation}
\label{eq:tau_tree_posterior}
p(\tilde{\bY}, \tilde{\bC} \vert T, \calB, \bcalE^{-}) =  \exp\left\{-\frac{1}{2}\left[\sum_{i = 1}^{n}{(r_{i}^{(\tau)} - c_{i} \times a_{i} \times g(\bx_{i}; T, \calB))^{2}}\right]\right\}.
\end{equation}

\textbf{Updating $\calE^{(\eta)}$.} To update the $m$-th tree $(T_{m}^{(\eta)}, \calB_{m}^{(\eta)})$ in $\calE^{(\eta)}$, we set 
\begin{equation}
\label{eq:eta_tree_likelihood}
p(\tilde{\bY}, \tilde{\bC} \vert T, \calB, \bcalE^{-}) = \exp\left\{-\frac{1}{2}\sum_{i = 1}^{n}{\left(r_{i}^{(\eta)} - g(\bx; T, \calB)\right)^{2}}\right\},
\end{equation}
where
\begin{equation}
\label{eq:partial_residual_eta}
r_{i}^{(\eta)} = R^{(c)}_{i} + g(\bx_{i}; T, \calB).
\end{equation}

\subsection{Updating splitting probabilities}
\label{app:gibbs_sampler_dart}

Once we update each tree in $\bcalE,$ we update the elements of $\bTheta$ conditionally on $\bxi$ and then update each element of $\bxi,$ conditionally on $\bTheta.$
Update $\btheta^{(\mu)} \vert \xi^{(\mu)}$ involves nothing more than a Dirichlet-Multinomial update.
To update $\xi^{(\mu)} \vert \btheta^{(\mu)},$ we use the Metropolis algorithm where proposals are drawn from the $\text{Beta}(0.5,1)$ prior on $\xi^{(\mu)}/(\xi^{(\mu)} + p).$
We similarly update the other components of $\bTheta$ and $\bxi.$

\setcounter{figure}{0}
\setcounter{equation}{0}
\setcounter{table}{0}
\section{Additional simulation results}
\label{app:simtables}
\reviseTwo{In the main paper, we found that running chains for longer allowed standard MCMC convergence metrics to reach usual thresholds. However, we found that these longer MCMC runs did not materially change our results. Thus, for supplementary results, we simply conducted MCMC with four chains, each with 1000 burn-in iterations followed by 1,000 posterior draws, unless otherwise noted}.

\subsection{Simulations on a simple data-generating process with continuous covariates}

\label{sec:simulation_largep}

\revise{This simulation study assessed how well \texttt{BCF-LATE} performs when $n$ and $p$ increase but when the underlying data-generating functions are relatively simple.
For this study, for each combination of $n \in \{500, 2000, 5000\}$ and $p \in \{2, 5, 10, 25, 50, 75, 100\},$ we generated 100 datasets of size $n$ using the functions
}
\begin{align*}
    \eta(\bx) &= 4 \times \ind{x_2 \ge 0.5}-2 &
    \mu(\bx) = \mu_c(\bx) &= 0 &
    \tau(\bx) &= 4 \times \ind{x_1 \ge 0.5}-2. 
\end{align*}

\begin{figure}[h!bt]
\centering
\begin{subfigure}{2.7in}
\centering
\includegraphics[width = 2.7in]{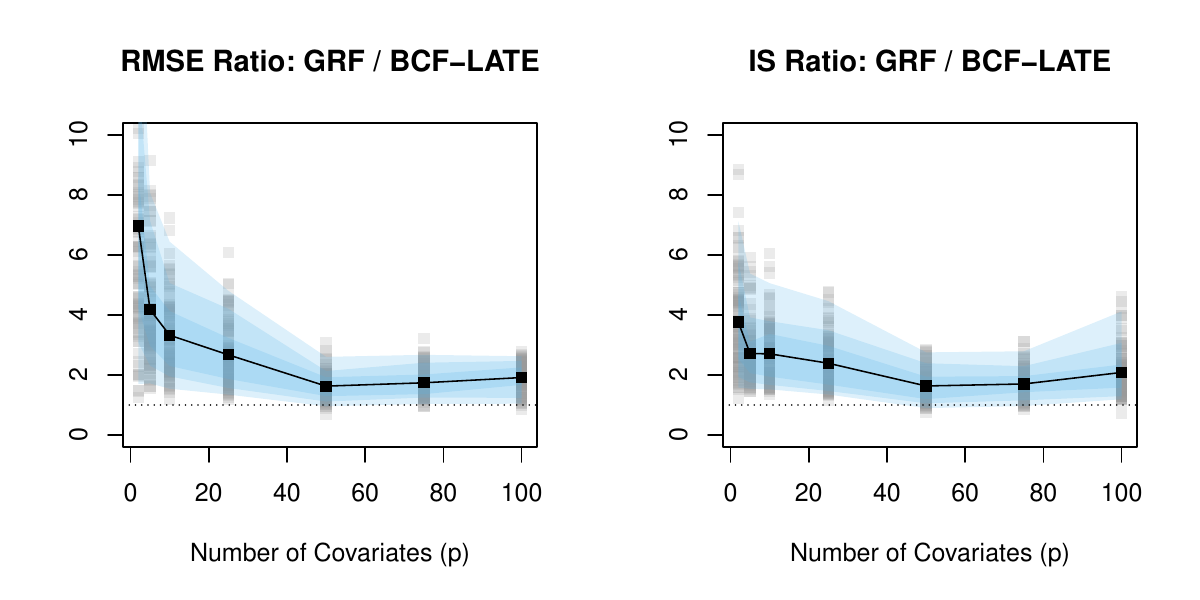}
\caption{}
\label{fig:sim-simpleDGP-increasingP}
\end{subfigure}
\begin{subfigure}{2.7in}
\centering
\includegraphics[width=2.7in]{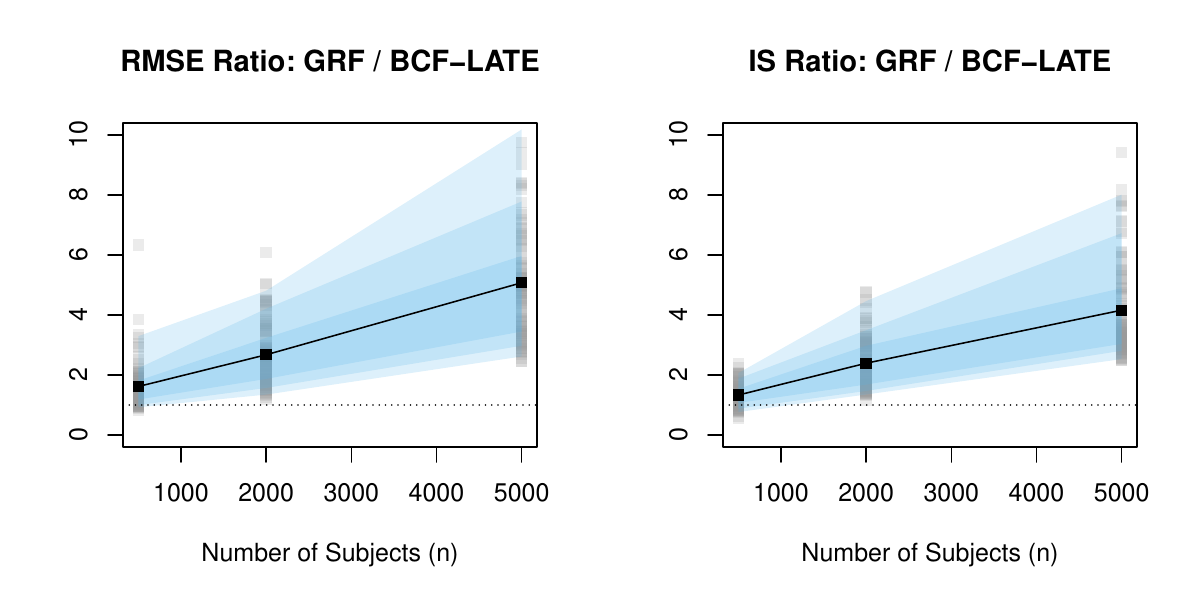}
\caption{}
\label{fig:sim-simpleDGP-increasingN}
\end{subfigure}
\caption{Comparison of \texttt{BCF-LATE} and \texttt{GRF} across 100 replications of the simple DGP when $n=2000$ is fixed and $p$ increases (a) and when $p=25$ is fixed and $n$ increases (b). Each simulation is represented with a transparent gray square. 
}
\label{fig:multi_simple}
\end{figure}

\revise{
\Cref{fig:sim-simpleDGP-increasingP} shows \texttt{GRF}'s RMSE (left) and interval score (IS, right) relative to \texttt{BCF-LATE}'s as a function of $p$ with $n = 2000$ fixed.
\Cref{fig:sim-simpleDGP-increasingN} is analogous and shows how \texttt{GRF}'s relative RMSE and interval score changes as $n$ increased while fixing $p = 25.$
The interval score combines interval coverage and interval width and is a proper scoring rule \citep{gneiting2007score}.
In the figures, values greater (resp. less) than one indicate that \texttt{BCF-LATE} performs better (resp. worse) than \texttt{GRF}. 
We see that for $n=2000$ and small $p$, \texttt{BCF-LATE} is markedly better than \texttt{GRF} at estimating $\clate(\bx)$.
While this performance gap shrinks as $p$ increases, it does not go away: for $p = 5,$ \texttt{BCF-LATE}'s RMSE is four times smaller than \texttt{GRF}'s and for $p = 100,$ it is two times smaller. 
On further inspection, we found that as $p$ increased, the width and coverage of \texttt{BCF-LATE}'s intervals increased while the width and coverage of \texttt{GRF}'s intervals decreased. 
Similarly, \texttt{BCF-LATE}'s edge in performance grows with the sample size $n$, for both point and interval estimation. 
\Cref{tab:sim-simpleDGP-increasingP,tab:sim-simpleDGP-increasingN} provide a more detailed tabulation of the results, specifically averages across simulations. 
}

\begin{table}[H]
\footnotesize
\centering
\caption{$\clate(\bx)$ estimation performance for \texttt{BCF-LATE} and \texttt{GRF} on a simple DGP. Metrics are averages for the 100 simulations, with n=2000. Coverage is 95\% interval coverage. IS is the interval score of \cite{gneiting2007score}, where smaller scores indicate better interval prediction. Ratios use \texttt{BCF-LATE} as a baseline, such that values greater than one indicate \texttt{BCF-LATE} performed better than \texttt{GRF} on average.}
\label{tab:sim-simpleDGP-increasingP}
\begin{tabular}{r|cccc|cccc|cc}
& \multicolumn{4}{c|}{\texttt{BCF-LATE}} & \multicolumn{4}{c|}{\texttt{GRF}} & \multicolumn{2}{c}{Ratio}  \\
  p & RMSE & Coverage  & Width & IS  & RMSE  & Coverage  & Width  & IS  & RMSE  & IS  \\ 
  \hline
   2 & 0.115 & 0.883 & 0.352 & 0.018 & 0.797 & 0.955 & 2.503 & 0.066 & 6.965 & 3.765 \\ 
     5 & 0.110 & 0.901 & 0.356 & 0.017 & 0.455 & 0.955 & 1.678 & 0.045 & 4.181 & 2.714 \\ 
    10 & 0.108 & 0.910 & 0.367 & 0.016 & 0.355 & 0.966 & 1.665 & 0.043 & 3.326 & 2.706 \\ 
    25 & 0.103 & 0.924 & 0.385 & 0.016 & 0.273 & 0.977 & 1.469 & 0.038 & 2.673 & 2.392 \\ 
    50 & 0.101 & 0.931 & 0.405 & 0.016 & 0.164 & 0.933 & 0.897 & 0.026 & 1.629 & 1.634 \\ 
    75 & 0.102 & 0.930 & 0.404 & 0.016 & 0.177 & 0.836 & 0.727 & 0.027 & 1.740 & 1.701 \\ 
   100 & 0.116 & 0.926 & 0.452 & 0.017 & 0.213 & 0.730 & 0.704 & 0.035 & 1.915 & 2.086 \\ 
   \hline
\end{tabular}

\end{table}

\begin{table}[H]
\footnotesize
\centering
\caption{$\clate(\bx)$ estimation performance for \texttt{BCF-LATE} and \texttt{GRF} on a simple data-generating process. Metrics are averages for the 100 simulations, with p = 25. Coverage is 95\% interval coverage, and ideally is 0.95. IS is the interval score of \cite{gneiting2007score}, where smaller scores indicate better interval prediction. Ratios use \texttt{BCF-LATE} as a baseline, such that values greater than one indicate \texttt{BCF-LATE} performed better than \texttt{GRF} on average.}
\label{tab:sim-simpleDGP-increasingN}

\begin{tabular}{r|cccc|cccc|cc}
& \multicolumn{4}{c|}{\texttt{BCF-LATE}} & \multicolumn{4}{c|}{\texttt{GRF}} & \multicolumn{2}{c}{Ratio}  \\
  n & RMSE & Coverage  & Width & IS  & RMSE  & Coverage  & Width  & IS  & RMSE  & IS  \\ 
  \hline
 500 & 0.210 & 0.730 & 0.447 & 0.037 & 0.337 & 0.886 & 1.535 & 0.048 & 1.616 & 1.332 \\ 
  2000 & 0.103 & 0.924 & 0.385 & 0.016 & 0.273 & 0.977 & 1.469 & 0.038 & 2.673 & 2.392 \\ 
  5000 & 0.069 & 0.965 & 0.326 & 0.011 & 0.350 & 0.989 & 1.752 & 0.044 & 5.076 & 4.161 \\ 
   \hline
\end{tabular}
\end{table}

\subsection{Simulations on a challenging data-generating process with continuous covariates}
\revise{
This subsection contains the tables that give greater detail about the simulation in \switchref{\Cref{sec:sim_complicated}}{Section 5.2}. \reviseTwo{MCMC for each simulation consisted of four chains, each with 1000 burn-in iterations followed by  50,000 iterations, thinned by  50 to have 1000 posterior samples per chain}. 
}

\begin{table}[ht]
\footnotesize
\centering
    \caption{$\clate(\bx)$ estimation performance for \texttt{BCF-LATE} and \texttt{GRF} on a data-generating process with continuous covariates.  Metrics are averages for 100 simulations, with n=2000 and only five covariates involved in the DGP.
Coverage is 95\% interval coverage, and ideally is 0.95. IS is the interval score of \cite{gneiting2007score}, where smaller scores indicate better interval prediction. Ratios use \texttt{BCF-LATE} as a baseline, such that values greater than one indicate \texttt{BCF-LATE} performed better than \texttt{GRF} on average. The last two columns report the percent of simulations where \texttt{BCF-LATE} performed better than \texttt{GRF}.}
    \label{tab:sim-hardDGP-increasingP}

\begin{tabular}{r|cccc|cccc|cc|cc}
& \multicolumn{4}{c|}{\texttt{BCF-LATE}} & \multicolumn{4}{c|}{\texttt{GRF}} & \multicolumn{2}{c}{Ratio} & \multicolumn{2}{c}{\% BCFL $<$ GRF}  \\
  n & RMSE & Coverage  & Width & IS  & RMSE  & Coverage  & Width  & IS  & RMSE  & IS & RMSE & IS  \\   \hline
  \hline
 5 & 0.084 & 0.928 & 0.295 & 0.392 & 0.116 & 0.880 & 0.448 & 0.724 & 1.411 & 1.930 & 0.960 & 0.970 \\ 
    10 & 0.087 & 0.921 & 0.306 & 0.411 & 0.114 & 0.944 & 0.572 & 0.687 & 1.330 & 1.718 & 0.970 & 0.990 \\ 
    25 & 0.090 & 0.925 & 0.321 & 0.428 & 0.111 & 0.945 & 0.580 & 0.705 & 1.256 & 1.725 & 0.960 & 0.980 \\ 
    50 & 0.092 & 0.928 & 0.330 & 0.430 & 0.112 & 0.946 & 0.604 & 0.732 & 1.216 & 1.743 & 0.980 & 0.990 \\ 
    75 & 0.095 & 0.918 & 0.333 & 0.451 & 0.111 & 0.942 & 0.587 & 0.732 & 1.179 & 1.672 & 0.910 & 0.980 \\ 
   100 & 0.097 & 0.921 & 0.341 & 0.455 & 0.113 & 0.933 & 0.576 & 0.740 & 1.184 & 1.671 & 0.920 & 0.980 \\ 
   \hline
\end{tabular}

\end{table}

\begin{table}[H]
\footnotesize
\centering
\caption{$\clate(\bx)$ estimation performance for \texttt{BCF-LATE} and \texttt{GRF} on a data-generating process with continuous covariates. Metrics are averages for the 100 simulations, with p = 25 but only five covariates involved in the DGP.
Coverage is 95\% interval coverage, and ideally is 0.95. IS is the interval score of \cite{gneiting2007score}, where smaller scores indicate better interval prediction. Ratios use \texttt{BCF-LATE} as a baseline, such that values greater than one indicate \texttt{BCF-LATE} performed better than \texttt{GRF} on average. The last two columns report the percent of simulations where \texttt{BCF-LATE} performed better than \texttt{GRF}.}
    \label{tab:sim-hardDGP-increasingN}
\begin{tabular}{r|cccc|cccc|cc|cc}
& \multicolumn{4}{c|}{\texttt{BCF-LATE}} & \multicolumn{4}{c|}{\texttt{GRF}} & \multicolumn{2}{c}{Ratio} & \multicolumn{2}{c}{\% BCFL $<$ GRF}  \\
  n & RMSE & Coverage  & Width & IS  & RMSE  & Coverage  & Width  & IS  & RMSE  & IS & RMSE & IS  \\   \hline
  \hline
500 & 0.105 & 0.926 & 0.398 & 0.525 & 0.135 & 0.954 & 0.723 & 0.841 & 1.298 & 1.659 & 0.870 & 0.970 \\ 
  2000 & 0.090 & 0.925 & 0.321 & 0.428 & 0.111 & 0.945 & 0.580 & 0.705 & 1.256 & 1.725 & 0.960 & 0.980 \\ 
  5000 & 0.073 & 0.939 & 0.267 & 0.332 & 0.100 & 0.944 & 0.534 & 0.654 & 1.390 & 2.027 & 0.990 & 1.000 \\ 
   \hline
\end{tabular}

\end{table}

\subsubsection{Different numbers of trees}
\label{sec:simulation_trees}
\begin{figure}[h!bt]
\centering
\begin{subfigure}{2.7in}
\centering
\includegraphics[width=2.7in]{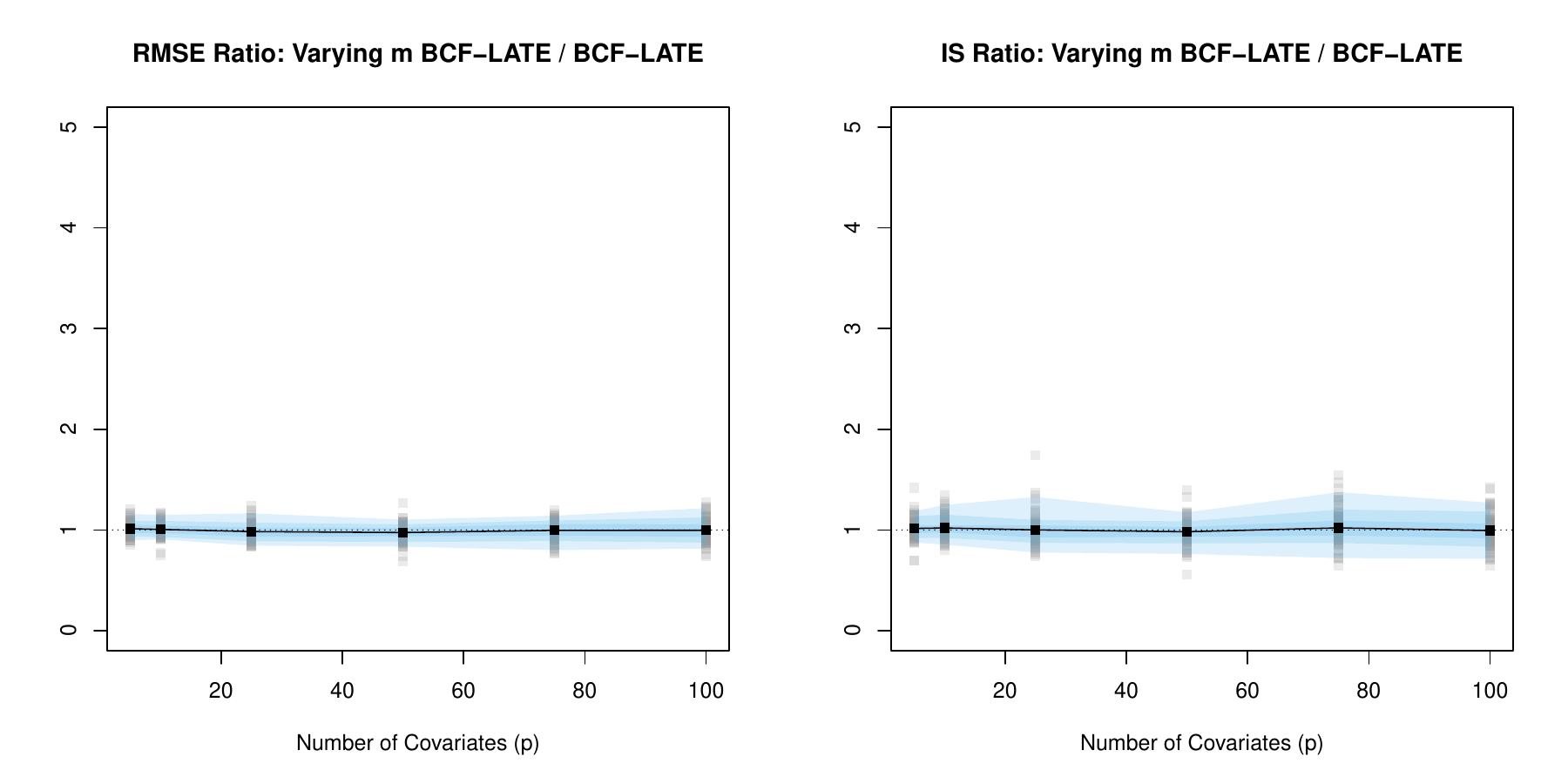}
\caption{}
\end{subfigure}
\begin{subfigure}{2.7in}
\centering
\includegraphics[width = 2.7in]{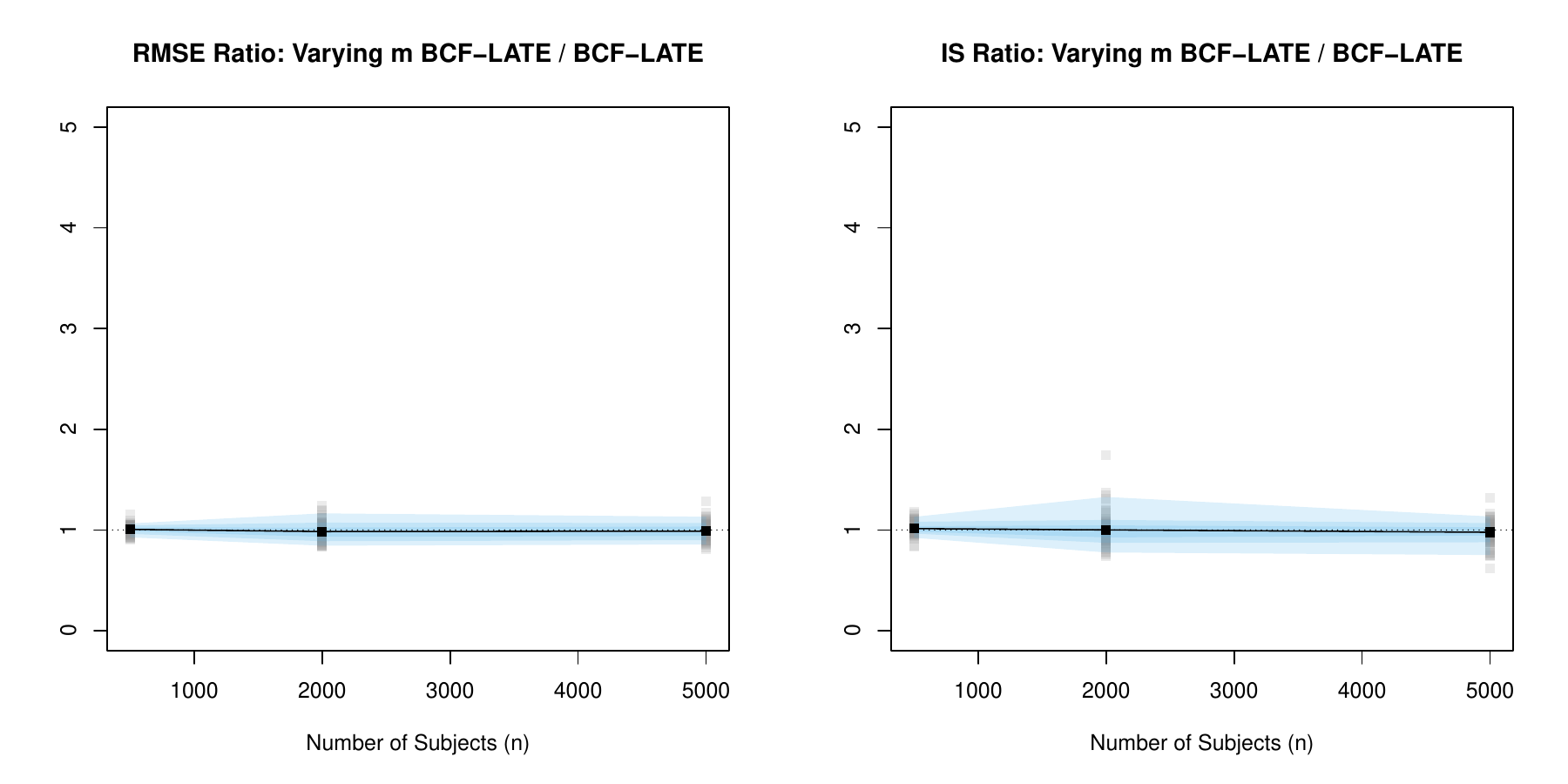}
\caption{}
\end{subfigure}
\caption{Comparison of \texttt{BCF-LATE} with default settings and \texttt{BCF-LATE} with different number of trees in the $\tau$ and $\eta$ ensembles. Results are averaged across 100 replications of the continuous-covariates DGP when $n=2000$ is fixed and $p$ increases (a) and when $p=25$ is fixed and $n$ increases (b). Each simulation is represented with a transparent gray square. 
}
\label{fig:change_M}
\end{figure}

\revise{
Using this simulation, we explored the effect of our particular hyperparameter choices. First, we changed the number of trees in the different ensembles. Our suggested default is to use the same number of trees in each ensemble: $M^{(\mu)} = M^{(\mu_{c})} = M^{(\tau)} = M^{(\eta)} = 50$. As $\eta(\bx)$ may be quite complicated, and as we already regularize $\tau(\bx)$ to be simple, we examine what happens if we set   $ M^{(\tau)} = 20$ and $M^{(\eta)} = 200$. The results are shown in \Cref{fig:change_M}, and we see that straying from our proposed defaults yields no visible change in point or interval accuracy. 
}

\newpage

\subsubsection{Different prior variances}
\label{sec:simulation_variances}
\begin{figure}[h!bt]
\centering
\begin{subfigure}{2.7in}
\centering
\includegraphics[width=2.7in]{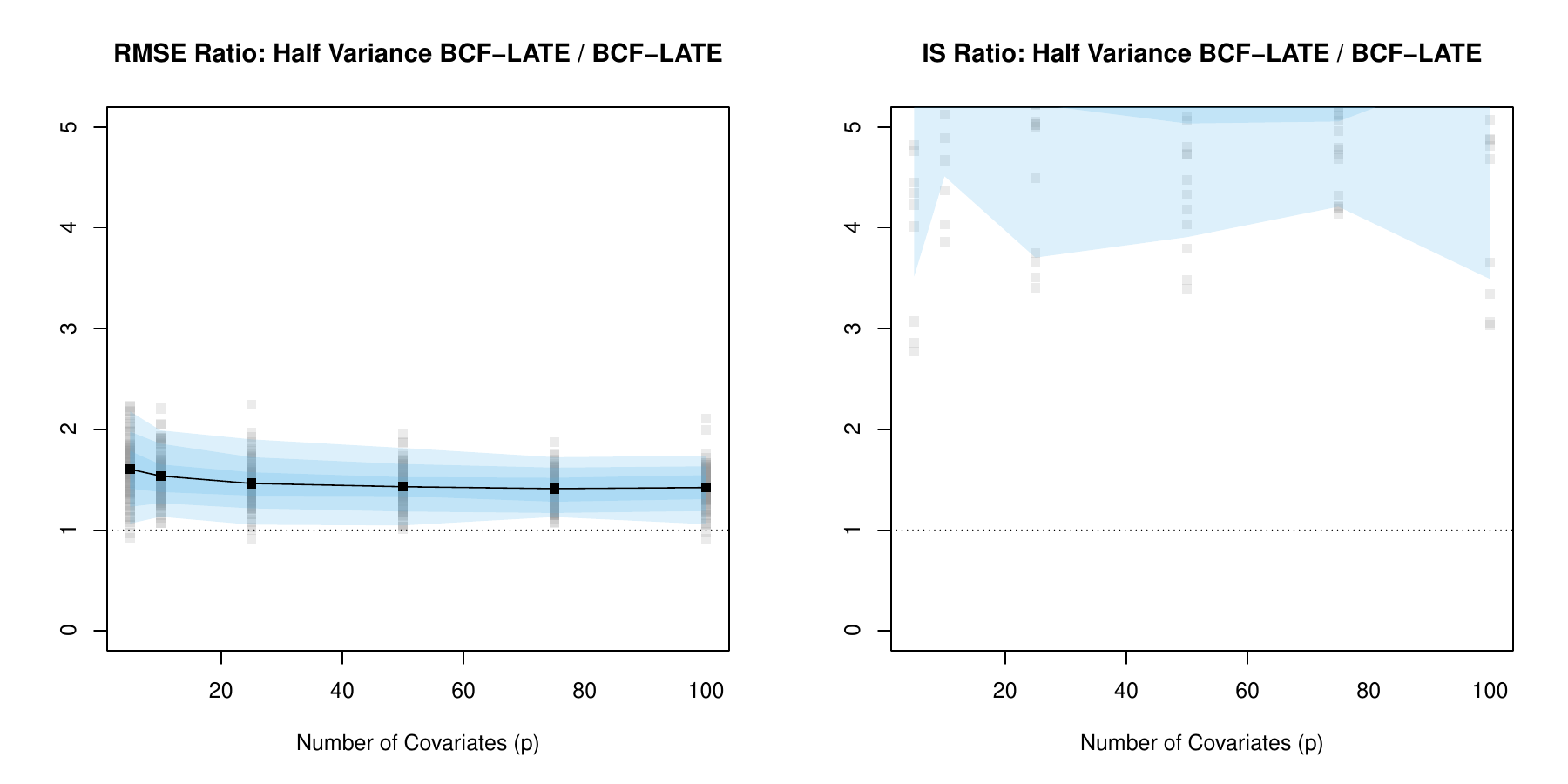}
\caption{}
\end{subfigure}
\begin{subfigure}{2.7in}
\centering
\includegraphics[width = 2.7in]{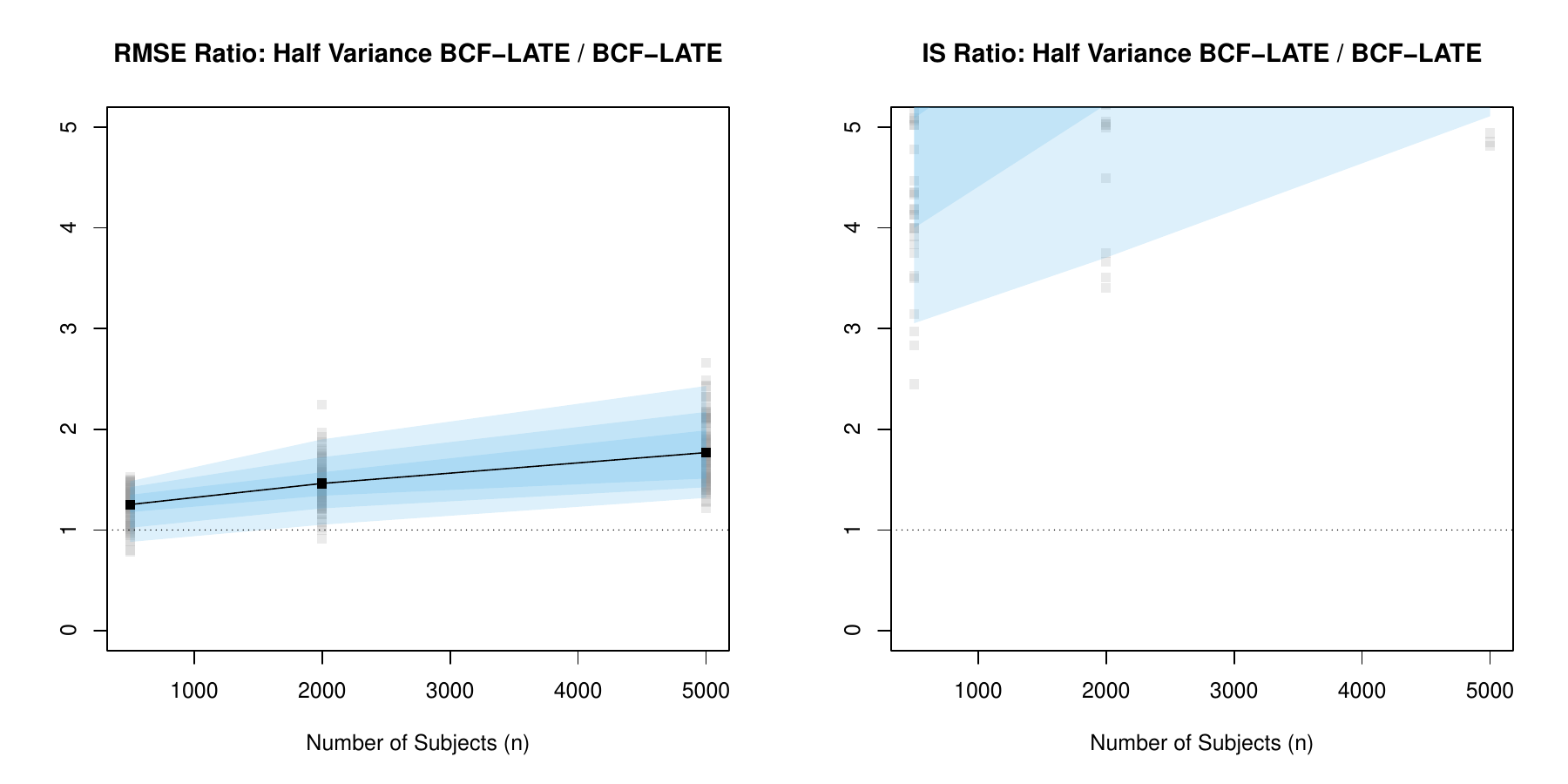}
\caption{}
\end{subfigure}
\caption{Comparison of \texttt{BCF-LATE} with default settings and \texttt{BCF-LATE} with leaf variances halved. Results are averaged across 100 replications of the continuous-covariates DGP when $n=2000$ is fixed and $p$ increases (a) and when $p=25$ is fixed and $n$ increases (b). Each simulation is represented with a transparent gray square. 
}
\label{fig:half_var}
\end{figure}

\begin{figure}[h!bt]
\centering
\begin{subfigure}{2.7in}
\centering
\includegraphics[width=2.7in]{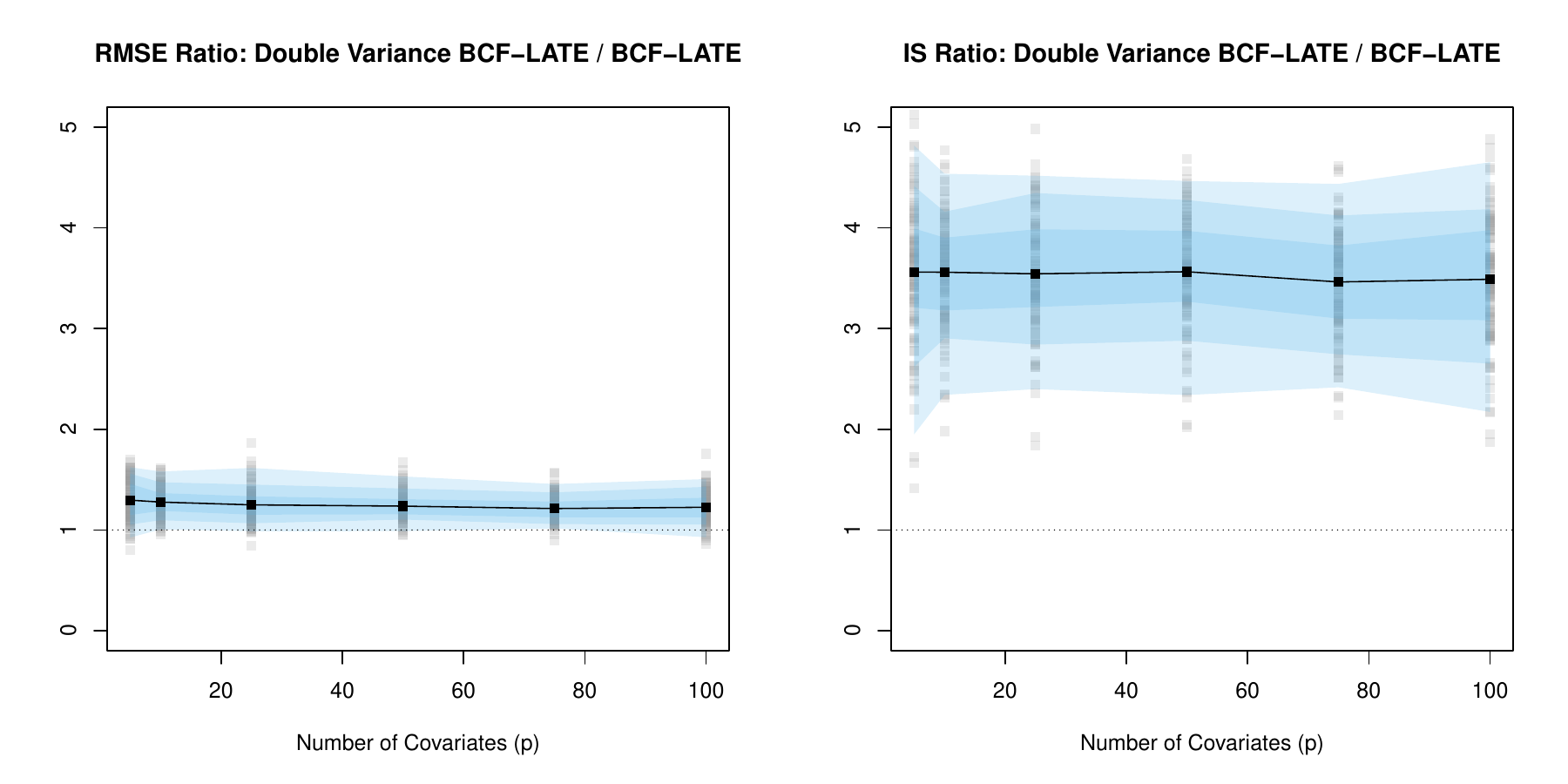}
\caption{}
\end{subfigure}
\begin{subfigure}{2.7in}
\centering
\includegraphics[width = 2.7in]{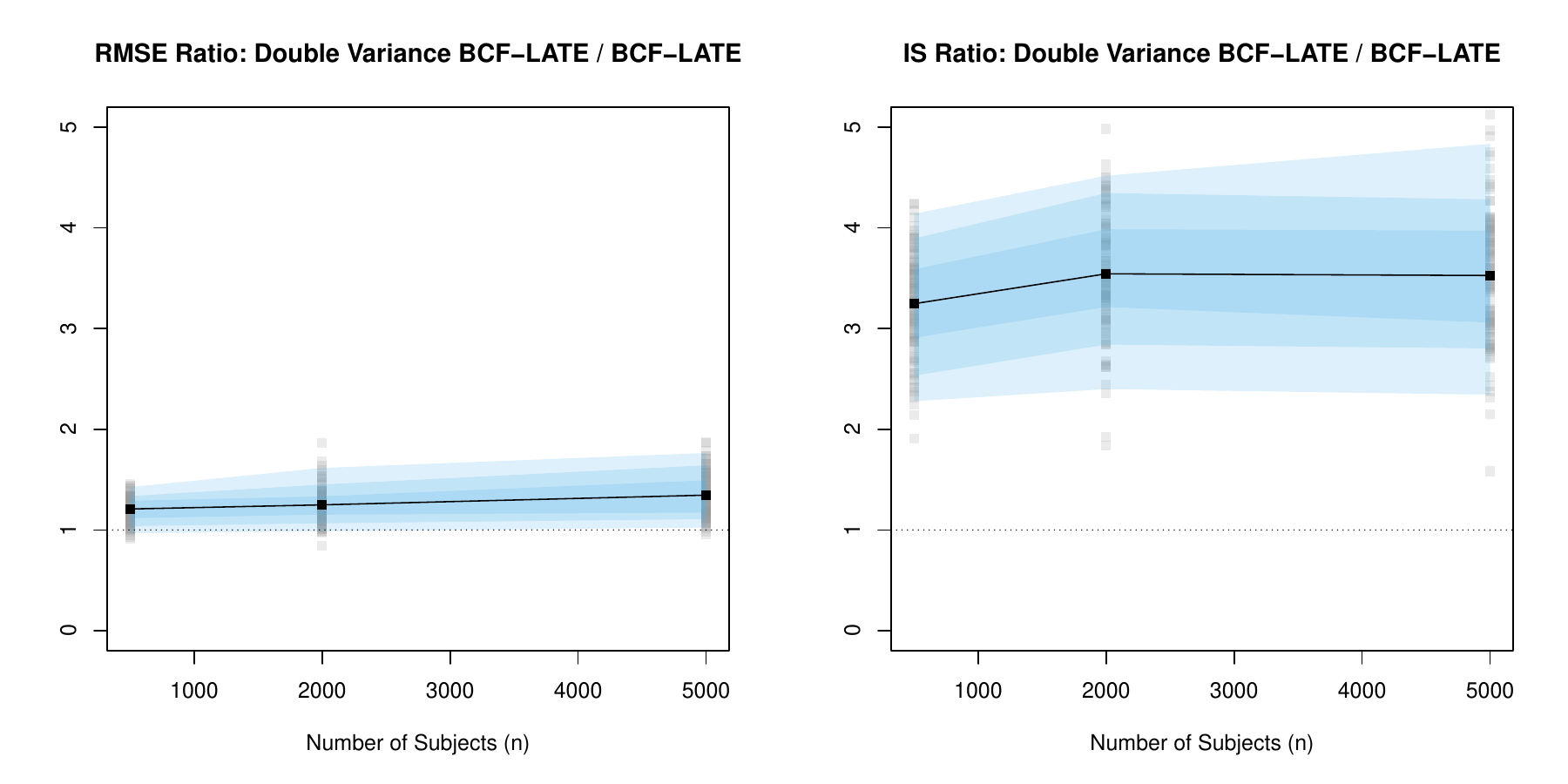}
\caption{}
\end{subfigure}
\caption{Comparison of \texttt{BCF-LATE} with default settings and \texttt{BCF-LATE} with leaf variances doubled. Results are averaged across 100 replications of the continuous-covariates DGP when $n=2000$ is fixed and $p$ increases (a) and when $p=25$ is fixed and $n$ increases (b). Each simulation is represented with a transparent gray square. 
}
\label{fig:double_var}
\end{figure}

\revise{
Next, we examine if our leaf node prior variance choices are appropriate. We propose $\sigma^{(\mu)} = \sigma^{(\eta)} =1.5 $ and $\sigma^{(\mu_c)} = \sigma^{(\tau)} =0.5 $ as we believe the effects of $\mu_c$ and $\tau$ to be smaller than the others.  \Cref{fig:half_var} depicts what happens when all four of the default variances are halved, while \Cref{fig:double_var} shows the results when the variances are doubled. In both cases, we see that RMSE is slightly worse, while interval scores are drastically worse. We thus conclude that, while there may be slight tweaks to our proposed prior variances that may be optimal, the current proposed defaults are reasonable, supporting our reasons for these choices in \switchref{\Cref{sec:prior_reg}}{Section 4.1}.
}

\newpage

\subsubsection{Sparsity through the DART prior}
\begin{figure}[h!bt]
\centering
\centering
\includegraphics[width=2.7in]{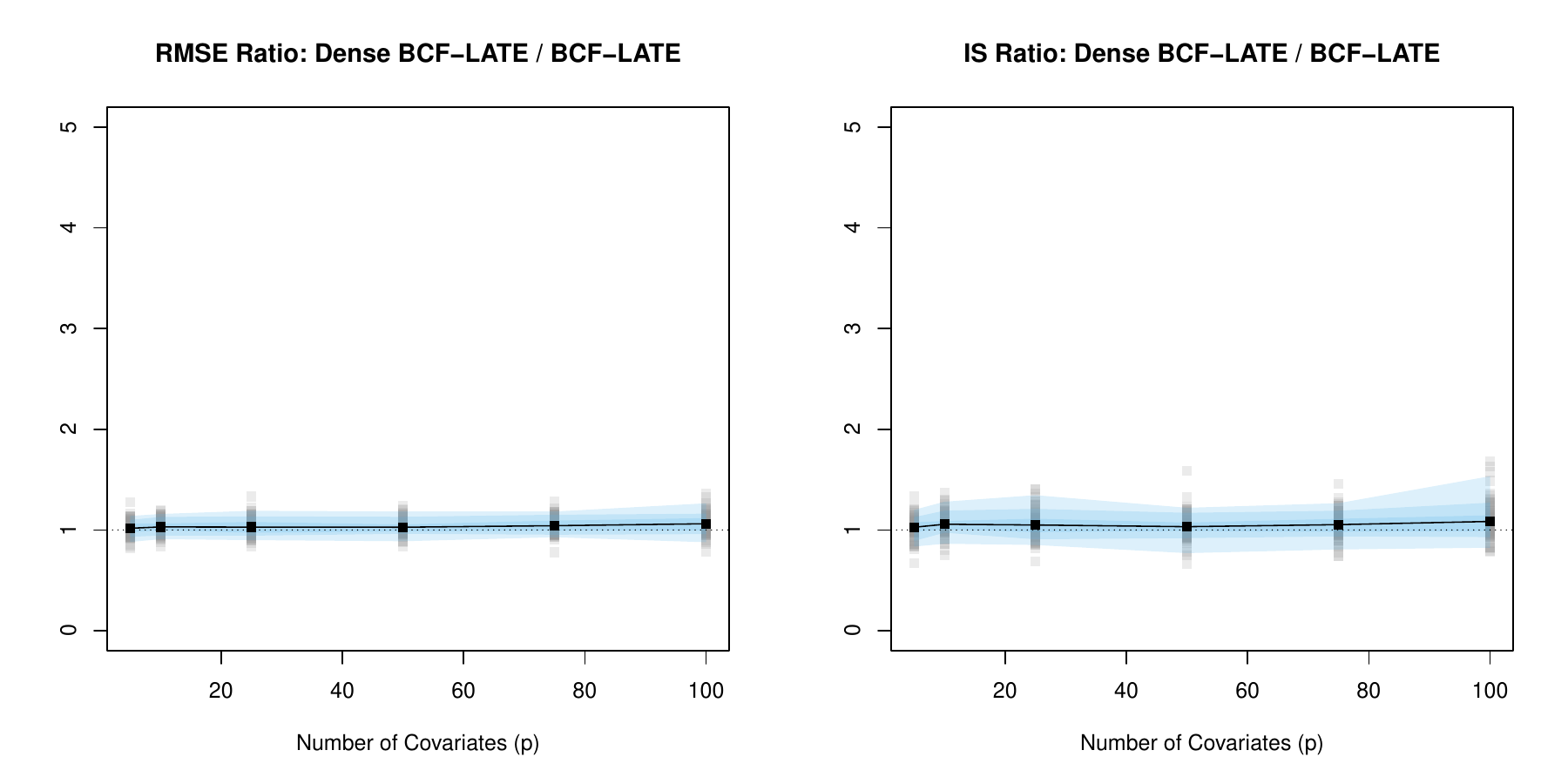}
\caption{Comparison of \texttt{BCF-LATE} with the DART prior (default setting) and \texttt{BCF-LATE} with a uniform prior on variable selection. Results are averaged across 100 replications of the continuous-covariates DGP when $n=2000$ is fixed and $p$ increases.
 Each simulation is represented with a transparent gray square. 
}
\label{fig:dart}
\end{figure}

\revise{
Lastly, we examine whether the use of the ``DART" prior of \citet{linero2018dart}, which encourages sparsity in which covariates are used, is reasonably different from a traditional uniform prior. 
\Cref{fig:dart} compares these two approaches. Recall that the data-generating process only uses five covariates. For when the two approaches are fit with $p=5$, i.e. the five useful covariates, their performances are comparable both in terms of RMSE and IS. As the number of spurious covariates increases (i.e. any $p>5$), we see the average ratios rise slightly above one, meaning the sparsity employed by DART is useful on average, as we would expect. Overall, there appears to be little difference between the two, so we suggest following the literature and using the DART prior as the default \citep{linero2018dart,vcbart}.
}

\newpage

\subsection{Simulations on a challenging data-generating process with binary covariates}
\label{sec:simulation_binary_appendix}
\revise{
This subsection contains the tables that give greater detail about the simulation in \switchref{\Cref{sec:simulation_binary}}{Section 5.3}.  \reviseTwo{MCMC for each simulation consisted of four chains, each with 1000 burn-in iterations followed by  50,000 iterations, thinned by  50 to have 1000 posterior samples per chain}. 
}

\begin{table}[ht]
\caption{$\clate(\bx)$ estimation performance for \texttt{BCF-LATE} and \texttt{GRF} on a data-generating process with categorical covariates. Metrics are averages for the 100 simulations, with p = 3. Coverage is 95\% interval coverage, and ideally is 0.95. IS is the interval score of \cite{gneiting2007score}, where smaller scores indicate better interval prediction. 
Ratios use \texttt{BCF-LATE} as a baseline, such that values greater than one indicate \texttt{BCF-LATE} performed better than \texttt{GRF} on average. The last two columns report the percent of simulations where \texttt{BCF-LATE} performed better than \texttt{GRF}.}
\centering
\begin{tabular}{r|cccc|cccc|cc|cc}
& \multicolumn{4}{c|}{\texttt{BCF-LATE}} & \multicolumn{4}{c|}{\texttt{GRF}} & \multicolumn{2}{c}{Ratio} & \multicolumn{2}{c}{\% BCFL $<$ GRF}  \\
  n & RMSE & Coverage  & Width & IS  & RMSE  & Coverage  & Width  & IS  & RMSE  & IS & RMSE & IS  \\   \hline
 500 & 0.068 & 0.984 & 0.349 & 0.365 & 0.087 & 0.910 & 0.368 & 0.583 & 1.297 & 1.548 & 0.720 & 0.590 \\ 
  2000 & 0.056 & 0.957 & 0.247 & 0.296 & 0.064 & 0.807 & 0.204 & 0.451 & 1.188 & 1.618 & 0.650 & 0.630 \\ 
  5000 & 0.047 & 0.938 & 0.188 & 0.216 & 0.054 & 0.790 & 0.147 & 0.388 & 1.193 & 1.833 & 0.590 & 0.740 \\ 
   \hline
\end{tabular}
\end{table}

\begin{table}[ht]
\caption{$\clate(\bx)$ estimation performance for \texttt{BCF-LATE} and \texttt{GRF} on a data-generating process with categorical covariates. Metrics are averages for the 100 simulations, with n = 2000. Coverage is 95\% interval coverage, and ideally is 0.95. IS is the interval score of \cite{gneiting2007score}, where smaller scores indicate better interval prediction. Ratios use \texttt{BCF-LATE} as a baseline, such that values greater than one indicate \texttt{BCF-LATE} performed better than \texttt{GRF} on average. The last two columns report the percent of simulations where \texttt{BCF-LATE} performed better than \texttt{GRF}.}
\centering
\begin{tabular}{r|cccc|cccc|cc|cc}
& \multicolumn{4}{c|}{\texttt{BCF-LATE}} & \multicolumn{4}{c|}{\texttt{GRF}} & \multicolumn{2}{c}{Ratio} & \multicolumn{2}{c}{\% BCFL $<$ GRF}  \\
  p & RMSE & Coverage  & Width & IS  & RMSE  & Coverage  & Width  & IS  & RMSE  & IS  & RMSE & IS  \\   \hline
 3 & 0.056 & 0.957 & 0.247 & 0.296 & 0.064 & 0.807 & 0.204 & 0.451 & 1.188 & 1.618 & 0.650 & 0.630 \\ 
     5 & 0.064 & 0.933 & 0.252 & 0.328 & 0.066 & 0.789 & 0.187 & 0.468 & 1.046 & 1.405 & 0.500 & 0.690 \\ 
     7 & 0.067 & 0.937 & 0.265 & 0.331 & 0.065 & 0.787 & 0.179 & 0.501 & 0.966 & 1.479 & 0.370 & 0.650 \\ 
   \hline
\end{tabular}
\end{table}

\setcounter{figure}{0}
\setcounter{equation}{0}
\setcounter{table}{0}
\section{Additional empirical results}
\label{app:grf_comp}
\begin{figure}[H]
\centering
\includegraphics[width = 4.2in]{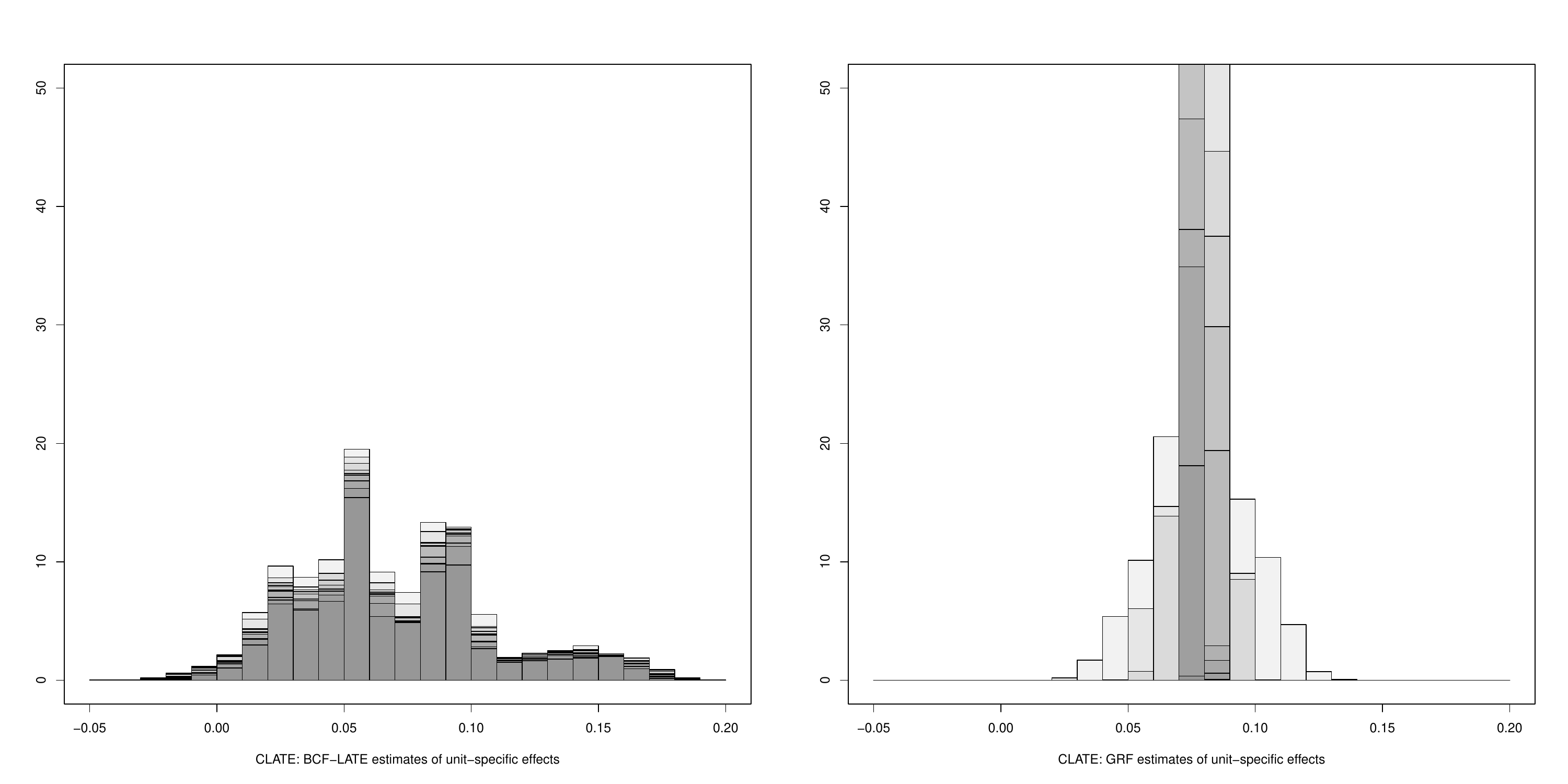}
\caption{Comparison of unit-level $\clate$ estimates across multiple runs of \texttt{BCF-LATE} and \texttt{GRF} for the outcome ``mgmtsafety\_0717'': whether subjects felt that their managers prioritized employee wellbeing.}
\label{fig:sens_mgmtsafety}
\end{figure}

\begin{figure}[H]
\centering
\includegraphics[width = 4.2in]{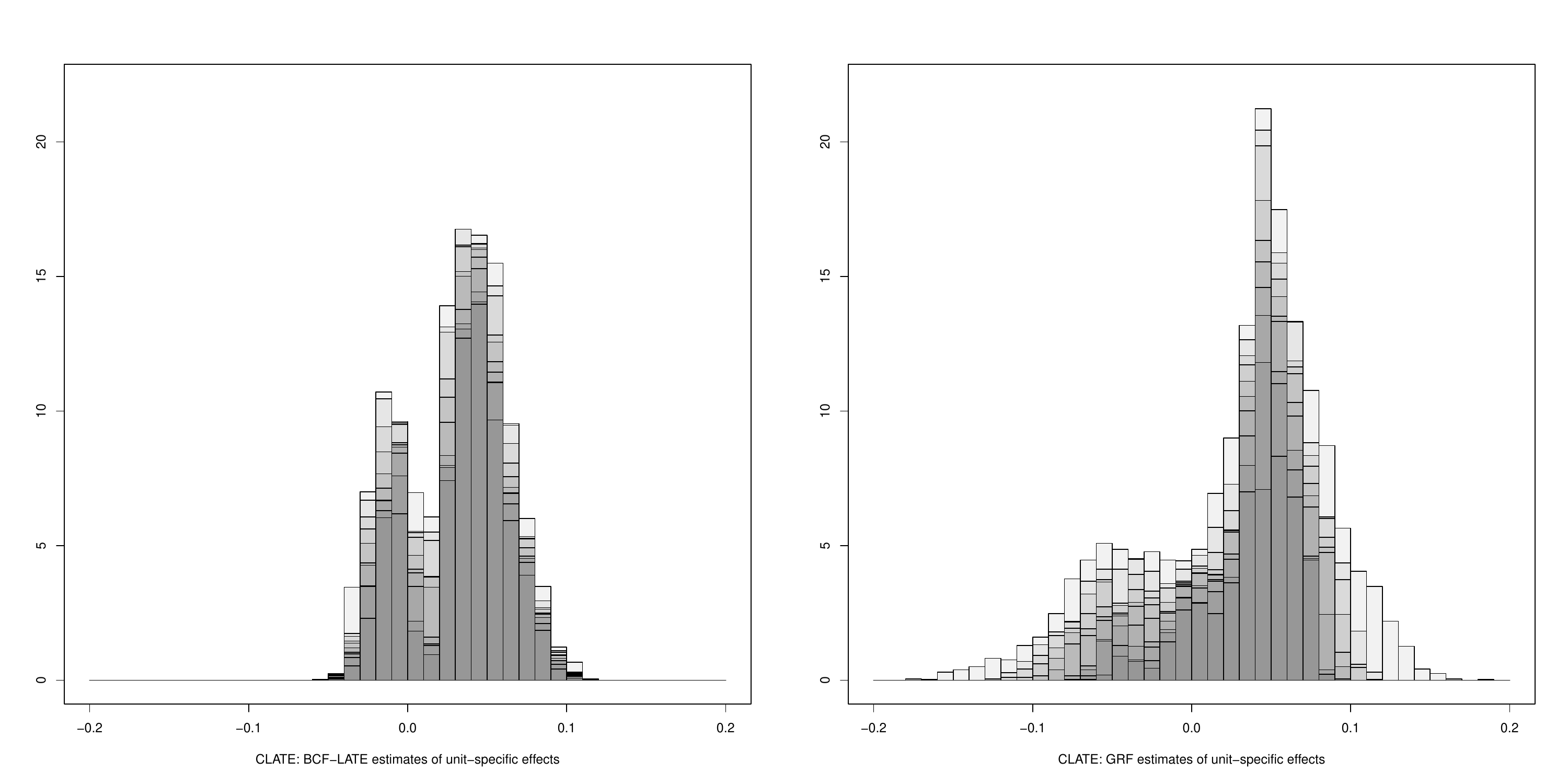}
\caption{Comparison of unit-level $\clate$ estimates across multiple runs of \texttt{BCF-LATE} and \texttt{GRF}  for the outcome ``badhealth\_0717'': whether subjects self report high blood pressure, cholesterol, or glucose.}
\label{fig:sens_badhealth}
\end{figure}

\Cref{fig:sens_mgmtsafety,fig:sens_badhealth} overlays histograms of the $\clate(\bx_{i})$'s values for two different outcomes obtained across ten different runs of \texttt{BCF-LATE} and \texttt{GRF}. 
Each run of \texttt{BCF-LATE} fit 10 chains, each with 1000 iterations of burn in and 1000 iterations saved. 
\texttt{GRF} was run with tune.parameters=``all'', tuning all available tuning parameters by cross-validation. 
\Cref{fig:sens_mgmtsafety} shows the CLATE estimates for the outcome measuring whether subjects felt that their managers prioritized employee wellbeing. \texttt{GRF} identified essentially no heterogeneity (the spike near 0.75 in \Cref{fig:sens_mgmtsafety}) in some runs while identified a wider range of effects in others (the light bars in the figure). 
\texttt{BCF-LATE} on the other hand regularly found effectively the same distribution of CLATE estimates, show by the dark bars all across the figure. 
\Cref{fig:sens_badhealth} shows the CLATE estimates for the outcome of self-reporting high levels of metabolic parameters. \texttt{BCF-LATE} again fairly consistently found two heterogenous subgroups, while \texttt{GRF} varies in how large the negative-CLATE subgroup is. 
Overall, \texttt{GRF} displayed much more variability run-to-run than \texttt{BCF-LATE}, which returned fairly consistent estimates of the effect heterogeneity.

\setcounter{figure}{0}
\setcounter{equation}{0}
\setcounter{table}{0}
\section{Background on the Illinois Workplace Wellness Study}
\label{app:realworld_preprocess}
 \begin{table}[H]
\caption{Variable names and summary statistics for the \textbf{covariates} considered.  These are the baseline survey values obtained in 2016.  The second column is the number of non-missing observations, the third through fifth are sample statistics, and the final column is a description extracted from the survey data codebook.  The mean statistics represent the average number of ``1's'' across available observations since all outcomes are binary variables except for the ``age'' covariate.}
\label{tab:covariate_tab}
\footnotesize
\centering
\begin{tabular}{l|ccccl}
  \vspace{-2mm}
  \\ 
   \textbf{Covariates} & N & Mean & Min & Max & Description \\ \vspace{-2mm} \\
  \hline
   male & 4834 & 0.43 &   0 &   1& Male indicator\\ 
   age & 4834 & 0.99 &   0 &   2& 0 = less than 37, 1 = 37-49, 2 = greater than 49 \\ 
   white & 4834 & 0.84 &   0 &   1& White race indicator \\ 
   everscreen\_0716 & 4834 & 0.89 &   0 &   1& Had at least 1 previous health screening (2016) \\ 
   active\_0716 & 4834 & 0.37 &   0 &   1&Physically active (2016) \\ 
   active\_try\_0716 & 4834 & 0.81 &   0 &   1&Trying to be more active (2016) \\ 
   cursmk\_0716 & 4833 & 0.07 &   0 &   1&Current smoker (2016) \\ 
   othersmk\_0716 & 4833 & 0.09 &   0 &   1&Other (non-cigarette) tobacco use (2016) \\ 
   formsmk\_0716 & 4833 & 0.20 &   0 &   1&Former smoker (2016) \\ 
   drink\_0716 & 4830 & 0.65 &   0 &   1&Drinker (2016)\\ 
   drinkhvy\_0716 & 4829 & 0.05 &   0 &   1&Heavy drinker (2016) \\ 
   chronic\_0716 & 4834 & 0.73 &   0 &   1& Has at least 1 chronic condition (2016) \\ 
   health1\_0716 & 4834 & 0.60 &   0 &   1&Health is excellent or very good (2016)\\ 
   health2\_0716 & 4834 & 0.99 &   0 &   1&Health is not poor (2016)\\ 
   problems\_0716 & 4834 & 0.39 &   0 &   1 & Problems with physical activities or pain (2016)\\ 
   energy\_0716 & 4834 & 0.32 &   0 &   1&Lots of energy (2016) \\ 
   ehealth\_0716 & 4834 & 0.29 &   0 &   1&Emotional health (2016) \\ 
   overweight\_0716 & 4834 & 0.54 &   0 &   1&Overweight status (2016) \\ 
   badhealth\_0716 & 4834 & 0.30 &   0 &   1&High blood pressure, cholesterol, or glucose (2016) \\ 
   sedentary\_0716 & 4833 & 0.54 &   0 &   1&Sedentary job (2016) \\ 
   druguse\_0716 & 4830 & 0.71 &   0 &   1&Taking 1+ prescription/OTC drugs (2016) \\ 
   physician\_0716 & 4833 & 0.76 &   0 &   1&Physician or ER utilization (2016) \\ 
   hospital\_0716 & 4833 & 0.03 &   0 &   1&Hospital utilization (2016) \\ 
  sickdays\_0716 & 4828 & 0.61 &   0 &   1&Sick days in past 12 months (baseline survey) (2016) \\ 
   hrsworked50\_0716 & 4831 & 0.18 &   0 &   1&Worked 50 or more hours per week (2016) \\ 
   jobsatisf1\_0716 & 4832 & 0.40 &   0 &   1&Very satisfied with job (2016)\\ 
   jobsatisf2\_0716 & 4832 & 0.84 &   0 &   1&Very or somewhat satisfied with job (2016) \\ 
   mgmtsafety\_0716 & 4831 & 0.78 &   0 &   1&Management places very high or some priority on \\ &&&&& health/safety (2016) \\ \hline
\end{tabular}
\end{table}

\begin{table}[ht]
\caption{Variable names and summary statistics for the \textbf{outcomes} considered.  The second column is the number of non-missing observations, the third through fifth are sample statistics, and the final column is a description extracted from the survey data codebook.  All outcomes have missing observations since the sample sizes are less than the entire experimental group of 4,834. The mean statistics represent the average number of ``1's'' across available observations since all outcomes are binary variables.} 
\label{tab:outcome_tab}
\footnotesize
\centering
\begin{tabular}{l|ccccl}
  \vspace{-2mm}
  \\ 
   \textbf{Outcomes} & N & Mean & Min & Max & Description \\ \vspace{-2mm} \\
  \hline
  chronic\_0717 & 3565 & 0.73 &   0 &   1 & Has at least 1 chronic condition (2017)\\ 
   badhealth\_0717 & 3567 & 0.32 &   0 &   1 & High blood pressure, cholesterol, or glucose (2017) \\ 
   mgmtsafety\_0717 & 3566 & 0.79 &   0 &   1& Management places very high or some priority on \\ &&&&&  health/safety (2017) \\ 
\hline
\end{tabular}

\end{table}

\end{document}